\documentclass{aa}  

\usepackage{graphicx}
\usepackage{txfonts}
\usepackage{booktabs}
\usepackage{hyperref}  
\hypersetup{linkcolor=blue,citecolor=blue,filecolor=black,urlcolor=blue} 

\begin{document} 

    \title{Fast multi-scale galaxy cluster detection with weak lensing:\\
    Towards a mass-selected sample}

    \titlerunning{Fast multi-scale weak lensing galaxy cluster detection}

   \author{G.~Leroy\inst{1},
           S.~Pires\inst{1},
          G.W.~Pratt\inst{1}
          \and
          C.~Giocoli\inst{2,3}
          }
   \authorrunning{G.~Leroy et al.}
   
   \institute{Université Paris-Saclay, Université Paris Cité, CEA, CNRS, AIM, F-91191, Gif-sur-Yvette, France
              \and 
              INAF - Osservatorio di Astrofisica e Scienza dello Spazio di Bologna, via Piero Gobetti 93/3, I-40129 Bologna, Italy 
              \and
              INFN - Sezione di Bologna, viale Berti Pichat 6/2, I-40127 Bologna, Italy\\
              \email{gavin.leroy@cea.fr}
}

   \date{Received XXXX ; accepted XXXX}

  \abstract{   
The sensitivity and wide area reached by ongoing and future wide-field optical surveys allows for the detection of an increasing number of galaxy clusters uniquely through their weak lensing (WL) signal. This motivates the development of new methods to analyse the unprecedented volume of data faster and more efficiently. Here we introduce a new multi-scale WL detection method based on application of wavelet filters to the convergence maps. We compare our results to those obtained from four commonly-used single scale approaches based on the application of aperture mass filters to the shear in real and Fourier space. The method is validated on Euclid-like mocks from the DUSTGRAIN-pathfinder simulations. We introduce a new matching procedure that takes into account the theoretical signal-to-noise of detection by WL and the filter size. We perform an analysis of the filters, and a comparison of the purity and the completeness of the resulting detected catalogues. We show that equivalent results are obtained when the detection is undertaken in real and Fourier space, and when the algorithms are applied to the shear and the convergence. We show that the multiscale method applied to the convergence is faster and more efficient at detecting clusters than single scale methods applied to the shear. We obtained an increase of 25$\%$ in the number of detections while maintaining the same purity compared to the most up-to-date aperture mass filter. We analyse the detected catalogues and quantify the efficiency of the matching procedure, showing that less than $5\%$ of the detections from the multiscale method can be ascribed to line-of-sight alignments. The method is well-adapted to the more sensitive, wider-area, optical surveys that will be available in the future, and paves the way to cluster samples that are as near as possible to being selected by total matter content.
}

   \keywords{Gravitational lensing : weak ;
                - 
                Galaxies: clusters : general ; 
                -
                methods : data analysis ;
                -
                Cosmology : dark matter, large-scale structure of Universe   
                -  
            }
   \maketitle
   
\section{Introduction}
\label{section_introduction}

Modern cosmological models show that gravitational collapse drives cosmic structure formation through a hierarchical assembly process in which 
objects merge into larger and larger structures. Clusters of galaxies sit at the endpoint of this process, so the formation and evolution of the cluster population directly traces the growth of cosmic structure over time. This information can be used to constrain cosmological models. For example, the number of clusters as a function of mass and redshift is highly sensitive to the underlying cosmological parameters \citep[e.g.][]{White1978, Perrenod1980, Voit2005, Allen2011}. 

The matter content of clusters -- composed of dark matter (DM; 85\%), ionised hot gas in the intracluster medium (12\%), and stars ($\sim3$\%) -- reflects that of the Universe. Clusters are typically detected through their baryonic components. Studies based on samples from X-ray and Sunyaev-Zeldovich (SZ) surveys have been very successful at providing cosmological constraints \citep[e.g.][]{Pacaud2006, Vikhlinin2009, Hasselfield2013, Bocquet2019, Salvati2021}, and increasingly-wide-field optical surveys have become competitive in recent years \citep[e.g.][]{Rozo2010, Hamana2015, Lesci2022}. However, leveraging the data from these surveys requires linking the baryonic observables (optical richness, X-ray luminosity, and SZ flux) to the underlying mass. Today, survey sample sizes are sufficiently large that the dominant uncertainties in cosmological parameter estimation with clusters lie in systematic effects in the mass estimates and the selection function \citep[e.g.][and references therein]{Pratt2019}. 

The vast amount of matter contained in clusters bends the light of background galaxies. This coherent deflection of the path of light from background sources by an intervening mass is termed gravitational lensing \citep[e.g.][]{Schneider1992, Pyne1993}. At large scales, the distortion in the shapes of the background galaxies is detectable only statistically, and is consequently termed weak gravitational lensing. The image distortions due to the weak lensing effect can be characterised by the shear (a warping of the background source images) and the convergence (a magnification effect on the same). Both shear and convergence can provide insights into the statistical properties of the weak lensing field, and has been shown that they contain precisely the same information \citep[e.g.][]{Schneider2002, Pires2020}. The weak lensing effect is directly sensitive to the total projected mass along the line of sight (LOS). As such, it is an attractive method for cluster detection, potentially paving the way towards true mass-selected samples.
However, the sensitivity of cluster detection through the weak lensing signal depends critically on the number of background sources that are available to be lensed by the intervening matter. Only recently have deep, wide-field surveys yielded the background source densities needed to detect substantial numbers of objects. 

A number of methods have been developed to detect galaxy clusters through their weak lensing shear or convergence signal in optical imaging data. 
Application of a simple Gaussian filter to the convergence, combined with thresholding, was widely used in early optical surveys \citep[e.g.][]{White2002, Miyazaki2002, Hamana2004, Tang2005, Gavazzi2006, Miyazaki2007, Fan2010, Shan2012, Shan2018}. Such methods demonstrated that it was possible to detect clusters through their weak lensing signal, while motivating the development of new, more efficient, approaches.
In the subsequent development of optimal filtering techniques, the filter kernel was adapted to include information on the shape of the expected halo profile, while excluding the shape noise and the contribution of large-scale structure (LSS) to the noise budget \citep[e.g.][]{Hennawi2005, Maturi2005, Wittman2006}.

The widely used aperture mass (AM) technique, introduced by \citet{Schneider1996}, consists of convolving the lensing signal with a filter function of a specific scale. A number of new filters have since been developed to maximise the effectiveness of the AM method; they have been tested on simulations and applied to various optical surveys \citep[e.g.][]{Schneider1996, Schneider1998, Jarvis2004, Schirmer2004, Hetterscheidt2005, Hennawi2005, Maturi2005, Maturi2007, Pace2007, Schirmer2007, Dietrich2010, Hamana2012, Lin2016, Miyazaki2017, Hamana2020, Oguri2021}.

The filter function is a key component of the AM method and must be designed to obtain the optimal signal-to-noise ratio (S/N)  at a given scale. A defining characteristic of the above approaches is that they all operate on a single scale. As such, their detection efficiency is highly dependent on the filter design and on its relation to the size of the structures we want to detect. While some studies have proposed using multi-scale wavelet filters to de-noise the convergence map \citep{Starck2006,Lanusse2016}, they were not optimised for cluster detection because they do not take the LSS contribution into account. Nevertheless, such multi-scale de-noising techniques have been shown to be a promising approach for cluster detection \citep{Leonard2015}.

Motivated by the potential of current and upcoming deep wide-field optical surveys with sufficient background source densities, such as the Hyper Supreme-Cam (HSC) survey (\citealt{Aihara2018}), the Legacy Survey of Space and Time (LSST\footnote{\url{https://www.lsst.org/}}; \citealt{Lsst2009,Ivezic2019}), \textit{Euclid}\footnote{\url{https://www.euclid-ec.org/}} (\citealt{Laureijs2011}), and the \textit{Roman} Space Telescope\footnote{\url{https://roman.gsfc.nasa.gov/}} (formerly WFIRST; \citealt{Spergel2015}), we revisit here the question of cluster detection through the weak lensing effect. We perform a quantitative comparison of existing single-scale detection methods, including a complete analysis of the filters. We introduce a new multi-scale detection approach based on the wavelet transform applied to the convergence. We chose to focus on the convergence because this quantity explicitly traces the total matter distribution integrated along the LOS and is computationally less expensive to analyse, making it ideal for application to upcoming large-scale survey data. We quantify the performance of our new multi-scale approach by applying it to the DUSTGRAIN-\textit{pathfinder} simulations detailed in \cite{Giocoli2018}, which feature source densities similar to those expected from the \textit{Euclid} survey. 
We find that the new multi-scale method operating on the convergence is faster and more efficient at detecting clusters than currently used single-scale methods operating on the shear. 

This paper is organised as follows. In Sect. 2 we summarise the key aspects of gravitational lensing. The mock dataset DUSTGRAIN-\textit{pathfinder} simulations are described in Sect. 3.
Section 4 introduces the AM formalism and several commonly used AM filters. A description of the wavelet formalism is also provided.
Section 5 details the detection procedure that we use to compare the different filters. An analysis of the different options in the implementation of the detection algorithm is provided in Sect. 6.
In Sect. 7 we provide the details of the matching procedure that we developed to allow for a fair comparison of the methods. Finally, the performance of the detection methods is evaluated in Sect. 8, and we conclude in Sect. 9.

\begin{figure*}[!ht]
\centering
\includegraphics[ clip, angle=0, keepaspectratio, width=1.\textwidth]{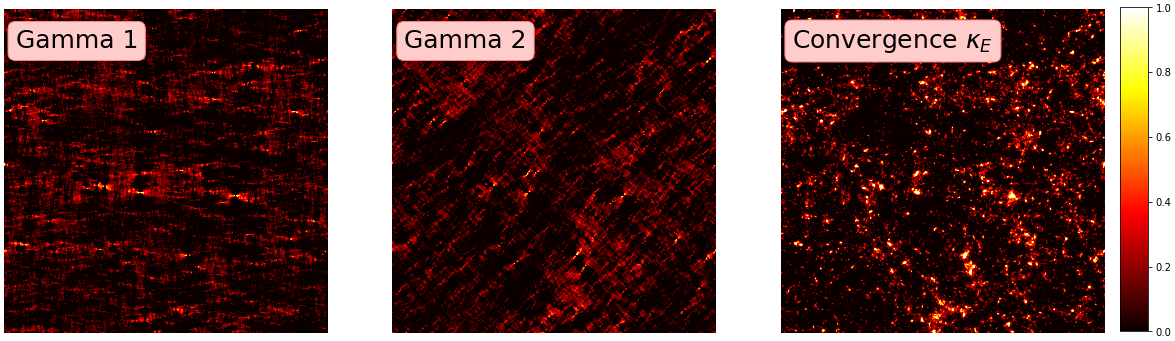}
\caption{Simulated shear maps covering a field  of $5^{\circ} \times 5^{\circ}$ (left and middle panels) and the corresponding E-mode convergence map (right). The E-mode convergence map directly traces the projected matter distribution, the overdensities appearing as bright structures.}
\label{Shearconv07}
\end{figure*}

\section{Weak gravitational lensing theory} 
\label{section_WGL_Theory}
\subsection{Weak lensing theory}
\label{subsection_WL_Theory}
\indent The gravitational field of massive objects affects the path of light in their vicinity.  Thus, the light from background galaxies is deflected as it travels towards us and their images appear distorted. These distortions, or (reduced) shear, are a direct observable and are an imprint of the intervening large-scale matter distribution. Different structures, such as clusters of galaxies, filaments, or even individual galaxies, can act as lenses and create this warping effect. 

We summarise here the gravitational lens theory that is sufficient for the treatment of lensing by galaxy clusters \cite[see e.g.][]{Bartelmann2001}. We consider 
a lens at angular position $\boldsymbol{\theta}$ and at distance $D_{\rm l}$ from the observer. Its surface mass density, $\Sigma(\theta)$, is integrated from its 3D mass density, $\rho(\boldsymbol{\theta},z),$ along the LOS:
\begin{equation}
    \Sigma(\boldsymbol{\theta}) = \int_{0}^{+\infty} \rho(\boldsymbol{\theta},z) dz . 
\end{equation}
From the surface mass density, we can define the lensing potential, $\psi(\boldsymbol{\theta}),$ according to the position of the background galaxy sources with distance $D_{\rm s}$ from the observer and distance $D_{\rm ls}$ from the lens: 
\begin{equation}
    \psi(\boldsymbol{\theta}) = \frac{4 G}{c^2}\frac{D_{\rm l} D_{\rm ls}}{D_{\rm s}} \int_{\rm \mathbb{R}^2}^{} \Sigma(\boldsymbol{\theta}') \ln{( \lvert \boldsymbol{\theta}'-\boldsymbol{\theta} \rvert )}\, d^2\boldsymbol{\theta}'.
    \label{Eqn:psi}
\end{equation}
When light rays travel close to a lens, they are bent, leading to image distortions and, potentially, to multiple images. These images appear with a deflection angle, $\alpha(\boldsymbol{\theta}),$ induced by the lensing potential, $\psi(\boldsymbol{\theta})$: 
\begin{equation}
    \alpha(\boldsymbol{\theta})  = \nabla \psi(\boldsymbol{\theta}).
    \label{Eqn:alpha}
\end{equation}
Here $\alpha(\boldsymbol{\theta})$ is the difference between the angular position, $\boldsymbol{\beta}$, where the images would be without the lens and the observed position, $\boldsymbol{\theta}$. This is summarised by the lens equation: 
\begin{equation}
    \alpha(\boldsymbol{\theta})  = \boldsymbol{\theta} - \boldsymbol{\beta}.
    \label{Eqn:lenseq}
\end{equation}

From Eqs. \ref{Eqn:psi}, \ref{Eqn:alpha}, and \ref{Eqn:lenseq}, the deviation, $\boldsymbol{\beta}$, of the observed image with respect to the undistorted image can be derived: 
\begin{equation}
    A \equiv \frac{\partial \boldsymbol{\beta}}{\partial \boldsymbol{\theta}} = \delta_{ij} - \frac{\partial^2 \psi(\boldsymbol{\theta})}{\partial \theta_i \partial \theta_j}  = 
    \begin{pmatrix} 
    1 - \kappa - \gamma_1 & - \gamma_2 \\
    - \gamma_2 & 1 - \kappa + \gamma_1 
    \end{pmatrix}.
\end{equation}
Here
\begin{gather}
    \gamma_1 = \frac{1}{2} ( \partial_1^2 - \partial_2^2) \psi \ \rm{and} 
    \label{Eqn:gam1}\\
    \gamma_2 = \partial_1\partial_2\psi 
    \label{Eqn:gam2}
\end{gather}
correspond to the two components of the shear $\gamma$, and 
\begin{equation}
    \kappa= \frac{1}{2} ( \partial_1^2 + \partial_2^2) \psi
    \label{Eqn:kappa}
\end{equation}
corresponds to the convergence that is the shape contraction or dilation of the image.

From the above, we can express the convergence as a function of the critical value of the surface mass density, $\Sigma_{\rm crit}$:
\begin{equation}
    \kappa(\boldsymbol{\theta}) = \frac{\Sigma(\boldsymbol{\theta})}{\Sigma_{\rm crit}},
\label{Eqn: convergence}
\end{equation}
with  
\begin{equation}
    \Sigma_{\rm crit} = \frac{c^2}{4\pi G} \frac{D_{\rm s}}{D_{\rm l} D_{\rm ls}}.
    \label{Sigma_crit}
\end{equation}
Thus, the convergence is a tracer of the total matter distribution integrated along the LOS, weighted by the redshift-dependent factor $1/\Sigma_{\rm crit}$.

The direct observable is the reduced shear, $g$, that is derived from the ellipticities of the observed background galaxies.
The reduced shear, $g$, is defined as follows:
\begin{equation}
g = \frac{\gamma}{1 - \kappa}
.\end{equation}
In the weak lensing regime ($\kappa \ll 1$), it approximates the shear, $\gamma$.

\subsection{Application}
\label{subsection_Application_KS}
In practice, we observe a distorted image, from which we measure the galaxy ellipticities and estimate the shear components, $\gamma_1$ and $\gamma_2$. \cite{KandS1993} introduced the mass inversion technique, which involves the computation of the convergence map $\kappa$ from the measured shear field. 

We can consider a complex notation to represent the shear field, $\gamma = \gamma_1 + \rm{i} \gamma_2 $, and the convergence field, $\kappa = \kappa_E + \rm{i}\kappa_B$, with $\kappa_E$ and $\kappa_B$, called E- and B-modes. 
Then, taking the Fourier transform of Eqs. \ref{Eqn:gam1}, \ref{Eqn:gam2}, and \ref{Eqn:kappa}, we obtain \begin{gather}
    \hat{\gamma} = \hat{P}\, \hat{\kappa},\\
    {{\rm with:} \, \,} \hat{P} = \hat{P}_1 + i\,\hat{P}_2 \rm{\,\,and\,\,}, \\
    \hat{P}_1 = \frac{ k_1^2 - k_2^2}{k_1^2 + k_2^2},\\
    \hat{P}_2 = \frac{ 2\,  k_1^2 k_2^2}{ k_1^2 + k_2^2},
\end{gather}
where the hat symbol refers to the Fourier transform,  and $k_i$ is the wave number at the angular position, $\theta_i$.

Considering the conjugate $\hat{P}* = \hat{P}_1 - i \hat{P}_2$, we can reconstruct $\kappa_E$ and $\kappa_B$ from the complex shear $\gamma$, obtaining  
\begin{gather}
    \hat{\kappa}_E = \hat{P}_1 \hat{\gamma}_1  + \hat{P}_2 \hat{\gamma}_2 \\
    \hat{\kappa}_B = -\hat{P}_2 \hat{\gamma}_1  + \hat{P}_1 \hat{\gamma}_2. 
\end{gather}
Given that $\psi$ is a scalar potential, it can be shown that weak lensing  does not in principle produce B-modes. 
Thus, B-modes can be used to estimate the level of the noise in the data.

Figure \ref{Shearconv07} shows an example of simulated shear and convergence maps without shape noise derived from the DUSTGRAIN-\textit{pathfinder} simulation (described in Sect. \ref{section_simulation}) covering a field of $5 ^{\circ} \times 5^{\circ}$. The left and middle panels correspond to the two components of the shear, $\gamma_1$ and $\gamma_2$ and the right panel to the corresponding E-mode convergence map, $\kappa_E$. Although  both the shear and the convergence can be used for cluster detection, the convergence is easier and computationally less expensive to analyse because it is a scalar field that is proportional to the projected matter distribution.

\section{Weak lensing simulations}
\label{section_simulation}
\subsection{The DUSTGRAIN N-body simulations}

We make use of a suite of cosmological N-body simulations called The DUSTGRAIN-\textit{pathfinder} \citep[see][for a detailed description]{Giocoli2018}. This DM-only simulation traces the collisionless evolution of $783^3$ DM particles with a mass $m_{\rm CDM} = 8.1 \times 10^{10}$ $h^{-1}$\,M$_\odot$\,, contained within a periodic cosmological box of side 750 $h^{-1}$\,Mpc. 
In the present work, we use a subset of the full DUSTGRAIN-\textit{pathfinder} runs consisting of 256 realisations sharing the same standard cosmological parameters in agreement with \citealt{Planck2016}: $\Omega_M = \Omega_{\rm CDM} + \Omega_{\rm b} + \Omega_{\nu} = 0.31345$, $\Omega_{\rm b} = 0.0481$, $\Omega_{\Lambda} = 0.68655$, $h = 0.6731,$ and $\sigma_8 = 0.847$. The different LOS realisations were obtained by randomising the stacked comoving cosmological boxes through combinations of the following procedures: (i) changing the sign of the Cartesian coordinates, 
(ii) redefining the position of the observer, and (iii) modifying the order of the axes in the coordinate system. By construction, these variations preserve the clustering properties of the particle distribution 
at the scale of the comoving simulation snapshot. 

\subsection{Mock dark matter halo catalogues}

The mock DM halo catalogues were generated by identifying the DM haloes in the DUSTGRAIN simulation through a friends-of-friends algorithm (\citealt{Davis1985}) with linking distance $\lambda = 0.16 \times d$, where $d$ is the mean separation distance between particles. The \verb?SUBFIND? algorithm (\citealt{Springel2001}) was then used to evaluate the standard parameters of each friends-of-friends-identified halo: the redshift, $z$, and the virial mass, $\rm M_{200c}$, and radius $\rm R_{200c}$, corresponding to the mass and radius of a spherical region around the fiducial centre of each halo enclosing 200 times the critical density of the Universe. 
Thus, for each of the 256 realisations, the catalogue is composed of the positions (i.e. the right ascension and the declination) of the identified DM haloes, their estimated redshift, $z_l$, and their virial mass, $\rm M_{200c}$, and radius, $\rm R_{200c}$. This DM halo catalogue is used in our study for the matching procedure to quantify the purity and completeness of each selected sample.

\subsection{Mock galaxy catalogues}
\label{mock}

The particles stored in 21 different snapshots were used to construct continuous past light cones from $z = 0$ to $z = 4$ using the \verb?MAPsim? pipeline \citep{Giocoli2014,Giocoli2017}. The routine extracts the positions of each particle to recreate the past light cone and at the same time, the particles are binned to build $27$ different lens planes to recompose the projected matter density distribution. Then, the shear of each galaxy is computed by projection, using the Born approximation, which assumes unperturbed light paths to integrate the lensing distortions \citep{Bartelmann2001}. The redshift distribution of the galaxies, $n(z)$, was built to follow a realistic distribution approaching the one expected for the \textit{Euclid} survey (e.g. \citealt{Cropper2013}). Each realisation contains around 3 million galaxies in a field of view of $5^{\circ} \times 5^{\circ}$, with a galaxy distribution extending up to $z = 4$. This leads to a galaxy density of about $n_{\rm{g}} = 30$ gal. arcmin$^{-2}$.

Noise was added to the shear to mimic realistic surveys. Uncertainties in the shear (referred to as `shape noise') arise from a combination of the unavoidable intrinsic shape of the galaxies (referred to as the `intrinsic shape noise'), and measurement errors that include, among other factors, uncertainties in the galaxy shape measurement and the point spread function correction. The intrinsic shape of the galaxies being the dominant component and the galaxies being randomly distributed, the shape noise can be modelled as an additive noise. This can be well approximated by a Gaussian distribution with a mean of $\mu = 0$ and a standard deviation of $\sigma_{\rm{\epsilon}} = 0.26$ \citep[e.g.][]{Leauthaud2007,Schrabback2015,Schrabback2018}. For each galaxy, the shape noise was therefore included in the catalogue by adding Gaussian noise to the two components of the shear. For each of the 256 realisations, the galaxy catalogue is composed of: (i) the positions (i.e. the right ascension and the declination) of the galaxies, (ii) the two components of the shear, and (iii) the redshift. These noisy galaxy catalogues are then used as the inputs to evaluate the detection algorithms described below.


\section{Filters}
\label{section_filters}

In this section we present several different filters that have been used for weak lensing cluster detection in previous studies, and we compare them to the wavelet filters that we use in our new multi-scale detection approach. 
A complete analysis and comparison of the different filters both in real and Fourier space is also provided.

Although used since the very first weak lensing detection algorithms \citep[e.g.][]{White2002, Miyazaki2002, Hamana2004, Tang2005, Gavazzi2006, Miyazaki2007, Fan2010, Shan2012, Shan2018}, we excluded the Gaussian filter from this analysis because it behaves as a low-pass filter and is thus not optimal for cluster detection.
Although low-pass filters appear to be natural tools for reducing the shape noise that dominates at high-frequencies, they are not adapted to diminish the low-frequency signal coming from LSS, which causes many spurious peaks \citep[see e.g.][]{Maturi2005}. Since the use of such low-pass filters is sub-optimal for the current application, we focus our analysis on pass-band filters that are better able to target the signal around a given scale.
\subsection{The aperture mass formalism}

\label{aperturemass}

The AM method \citep{Schneider1996} is commonly used to reconstruct maps of the projected matter distribution at a given scale. The AM map can be evaluated at a position $\boldsymbol{\theta_{\rm o}}$ by convolving the convergence field, $\kappa$, with a filter, $U$, such that
\begin{equation}
    M_{\rm ap}(\boldsymbol{\theta_{\rm o}}) = \int_{\mathbb{R}^2} \kappa(\boldsymbol{\theta})\, U(\lvert\, \boldsymbol{\theta} \, -\, \boldsymbol{\theta_{\rm o}} \,\rvert)\, d^2 \boldsymbol{\theta}, 
    \label{Eqn:Map}
\end{equation}
where $U$ must be compensated; that is,
\begin{equation}
    \int_{0}^{\rm \theta_{\rm ap}}\rm \theta' \, U(\rm \theta')\,d\theta' = 0
\end{equation}
must be fulfilled within the aperture of radius $\theta_{\rm ap}$. Alternatively, the AM map can also be computed from the tangential component of the shear, $\gamma_{\rm t}$, which is a quantity that can directly be computed from $\gamma_1$ and $\gamma_2$:
\begin{equation}
    \gamma_{\rm t} = \gamma_1 \cos (\rm \phi) + \gamma_2 \sin (\rm \phi),
\end{equation}
where $\phi$ is the polar angle $\phi(\boldsymbol{\theta}, \boldsymbol{\theta_0)}$ relative to the centre of the aperture, $\boldsymbol{\theta_0}$. The AM map at position $\boldsymbol \theta_{\rm o}$ can then be obtained by convolving the tangential shear with a filter, $Q$, such that
\begin{equation}
    M_{\rm ap}(\boldsymbol{\theta_o})= \int_{\mathbb{R}^2} \gamma_t(\boldsymbol{\rm \theta})\, Q(\lvert\, \boldsymbol{\rm \theta} \, -\, \boldsymbol{\rm \theta_{\rm o}} \,\rvert)\, d^2 \boldsymbol{\rm \theta}, 
\end{equation}
the filters $Q$ and $U$ being linked by the relations \citep[]{Schneider1997} 
\begin{equation}
    Q(\rm \theta) = \frac{2}{\rm \theta^2} \int_{0}^{\rm \theta} \rm \theta' U(\rm \theta')\,d\theta' - U(\rm \theta)
\label{Eqn:UtoQ}
\end{equation}
and 
\begin{equation}
    U(\rm \theta) = 2 \int_{\theta}^{ +\infty} \rm \frac{Q(\rm \theta')}{\theta'} \,d\theta' - Q(\rm \theta).
\label{Eqn:QtoU}
\end{equation}

The performance of the AM technique at finding clusters on a map depends on the exact choice of filter, and on its capacity to isolate the lensing signal from the shape noise and the LSS component. The optimal filters $U$ or $Q$ should be designed to fulfil two properties. In real-space, the filters should be local, that is, decrease smoothly to zero within a finite radius. Ideally, they should fit the shape of the structures of interest. 
Moreover, they should be local in Fourier space, focusing on a particular angular scale. 
Ideally, their size should match as closely as possible the size of the structures of interest.

A number of functions or families of functions have been proposed \citep[e.g.][]{Schneider1996, Schirmer2004Thesis, Schirmer2007, Jarvis2004, Hamana2012, Miyazaki2017}. In the present study, we consider the following four well-known AM filters. 

The first is the original AM family of filters defined in the paper by \citet[hereafter $\rm S96$]{Schneider1996}:
\begin{equation}
\label{Eqn:USchneider}
    U_{\rm S96}(\theta) = \left\{
    \begin{array}{ll}
        1 & \mbox{x $\in$ [0, $\nu_1R$],}\\
        \frac{1}{1-c} \big[\frac{\nu_1R}{\sqrt{(x-\nu_1R)^2+(\nu_1R)^2}} \big] & \mbox{x $\in$ [$\nu_1R$, $\nu_2R$],} \\
         \frac{b}{R^3} (R-x)^2\,(x-\alpha R) & \mbox{x $\in$ [$\nu_2R$, $R$],} \\
        0 & \mbox{x > $R$},
    \end{array}
\right. 
\end{equation}
where parameters $\alpha$, $b,$ and $c$ ensure that the filter is compensated and continuous in real space. The parameters $\nu_1$, $\nu_2$, and $R$ define the angular extent of the filter and must be adapted to the size and shape of the clusters we want to detect. We implemented the different parameterisations presented in \citet[]{Schneider1996} and we use $\nu_1 = 0.1$, $\nu_2 = 0.9$, $R = 9\arcmin$, $\alpha = 0.8531$, $b = -329.8,$ and $c= 0.2415$, which we found to maximise the purity and completeness. The corresponding $Q_{\rm S96}$ can be calculated analytically with Eq.~\ref{Eqn:UtoQ}.

The second is the filter developed by \citet[hereafter $\rm TANH$]{Schirmer2004Thesis}:
\begin{equation}
    Q_{{\rm TANH}}(\theta) = \frac{1}{1+ {\rm e}^{6-150\,\frac{\theta}{R}}+ {\rm e}^{-47+50\,\frac{\theta}{R}}} \frac{\tanh\left(\frac{\theta}{x_c R}\right)}{\frac{\theta}{x_cR}} ,
    \label{Eqn:QSchirmer}
\end{equation}
where $R$ is the truncation radius and $x_c$ (dimensionless) defines the width of the filter. This filter was directly derived from the shape of the tangential shear of the Navarro-Frenk-White (NFW) profile \citep[]{Navarro1996}. Following \cite{Schirmer2007}, we used $x_c = 0.1$. We fixed the truncation radius $R = 7\arcmin$ as the optimal choice to maximise the number of detections, following the procedure described by \cite{Hetterscheidt2005}. 
The corresponding $U_{\rm TANH}$ can be computed numerically using Eq.~\ref{Eqn:QtoU}. 

The third is the filter proposed by \citet[hereafter J04]{Jarvis2004}:
\begin{equation}
    U_{\rm J04}(\rm \theta) = \frac{1}{\rm 2\pi \sigma^2}\, \left(1-\frac{\rm \theta^2}{\rm 2 \sigma^2}\right)\, \exp{\left(-\frac{\rm \theta^2}{2 \sigma^2}\right)},
    \label{Eqn:UJarvis}
\end{equation}
\begin{equation}
    Q_{\rm J04}(\rm \theta) = \frac{1}{\rm 2\pi \sigma^2}\frac{\rm \theta^2}{\rm 2 \sigma^2}  \exp{\left(-\frac{\rm \theta^2}{\rm 2 \sigma^2}\right)},
    \label{Eqn:QJarvis}
\end{equation}
where $U_{\rm J04}$ was defined following \cite{Waerbeke1998} as the second derivative of a Gaussian function with standard deviation $\sigma$. The $U_{\rm J04}$ is also known as the Mexican hat wavelet filter.
In the following mass aperture computation, we use $\sigma = 4 \arcmin$, corresponding to an apparent angular radius of about $4^\prime$, and a truncation radius $R = 20 \arcmin$, as suggested by \citet{Leonard2012} to minimise oscillations and high frequency mode contamination. 

The fourth is the filter introduced by \cite{Hamana2012} and used by \citet[hereafter M18]{Miyazaki2017}, which is defined as a truncated Gaussian filter:
\begin{equation}
    U_{\rm M18}(\rm \theta) = \frac{1}{\rm \pi\,\theta_s^2}\,\exp{\left(\frac{\rm \theta^2}{\rm \theta_s^2}\right)}\,-\, U_{\rm o},
    \label{Eqn:UHamana}
\end{equation}
where $U_0$ is a parameter to ensure that the filter $U$ is compensated and
\begin{equation}
    Q_{\rm M18}(\theta) = \frac{1}{\rm \pi\,\theta^2}\bigg[1\,+\,\left(1-\frac{\rm \theta}{\rm \theta_s}\right)^2\bigg]\,\exp{\left(\frac{\rm \theta^2}{\rm \theta_s^2}\right)},
    \label{Eqn:QHamana}
\end{equation}
with $\theta_s$ being defined as the angular scale of the aperture. In practice, the choice of $\theta_s$ influences the angular radius of the cluster the filter targets, but the relation is not direct. Following \citet{Miyazaki2017}, we use $\theta_s = 1.5 \arcmin$, and  $U_{\rm M18}(\theta) = 0$ if $\theta > R$, with a truncation radius $R = 15 \arcmin$. We fixed this value of $R$ to reduce the impact of the truncation on the filter behaviour as it derives from a Gaussian filter. Indeed, the filter $U_{\rm M18}$ reaches negative values at $U_{\rm M18}(R)$ and is subsequently forced to $0$ by the truncation. This step in the function results in contamination in the signal. As the choice of truncation radius strongly impacts the computation time, there is a clear trade off between the computation time and the filter behaviour.

\subsection{The wavelet formalism}
\label{subsection_wavelet_formalism}

The wavelet formalism can also be used to reconstruct maps of the projected matter distribution at a given scale \citep[see e.g.][]{Leonard2012, Pires2012}. In \cite{Leonard2012}, the authors showed that wavelet filter functions at a given scale are formally identical to AM filter functions at that scale. Similar to the AM functions, many different wavelet functions exist, including starlet, Mexican hat, Morlet, and biorthogonal \citep[see e.g.][for reviews]{starck1998image,Starck2006}. Many wavelet functions have been specifically designed to fulfil the properties of localisation in real and Fourier spaces, which is a distinct advantage compared to the AM approach.

A wavelet map of the convergence $\kappa$ at position $\theta$ and scale $a$ can be computed as%
\begin{equation}
    W_{a}(\theta) = \frac{1}{\sqrt{a}} \int \kappa(t)\, \Psi_{a}(\,t-\theta\,) \,dt,
\end{equation}
where $\Psi_{a}$ is the wavelet filter of scale $a$. All wavelet filters are defined such that they respect the condition of zero mean. Therefore, they are compensated by definition. 

\begin{figure*}[!h]
\centering
\includegraphics[width=\hsize]{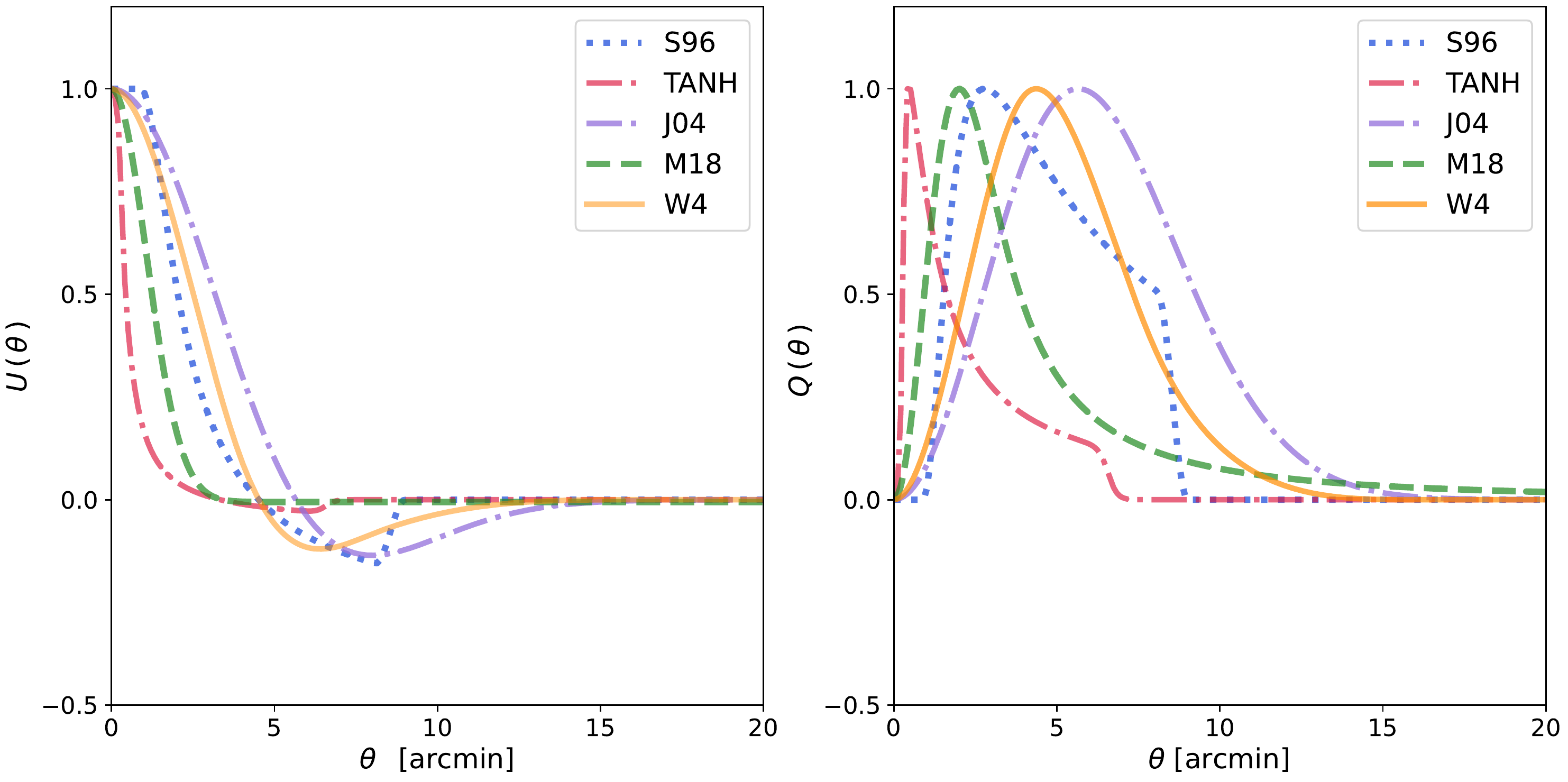}
\caption{Comparison of different filter functions, $U$ (left) and $Q$ (right), for AM map reconstruction. 
The figure shows in blue (dotted lines) the AM filter functions defined by \citet[][S96]{Schneider1996} with $R = 9 \arcmin$; in red (dash-dotted lines) the AM filter functions defined in \citet[][TANH]{Schirmer2004Thesis} with $R = 7 \arcmin$; in purple (dash-dotted lines) the AM filter functions defined in \citet[][J04]{Jarvis2004} with $\sigma = 4 \arcmin$; and in green (dashed lines) the AM filter functions defined by \citet[][M18]{Miyazaki2017}. The filter $U_{\rm M18}$ is also a compensated filter thanks to the $U_0$ term, but due to the high truncation radius we used, this is not distinctly visible. Finally, the AM filter functions corresponding to the wavelet filter functions at scale $i = 4$ are displayed in orange (solid lines). All the filters are normalised to have a maximum amplitude of 1.}
\label{U and Q AM all}
\end{figure*}

Another advantage of the wavelet formalism is that it can decompose a given image into several complementary scale components. This is possible using the wavelet transform, for which fast algorithms exist. For our study, we use the isotropic un-decimated wavelet transform, also called the starlet transform \citep[e.g.][]{starck1998image,Starck2006}, which is able to simultaneously compute several wavelet maps on dyadic scales. The starlet functions are isotropic, which makes them well suited to extracting galaxy clusters, which appear roughly circular in the convergence maps. As such functions decompose an image into several complementary scales (see Sect.~\ref{subsection_filterAnalysis}), they are ideal for undertaking  a multi-scale analysis and for targeting structures of different angular sizes. 
 The starlet transform decomposes the convergence as follows:
\begin{equation}
\kappa(\theta) = C_J(\theta) + \sum_{i=1}^J W_{i}(\theta)
\label{waveletdecomposition}
,\end{equation}
where $J$ is the number of scales of the decomposition, $C_J$ is the corresponding smoothed version of the convergence $\kappa$, and $W_i$ are the wavelet maps targeting clusters with an apparent angular radius of $2^{i-1}$ times the size of the pixel. As a consequence, the wavelet scales are fixed by the pixel size, and, by definition, increase as a power of two.

In the starlet transform, the wavelet functions are defined as the difference between two $B^3$-spline functions at different resolutions. Their application to the convergence map was shown to be equivalent to applying the following AM filter \citep{Leonard2012}:

\begin{multline}
    U_{W_i}(u) = \frac{1}{9}\, \bigg[\,93\, \left\lvert u\right\rvert^3\, - \,64\, \left(\,\left\lvert\frac{1}{2}\,-\,u\,\right\rvert^3\,+\,\left(\,\frac{1}{2}\,+\,u\,\right)^3\right) \\
    + 18\,\left(\, \left\lvert\,1\,-\,u\,\right\rvert^3\, +\, \left\lvert\,1+ \,u\,\right\rvert^3\,\right) - \frac{1}{2}\,\left(\,\left\lvert\,2-\,u\,\right\rvert^3+\left\lvert\, 2+\,u\,\right\rvert^3\right) \bigg],
    \label{equation_ondelette}
\end{multline}
where 
\begin{equation}
u = \frac{\theta}{2^i x}, 
\label{Eqn:wavelet_scales}
\end{equation}
and $x$ is the pixel size. The new multi-scale detection method that we introduce in Sect.~\ref{subsection_multi-scale_detection_procedure} takes advantage of this multi-scale decomposition, the wavelet transform being able rapidly to decompose the convergence into different scales.

\subsection{Filter analysis}
\label{subsection_filterAnalysis}

The performance of each detection method is closely linked to the capacity of the filter functions to reduce the shape noise and the contribution from LSS, while simultaneously minimising the signal loss. In the following, we analyse the properties of the filters described above both in real and Fourier spaces to help in the interpretation of the results presented in Sect.~\ref{section_results}.

\begin{figure}[!ht]
\centering
\includegraphics[width=\hsize]{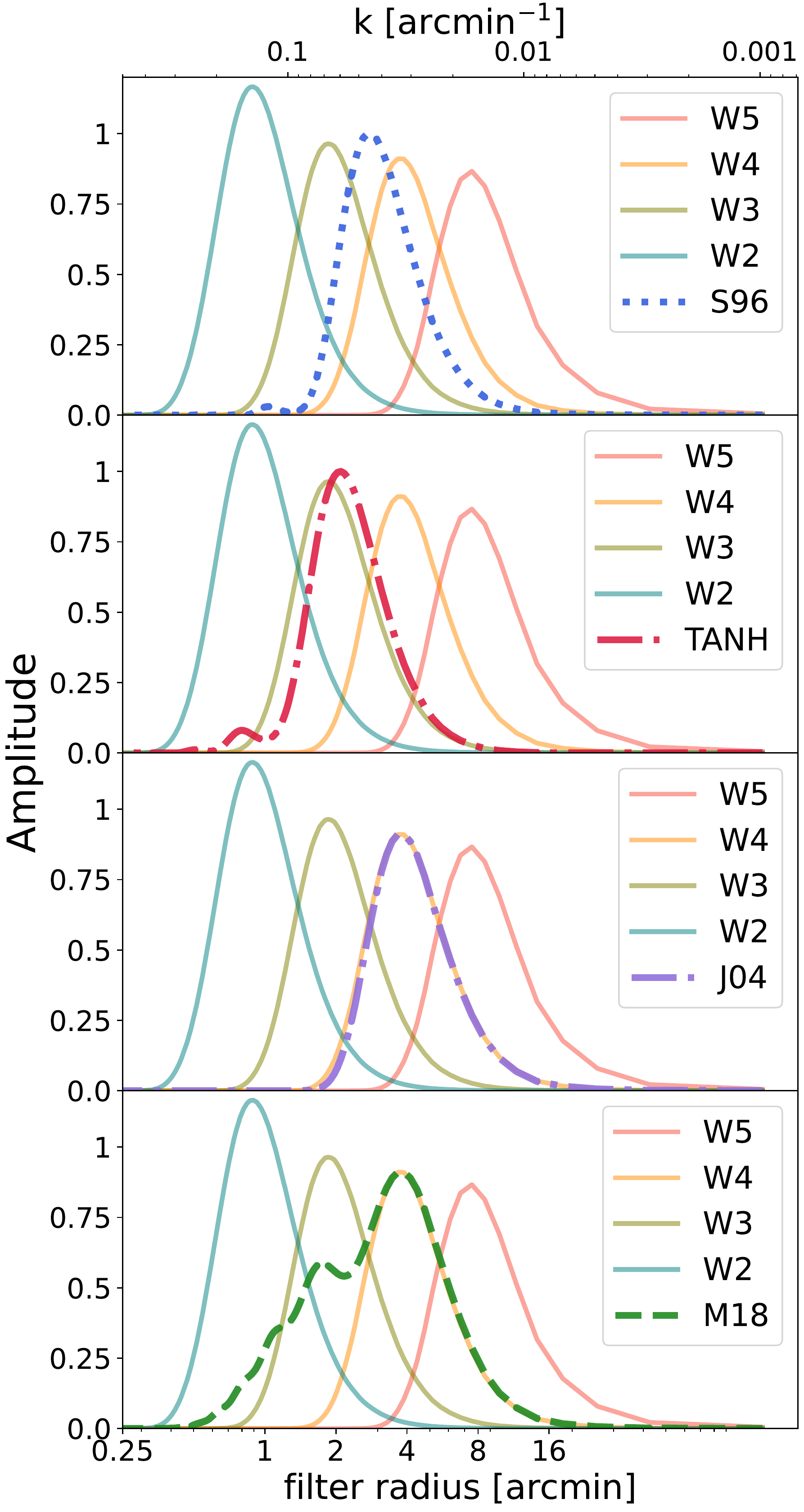}
\caption{Fourier space comparison of several AM filters and wavelet filters, expressed as a function of wave number $k$ in arcmin$^{-1}$ and filter radius in arcmin. The dotted blue line (upper panel) represents the S96 filter, the dash-dotted red and purple lines (middle panels) represent the TANH and J04 filters, and the dashed green line (lower panel) corresponds to the M18 filter. The solid lines show the different wavelet filters: the teal, green, orange, and red light lines correspond to the W2, W3, W4, and W5 wavelet filters with radii of $1\farcm17$, $2\farcm32$, $4\farcm68$, and $9\farcm37,$ respectively. For comparison purposes, the J04 and M18 filters are forced to the same amplitude as the W4 function to illustrate their behaviour in terms of filter radius.}
\label{CompareFall}
\end{figure}

In Fig.~\ref{U and Q AM all} we compare the different filter functions $Q$ and $U$, defined in Sects.~\ref{aperturemass} and~\ref{subsection_wavelet_formalism}, in real space. This comparison underlines the differences in filter shape. In particular, we focus our attention on their compensated and local properties. 
The dotted blue curves correspond to the AM functions defined in \citet[][]{Schneider1996}. The filter $U_{\rm S96}$ is compensated and local by design.
The dash-dotted red curves correspond to the AM functions defined in \citet[][]{Schirmer2004Thesis}. The filter $U_{\rm TANH}$ drops rapidly to zero, but it is not local and is only compensated at infinity. The localisation is imposed by a truncation radius at $R = 7'$.
The dash-dotted purple curves correspond to the AM functions defined in \cite{Jarvis2004}. This filter $U_{\rm J04}$ is also not local, and is also only compensated at infinity since it tends to zero when $\theta$ tends to $+\infty$. In practice, the filter is truncated at R=$20\arcmin$ to have a local support. The dashed green curves correspond to the AM functions defined in \cite{Miyazaki2017}. In the definition of $U_{\rm M18}$, $U_0$ ensures that the filter is compensated within an aperture of radius $R$. 
However, the $Q_{\rm M18}$ function is non-local, which can introduce errors when  truncating the aperture. 
The solid orange curves correspond to the wavelet filter function $U_{W_4}$ defined in Eq.~\ref{equation_ondelette} for $i = 4$ (hereafter W4).
The wavelet filter is shown to be compensated within an aperture radius equal to two times the filter radius. 

It is also instructive to study the representation of the filter functions in Fourier space as this allows us to fully understand how the filters are designed to locate structures. In Fourier space, an optimal band-pass filter should be local, in order to select only a specific range of frequencies around the scale of interest, and thus reduce the noise and the contribution from LSS.
The design of AM filters can be delicate, and some such filters suffer from oscillations in Fourier space owing to truncations applied in real space that deteriorate the band-pass ability of the filter \citep[see e.g.][for more details]{Leonard2012}. In contrast, many wavelet functions have been specifically designed to fulfil the properties of localisation in real and Fourier space. 

To analyse the filters precisely in Fourier space, we studied their impulse response. For this purpose, we simulated a null convergence map with a single peak at the centre. Then we passed this convergence map through the four AM filters and each of the wavelet functions, and computed the corresponding power spectra in Fourier space. The resulting power spectra are compared in Fig.~\ref{CompareFall}. 
This representation highlights each characteristic filter radius, corresponding to the characteristic size of the structures the filter targets. 
The 3\arcmin, radius of the S96 filter falls
between that of the W3 and W4 wavelet filters.
The TANH filter, with a radius of about 2\arcmin, matches that of the W3 wavelet filter.
The J04 filter, which has a radius of about 4\arcmin, matches that of the W4 wavelet filter.
Finally, the M18 filter can be compared to a combination of the W3 and W4 wavelet filters.
In the Fourier domain, we can clearly see the band-pass behaviour of the different functions. 
In particular, the S96 and TANH filters show oscillatory behaviour, which can be explained by their design.

\begin{figure*}[!ht]
\centering
\includegraphics[ clip, angle=0, keepaspectratio, width=1.\textwidth]{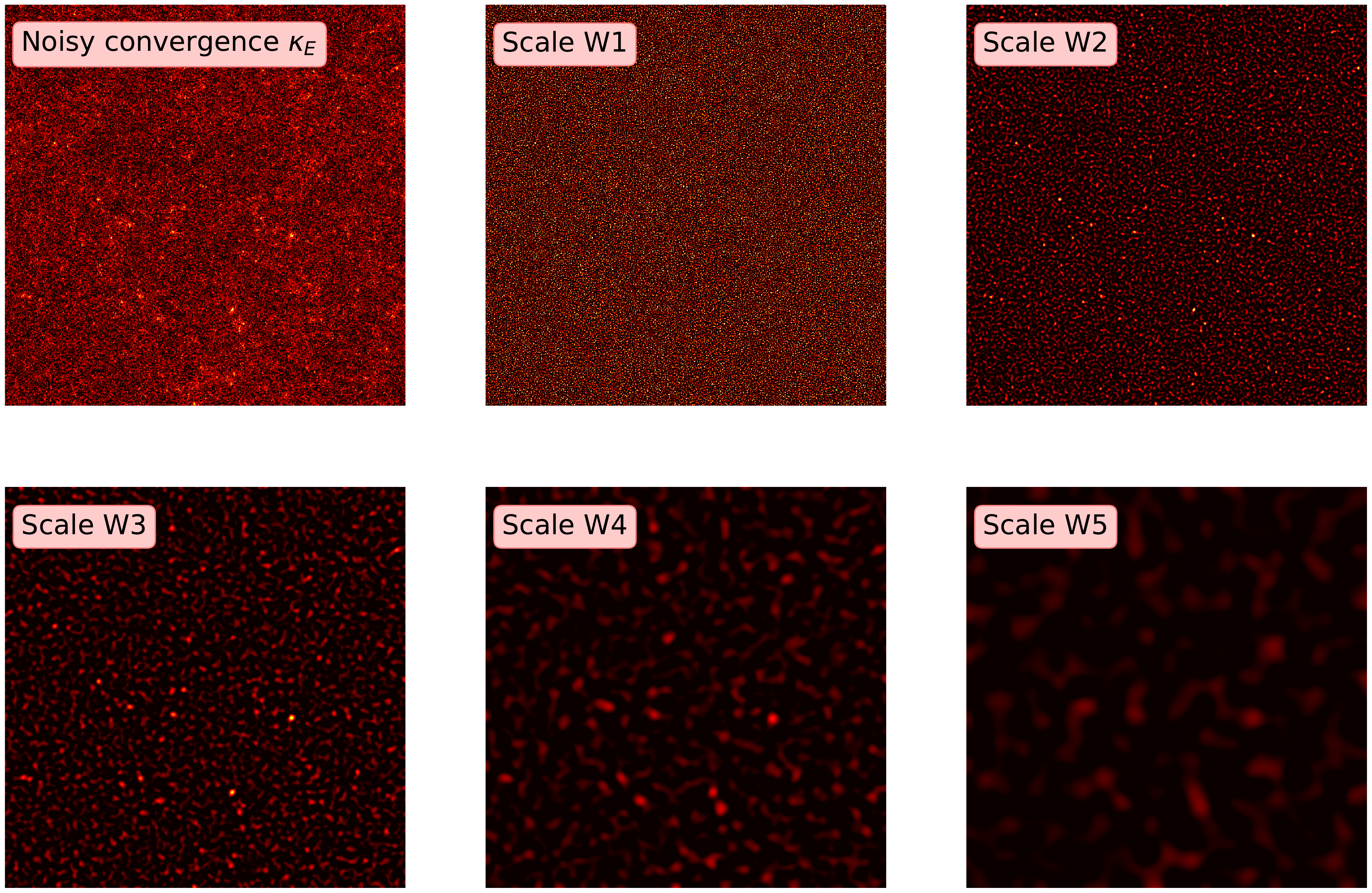}
\caption{Wavelet decomposition of an E-mode convergence map (including shape noise), covering a simulated field of $5^\circ\times5^\circ$. The upper-left panel corresponds to the E-mode convergence map to be decomposed. The other panels correspond to the wavelet maps at scale W1 = $1\farcm17$ (upper-middle panel), scale W2 = $2\farcm32$ (upper-right panel), scale W3 = $4\farcm68$ (lower-left panel), scale W4 = $9\farcm37$ (lower-middle panel), and scale W5 = $18\farcm75$ (lower-right panel). The wavelet maps are sensitive to structures of different apparent angular sizes. The W1 map (upper-middle panel) appears to be dominated by the shape noise. The wavelet decomposition also includes the smoothed version, which we excluded from this representation.
}

\label{Wavelets}
\end{figure*}

This Fourier space representation further shows that the selected scales of the wavelet filter appear to be well defined. In the wavelet formalism, the wavelet filter functions have complementary dyadic scales, defined by the pixel resolution as explained in Sect. \ref{subsection_wavelet_formalism}. 
Their comparison with the four AM filters highlights that a wavelet filter at a given scale can directly be compared to an AM filter. However, unlike AM filters, which focus essentially on one scale, the wavelet formalism allows us to extract signal at more than one complementary scale. 
Generally, the standard AM methods target clusters located at relatively low redshift with a large apparent angular size. Thus, typical AM filter radii are taken between $1\arcmin$ and $5\arcmin$ \citep[e.g.][]{Hamana2012} to maximise the S/N for clusters with a similar angular extent. 
In contrast, the lensing signal from clusters with different apparent angular sizes can easily be targeted using the wavelet formalism. 

\section{Detection algorithms}
\label{section_detectionAlgorithms}

In this section we discuss the implementation of the detection approach we used to compare the filters. The detection algorithms are all based on the same basic principles. The lensing signal is first convolved with a filter function of a specific scale (described in the previous section) to reconstruct E- and B-mode AM maps. Thresholding is then applied to the E-mode AM map to locate the overdensity peaks corresponding to clusters.

\subsection{Binning and filtering}
\label{subsection_filtering}

In the following, we describe the steps used to produce the E- and B-mode AM maps on which the detection is performed. As explained in Sect. \ref{mock}, the noisy galaxy catalogues are the basic inputs for the detection algorithms. These contain the shear at the discrete positions of each of the galaxies. 

At this stage the convolution by the filter function can be undertaken in real space, on the shear directly at the position of the galaxies, or in Fourier space, by binning the shear on a regular grid to build a shear map. A third possibility is to reconstruct the convergence from the shear maps and then to perform the filter convolution on the resulting convergence map. A comparison of these approaches is provided in Sects. \ref{subsection_realvsfourier} and \ref{subsection_shearvsconvergence}, where we show that we obtain statistically similar results irrespective of approach. 
For the following, we use the third approach because it is computationally faster, in particular for wavelet filters, with the use of the wavelet transform.

We first binned the observed galaxy ellipticities on a regular grid to create what we refer to as the noisy shear maps. In our study, the simulated field being $5^{\circ} \times 5^{\circ}$, we decided to bin the galaxies in a grid of $512 \times 512$ pixels, yielding a pixel size of $0\farcm58$. On average, with a galaxy density of 30 galaxies per arcmin$^{2}$, about 10 galaxies fall into each pixel. 
The standard method for binning the shear on a regular grid consists of simply calculating the average shear per pixel \citep[see e.g.][]{KandS1993, VanWaerbeke2000, Pires2020}. 
In this paper, we introduce a new binning strategy to be closer to the real space approach. This consists of summing the shear of all the galaxies that fall into each pixel. Thus, the shear in each pixel is effectively weighted by the number of galaxies in that pixel. This binning strategy is more adapted to weak lensing detection because additional information on the galaxy density is included in each pixel. 
The resulting shear map can then be normalised by the mean number of galaxies per pixel to facilitate the comparison with other approaches. However, this global normalisation has no impact on the detection because it affects the E- and B-modes in a similar way. The impact of these two different binning approaches is discussed in more detail in Sect.~\ref{subsection_realvsfourier}. The disadvantage of this approach is that the weak lensing signal depends significantly on the number of galaxies per pixel, complicating its interpretation in terms of mass. However, such calculations can be undertaken in a second step, after detection.

We then applied the Kaiser \& Squires inversion described in Sect. \ref{subsection_Application_KS} to the noisy shear maps $\gamma_1$ and $\gamma_2$, to reconstruct the E- and B-mode convergence maps. Convolution by the different filters was then performed both on the E- and B-mode convergence maps, to produce the corresponding AM maps. 

In the following, we refer to maps of the projected matter distribution at a given scale as AM maps. In practice, these can be obtained by applying AM filters or wavelet filters to the shear or to the convergence. The procedure to detect the haloes on these E- and B-mode AM maps using a single scale is identical, regardless of the method that has been used to produce them.

\subsection{Single-scale detection procedure}
\label{subsection_detectionProcedure_1scale}

Once the E- and B-mode AM maps at a specific
scale are produced, the detection procedure starts by applying a thresholding step. Since the weak lensing effect produces only E-modes, the B-modes are simply due to the noise and can therefore be used to evaluate the level of the noise in the E-mode AM map. 
As proposed by \cite{Miyazaki2017}, we used the maximum pixel value in the B-modes to define the threshold that is then applied to the E-mode AM map. We then only keep the pixels in the E-mode AM map that are greater than this threshold, setting the other pixels to zero.

In their study of the HSC data, \cite{Miyazaki2017} find that this threshold corresponds to $4.7$ times the standard deviation of the noise. The standard deviation depends on the filter under consideration and is computed from the B-mode AM map. For the DUSTGRAIN-\textit{pathfinder} simulation we use here, the equivalent average thresholds for the 256 realisations correspond to $4.09$, $4.37$, $4.01$, $4.35$, $4.57$, $4.35,$ and $4.04$ times the standard deviation for the S96, TANH, J04, M18, W2, W3, and W4 filters, respectively.

The final detection step is to identify the peaks in each of the thresholded E-mode AM maps. A peak is defined as a pixel whose value is greater than its eight nearest neighbours. The peaks at the border of the map over a width equal to the filter radius are discarded to avoid contamination due to boundary effects introduced by the convolution step. Following this procedure, we obtain a list of detections and associated peak coordinates. While this procedure is sufficient for single-scale methods, it needs to be further developed in the multi-scale case to deal with multiple detections, as we describe below.

\subsection{Multi-scale detection procedure}
\label{subsection_multi-scale_detection_procedure}

In the detection method based on the wavelet transform, we chose to decompose the lensing signal into $J = 5$ scales (see Eq.\,\ref{waveletdecomposition}), as this will encompass a maximum of possible galaxy cluster apparent angular sizes. Figure~\ref{Wavelets} shows the result of such a decomposition applied to the noisy convergence map of a $5^\circ \times 5^\circ$ simulated field. 
The top left panel shows a noisy version of the E-mode convergence map displayed on the right panel of Fig. \ref{Shearconv07}, and the other panels show the wavelet maps corresponding to scales $i = 1$ to 5 (W1-W5). Each wavelet map gives details of the original convergence map at different scales. The first scale (W1) is mostly dominated by the noise. In contrast, the scales from W2 to W5 shed light on different signals within the input E-mode convergence map. In our analysis, we decide to only keep the scales W2, W3 and W4. We remove scale W1 because it is dominated by noise. Scales W5 and above have also been removed because these filter sizes greatly exceed the expected angular size of the clusters we want to detect.

On the remaining scales W2, W3, and W4, we apply the single-scale detection procedure described above to extract the local maxima at each individual scale (hereafter W234). We now have to deal with an issue that is specific to the multi-scale approach. A signal from a given object can be detected at several scales, and with a position that varies slightly from one scale to another. We refer to these henceforth as multiple detections. Once we have obtained a peak candidate list for each scale, it is important to identify such multiple detections and to recombine them. This is achieved by computing the separation distance between all the pairs of detections on consecutive scales. If two peaks on different scales have a separation distance smaller than the larger of the two filter radii, we consider them to be a multiple detection. These are then recombined, and the position of the recombined peak is set by the position of the detection at the finest scale. If there are more than two peaks, we repeat this procedure for all the detection peaks and all the scales consecutively. We keep track of the index $i$ of the finest scale where the recombined detection appears, as this will be considered as the scale at which the signal was detected for the matching procedure described below in Sect.~\ref{section_matching}. We also saved the detections at each individual wavelet scale to study their complementarity before combination and to allow a comparison of their performance with respect to the corresponding AM filters.

\begin{figure*}[!ht]
\centering
\includegraphics[width=1.025\columnwidth]{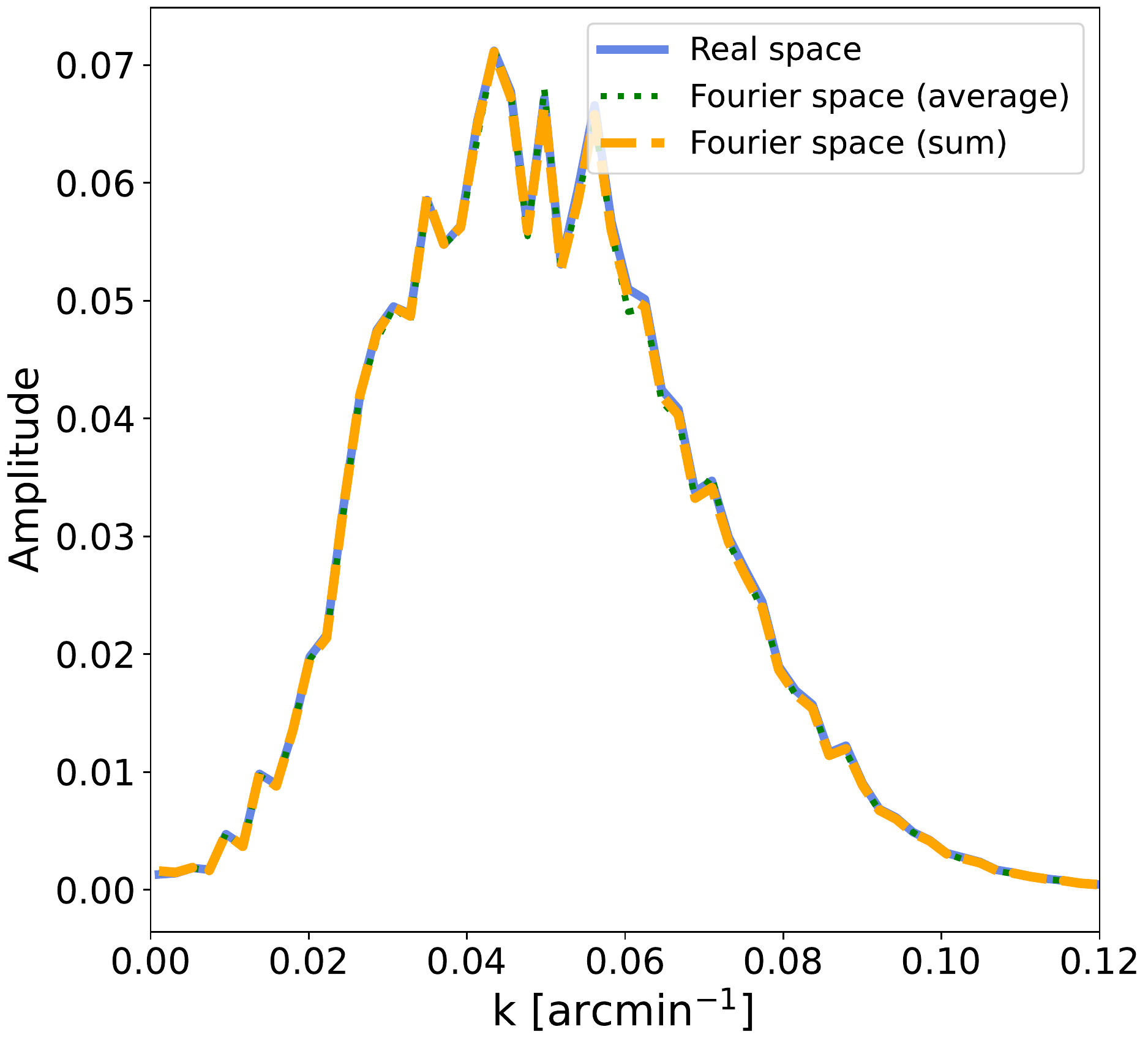} 
\hfill
\includegraphics[width=0.935\columnwidth]{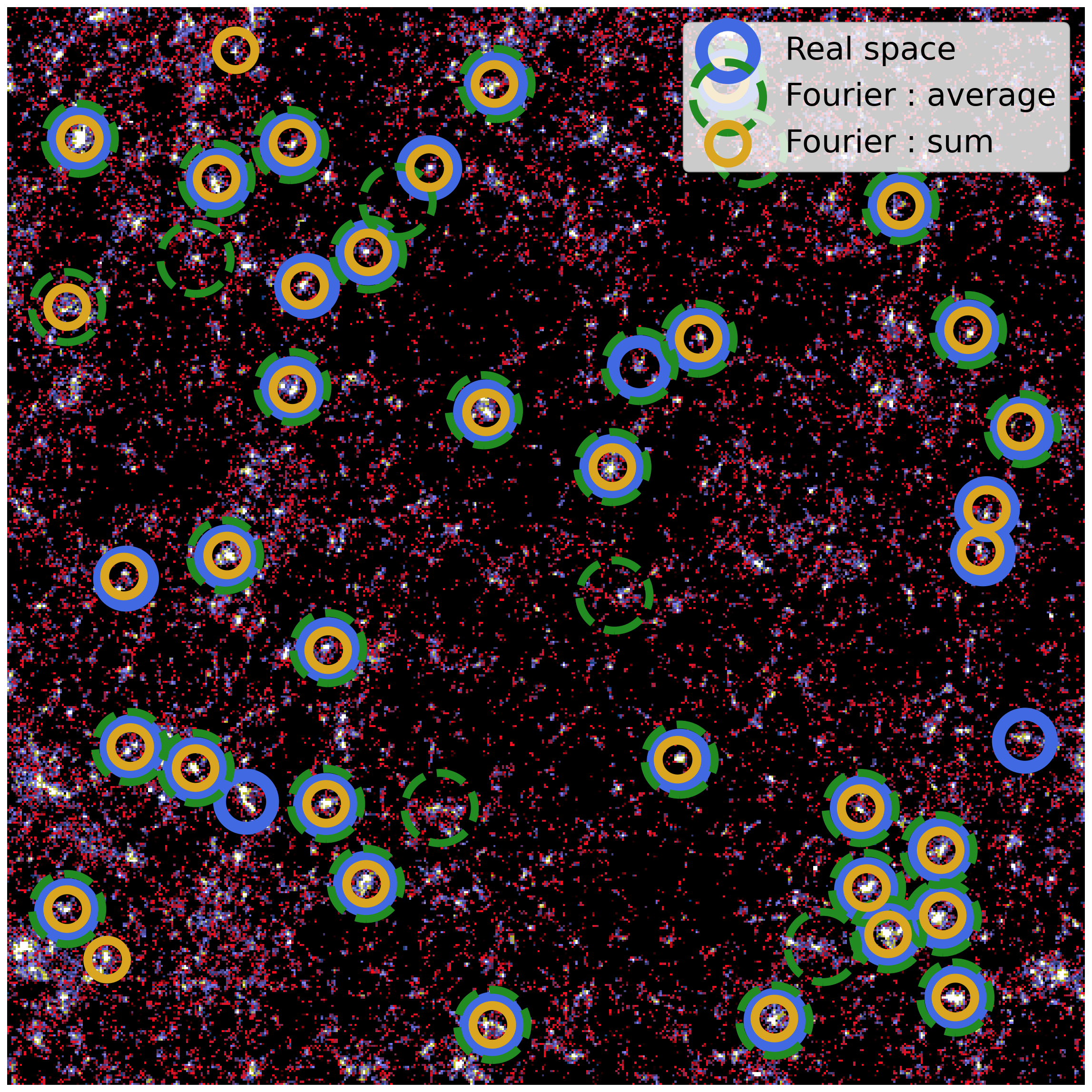}
\caption{Comparison of the real and Fourier space implementations. {Left:} Power spectra of a $5^\circ \times 5^\circ$ AM map with $512 \times 512$ pixels obtained using the real space implementation (blue) and using the Fourier space implementations (orange and green lines for summed and averaged pixel binning, respectively). 
Compared to the real space implementation, there is a $1.2\%$ and $1.5\%$ loss of power for the Fourier space implementation with summed and averaged pixel binning, respectively.
{\it Right:} Detections obtained on the same field when the computation is undertaken in real space (blue circles) or in Fourier space using either the summed pixel binning (golden circles) or the averaged pixel binning (green circles). The Fourier space implementation with summed pixel binning is very close to the real space implementation in terms of detections. When the binning is performed by averaging the shear values, there are fewer detections, and about 30\% of these detections are different from those obtained with the real space implementation.}
     \label{BB_RvsFspace}
\vfill
\vspace{0.5cm}
\vfill
\centering
\includegraphics[width=1.025\columnwidth]{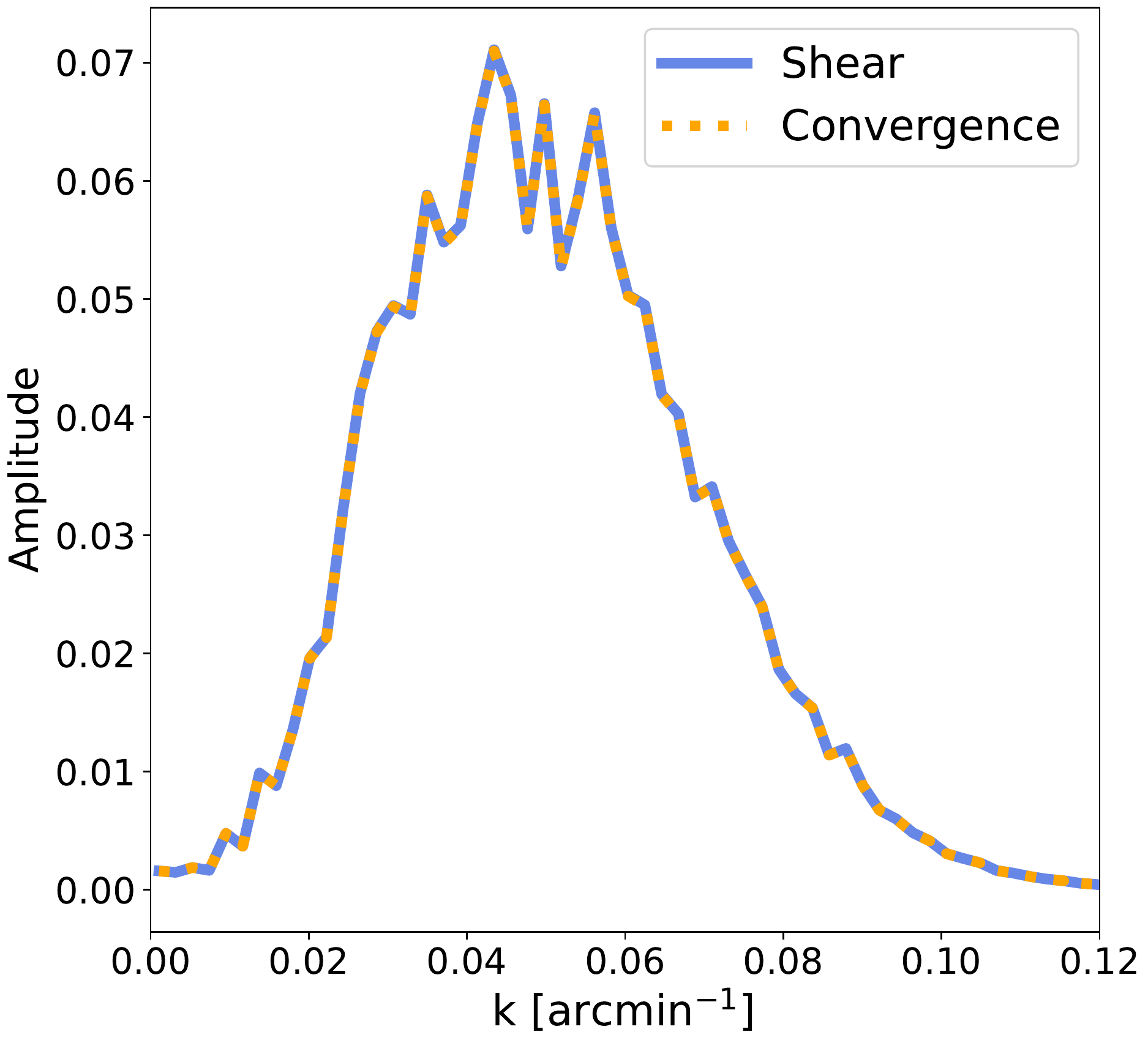} 
\hfill\includegraphics[width=0.935\columnwidth]{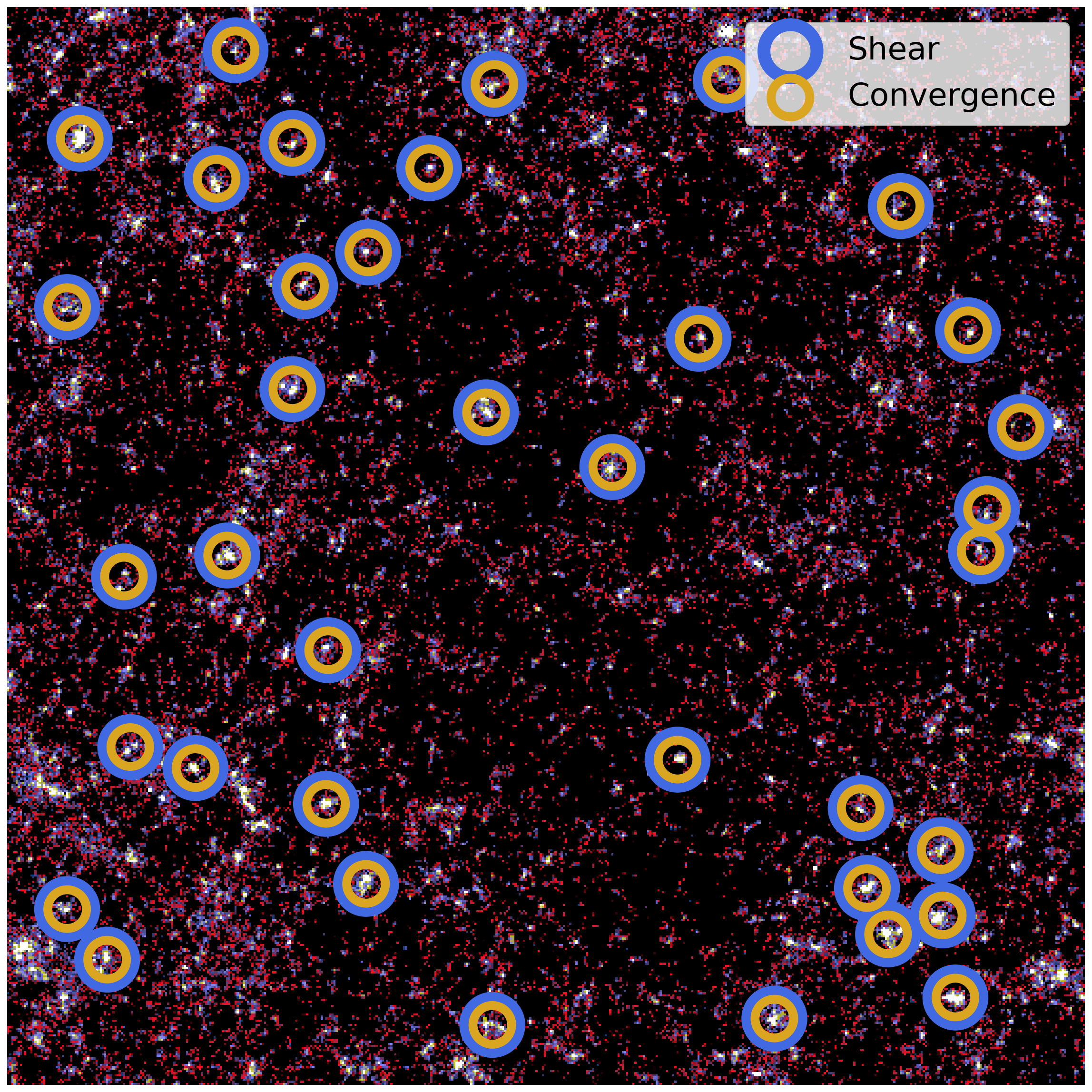}
  \caption{Comparison of the shear and convergence approaches. 
    {\it Left:} Power spectra of a $5^\circ \times 5^\circ$ AM map with $512 \times 512$ pixels obtained by applying the AM J04 filter to the shear maps (blue line) and to the E-mode convergence map (orange dots) computed from the same data as in Fig.~\ref{BB_RvsFspace}. The power spectra agree perfectly. 
    {\it Right:} Detections obtained on the same field when the convolution by the J04 AM filter is performed on the shear map (blue circles) and on the convergence map (golden circles). There is a one-to-one correspondence between the detections.}
     \label{BBconvshear}
\end{figure*}

\section{Analysis of implementation options for the detection algorithms}
\label{section_Applications}

In practice, the detection is undertaken on a catalogue of galaxies, for each of which a noisy shear measurement is available. As described above, this is typically undertaken through the convolution of the shear with a compensated filter. This convolution can be performed in real space, directly at the position of the galaxies \citep[e.g.][]{Maturi2005, Wittman2006, Miyazaki2017, Hamana2020}, or in Fourier space, binning the shear on a regular grid \citep[e.g.][]{Hennawi2005, Leonard2015, Oguri2021}. 
Moreover, some cluster weak lensing detection algorithms are based on the convolution of the convergence (derived from the shear and also defined on a regular grid) by a compensated filter \citep[e.g.][]{White2002, Hamana2004, Tang2005, Miyazaki2007}. In this section we explore how binning impacts the signal, and whether the choice to use the shear or the convergence has any effect on the detection efficiency.

\subsection{Detection in real space versus Fourier space}
\label{subsection_realvsfourier}

In the limit of a perfectly homogeneous distribution of background sources, the detection of convergence peaks, whether in real space or in Fourier space, depends only on the number density of these sources. However, as the background sources are not homogeneously distributed, an additional uncertainty is introduced into the method by the binning. This can impact the position and the S/N of the detections. 
This point has not been fully studied here because the distribution of the background sources is homogeneous in the simulations. However, as the number of background sources per pixel varies slightly for statistical reasons, our study already provides some insights on the impact of the background source distribution.

\subsubsection{Binning strategy}

The computation time for AM-based detection algorithms is considerable for the real space approach.
Computation time is significantly reduced for AM-based detection algorithms when using the Fourier approach. However, there is a loss of resolution compared to the real space approach because the galaxies must be binned on a shear map before performing the Fourier transform, instead of using their exact position. This loss of resolution introduces a loss of power for the high S/N pixels in the shear map, affecting the efficiency of the detection method. An additional disadvantage of the Fourier approach is that there are artefacts at the border of the field due to the periodic assumption of the discrete Fourier transform. Such effects can be significantly mitigated by removing the image boundaries or by dealing correctly with borders during the mass inversion, as proposed in \cite{Pires2020}.

The impact of the loss of resolution when working in Fourier space depends on the 
map resolution (i.e. the number of pixels of the map), and can be mitigated by using an appropriate value. Indeed, there is a trade-off to be made on the map resolution. Decreasing the resolution too much leads to power leakage, which can affect the detection method. On the other hand, finer binning can oversample the underlying data, and can introduce missing data into the map. 

Both real space and Fourier space approaches can be found in the literature \citep[e.g.][]{Hennawi2005, Gavazzi2006, Miyazaki2017, Hamana2020, Oguri2021}. In this paper we have introduced an additional Fourier space approach with a different binning strategy as explained in Sect. \ref{subsection_filtering}.

\subsubsection{Impact on the signal}

We first assessed the impact of the choice of a real or Fourier space approach on the signal conservation, when using an appropriate map resolution. We considered three different implementations: (i) application of an AM filter to the shear at the positions of the galaxies, directly in real space;
(ii) application of an AM filter in Fourier space to the binned shear obtained by averaging the signal from the galaxies that fall in each pixel (averaged pixel binning); and (iii) application of an AM filter in Fourier space to the binned shear obtained by summing the signal from the galaxies that fall in each pixel (summed pixel binning). 

First, we studied the differences between the three implementations through their spectra in the Fourier domain. For this test, we used the filter proposed by \cite{Jarvis2004} because it can be easily applied to the shear, either at the position of the galaxies or binned on a regular grid. From the shear catalogue of a $5^\circ \times 5^\circ$ field without noise, we computed the E-mode AM maps for the three different implementations and derived the corresponding power spectrum.

The results are presented in the left-hand panel of Fig.~\ref{BB_RvsFspace}, with the power spectra shown for implementations 1, 2, and 3. 
The power spectra are very similar, showing that on average the signal is equally well preserved by all three approaches. To quantitatively compare the implementations, we computed the integral of the power spectra. Compared to the real space approach, we find that the power spectrum integral of the Fourier space implementation suffers from a $1.2\%$ loss for summed pixel binning, and a $1.5\%$ loss for averaged pixel binning.

\subsubsection{Impact on the detections} 

We now analyse the impact of the three implementation choices discussed above on the detection efficiency. For each of the three implementations, we built the E- and B-mode AM map and applied the detection procedure described in Sect.~\ref{subsection_detectionProcedure_1scale} using the \citet{Jarvis2004} AM filter. This was undertaken on all 256 realisations. 
The right-hand panel of Fig.~\ref{BB_RvsFspace} compares the detections obtained with the three different implementations for one of the realisations.

When averaging these results over the 256 realisations, we found that the mean number of detections using the real space approach ($49.2 \pm 0.6$) is very close to that from the Fourier space approach with summed pixel binning ($47.9 \pm 0.5$). In contrast, the Fourier space approach with averaged pixel binning gives a smaller mean number of detections ($43.4 \pm 0.5$). 
Moreover, a closer look on the detections made by the different approaches reveals noticeable differences. We found that $28.2\% \pm 2.1\%$ of the detections obtained from the averaged pixel binning approach differ from those made by the real space approach. In contrast, with the summed pixel binning approach only $7.6\% \pm 1.3\%$ of the detections are different.

These results show that the binning strategy has a non-negligible impact on the detection efficiency, both in terms of number and distribution of the detections. In light of these results, we conclude that the Fourier space approach with summed pixel binning strategy agrees very well with the real space approach, producing results that are statistically similar. 
In the following, we decided to perform all tests and comparisons using the third implementation (i.e. the Fourier space approach using a summed pixel binning strategy) because it is computationally less demanding. It has the additional advantage of providing a substantial speed increase with respect to detection in real space: for our applications, we find a speed increase of two orders of magnitude, in agreement with the study by \citet{Leonard2012}.

\subsection{Detection using shear map versus convergence map}
\label{subsection_shearvsconvergence}
Both shear and convergence can give insights into the statistical properties of the weak lensing field, and indeed, it can be shown that they contain precisely the same information \citep[e.g.][]{Schneider2002, Pires2020}. The (reduced) shear is a direct observable and is usually preferred for reasons of simplicity. However, the convergence has the key advantage that $\kappa_E$ encapsulates all the lensing signal, while it is inevitably shared between $\gamma_1$ and $\gamma_2$ in the shear. In this connection, the convergence is more adapted for galaxy cluster detection because it explicitly traces the total matter distribution integrated along the LOS. A further advantage of using the convergence is that it is computationally less expensive to analyse.

\begin{figure*}[!ht]
\centering
\includegraphics[width=1.\columnwidth]{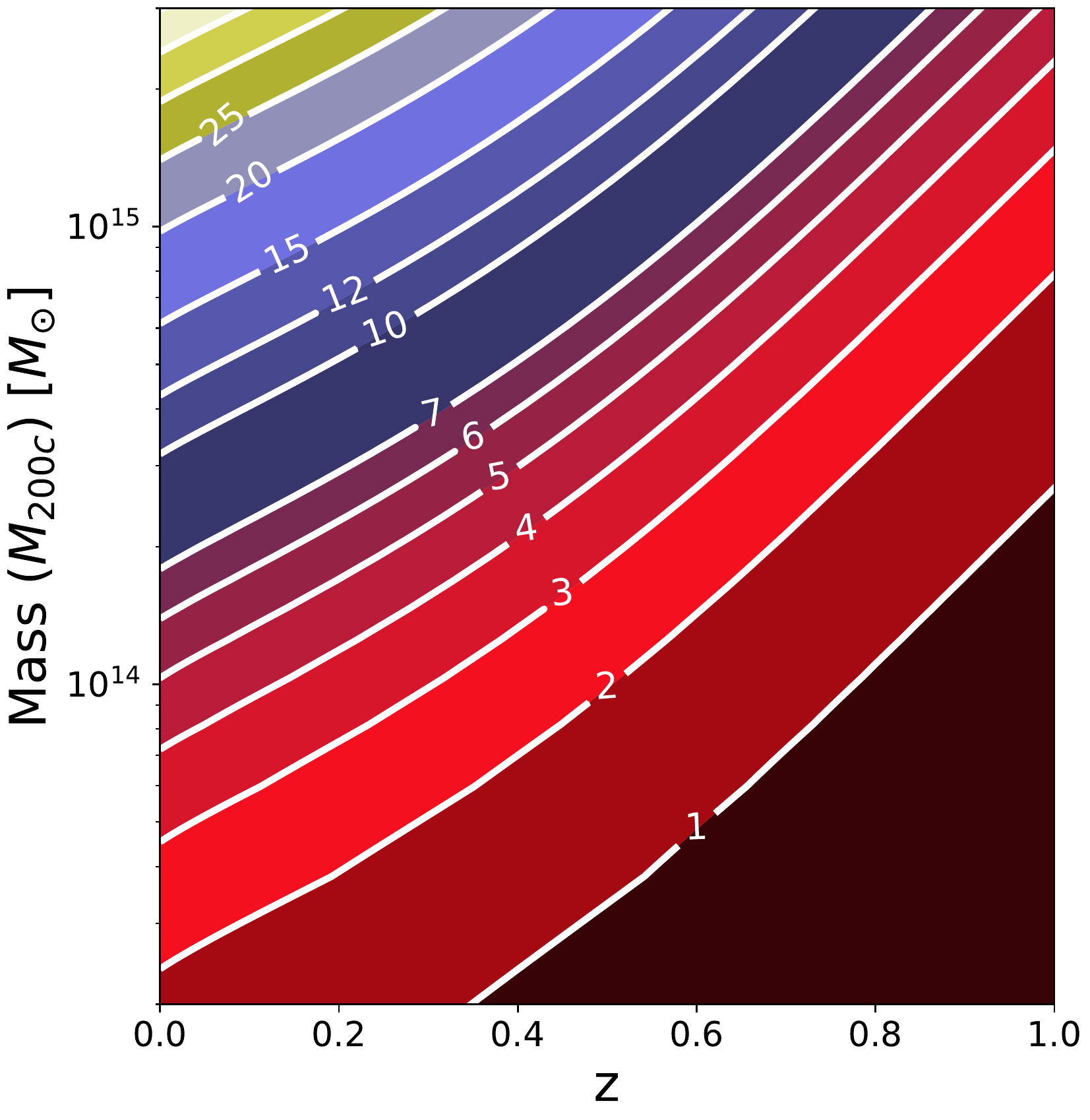}
\hfill
\includegraphics[width=1.\columnwidth]{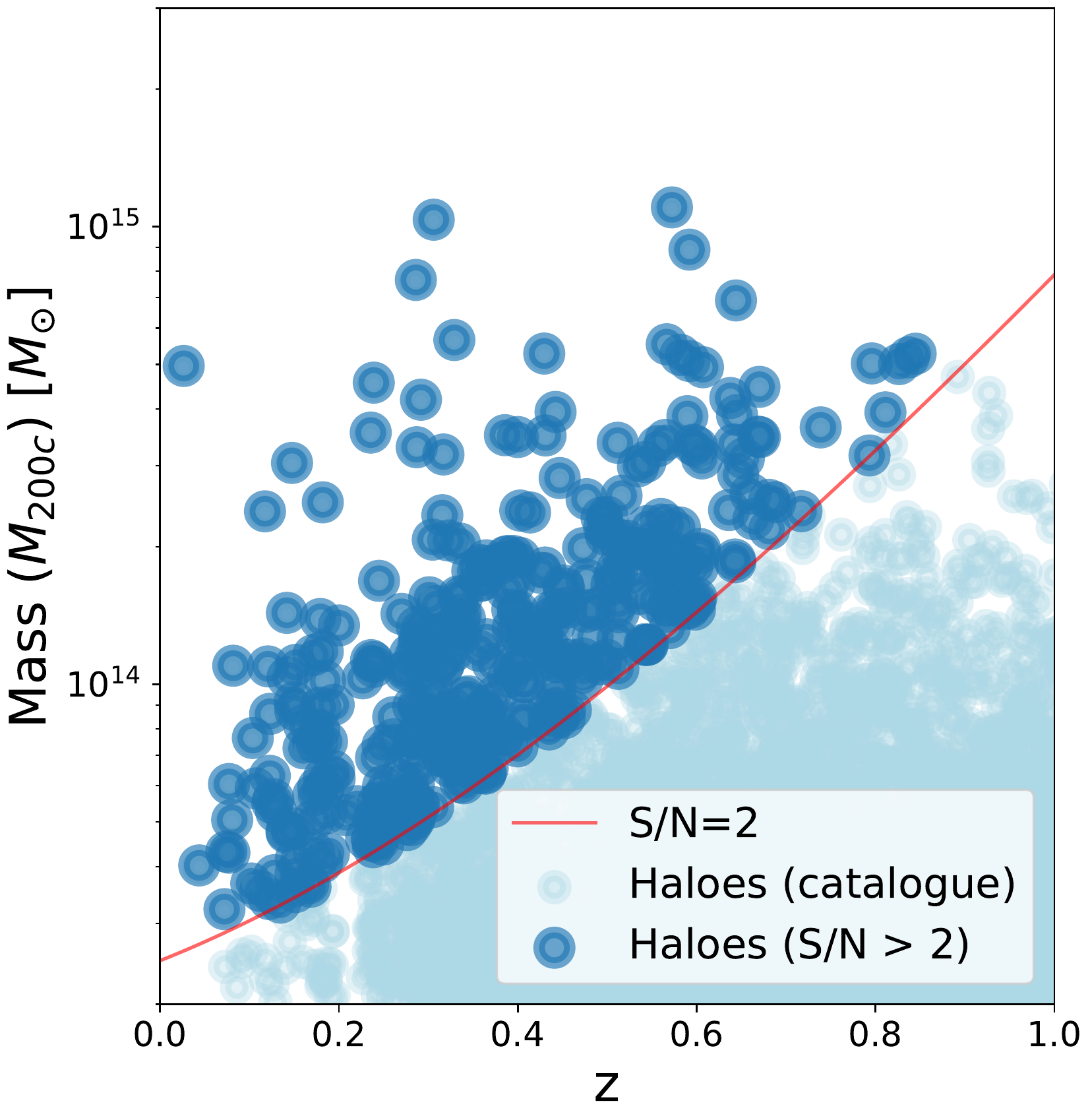}
\caption{Application of a theoretical S/N pre-selection to the halo catalogue. {\it Left:} Weak-lensing selection function for clusters of galaxies, in the redshift–mass plane.The selection function has been computed assuming a {\it Euclid}-like redshift distribution with $\rm n_{\rm g} = 30 \, gal.{arcmin}^{-2}$ and $\sigma_{\epsilon} = 0.26$. Contours denote the theoretical S/N for an NFW profile.
{\it Right:} Example of halo catalogue pre-selection using the theoretical S/N, for one $5^{\circ} \times 5^{\circ}$ field. The light blue dots correspond to all the haloes in the field, and the dark blue dots are the selected haloes with S/N$ > 2$.}
\label{FigSNR}
\end{figure*}

Although the reconstruction of the convergence field from survey data is a difficult task, \cite{Pires2020} show that the lensing signal is preserved in the convergence maps provided the mass inversion is performed without noise regularisation, and systematic effects such as irregular sampling and complex survey geometries are well controlled.

The AM formalism can be expressed either on the shear or on the convergence maps. We now assess the difference between these two approaches. We first compared them in Fourier space. For each realisation of the noisy shear catalogue, we produced the E- and B-mode AM maps using the J04 AM filter applied to either the shear or the convergence maps. Then, we computed and compared the power spectra. The result of this comparison for one realisation is shown in the left-hand panel of Fig. \ref{BBconvshear}, where we see that the power spectra obtained from the shear aperture map and the convergence aperture map are in good agreement.
Averaging over all 256 realisations, we measure a discrepancy of $0.03\%$ in the power spectrum integral between the shear and the convergence approaches. This difference can be explained by residual border effects that are not totally suppressed when removing the image boundaries.

We then compared the two approaches in terms of detections, applying the detection procedure described in Sect.~\ref{subsection_detectionProcedure_1scale}. The results, shown in the right-hand panel of Fig.~\ref{BBconvshear}, highlight the perfect agreement between the approaches. This result remains valid for wavelet filters that are formally identical to AM filters (see \citealt{Leonard2012} for more details). 
However, while the AM formalism can be applied to either the shear or the convergence, the wavelet transform is not designed to process spin-2 fields (i.e. with two independent components) such as the shear field. So, the shear and convergence approaches being equivalent, we decided to perform all the tests and comparisons applying the AM and Wavelet filters to the convergence maps, to be able to use the wavelet transform.

\section{Matching}
\label{section_matching}
Once the peaks are detected in the E-mode AM maps, they must be cross-matched with the position of the haloes in the DM halo catalogue. 
This matching procedure is essential to characterise the performance of the different methods, and the results can be biased if not undertaken correctly. This is particularly true when the filter characteristics differ significantly between methods.
We first identify two main points in the matching procedure that can significantly increase the number of false detections.

The first concerns the characteristics of the halo catalogue used for matching.
The DM halo catalogue contains objects down to a mass of $\rm M_{200c}=10^{12}\,h^{-1}\,M_{\odot}$ and up to a redshift of $\rm z = 3.6$. For the cosmology under consideration, a halo catalogue with these characteristics contains about $15,000$ haloes in a field of $5^{\circ} \times 5^{\circ}$. Obviously, only a small fraction of these haloes will be detectable through the weak lensing effect, which is only sensitive to the most massive clusters with sufficient galaxy sources behind them to trace their mass distribution. Changing the characteristics of the halo catalogue, such as imposing limits in mass and/or redshift, changes the performance of the methods with respect to a number of key points (e.g. matching, false associations).

The second is the association distance. Typically, the matching is performed within a fixed physical, comoving, or angular radius centred on the candidate peak, the association distance of which is usually optimised for a given filter \citep[e.g.][]{Hennawi2005,Gavazzi2006,Miyazaki2017,Hamana2020,Oguri2021}. However, when comparing different filters, this association distance must be adapted for each detection method to allow a fair comparison.

In the following, we address the first issue by applying a pre-selection to the halo catalogue based on the theoretical S/N of the detection of an NFW halo profile by weak lensing.
Regarding the second issue, we developed a method that allows us to adapt the matching distance to the filter characteristics used in the detection method.

\subsection{Halo catalogue pre-selection by S/N}
\label{subsection_S/N}

Imposing a pre-selection on the DM halo catalogue, by removing those haloes that are unlikely to be detected through weak lensing, reduces significantly the number of false associations. We defined our pre-selection by deriving a measure of the halo detectability in the form of the theoretical S/N of the detection by weak lensing \citep[e.g.][]{Hamana2004,Hetterscheidt2005,Berge2010,Andreon2012}. We followed the approach proposed by \cite{Berge2010}. 
Assuming the filter perfectly represents the signal $\kappa$, the S/N of an halo of mass $\rm M_{200c}$ and redshift $z_l$ can be expressed as
\begin{equation}
    \nu = \frac{\sqrt{\rm n_{\rm g}}}{\sigma_{\epsilon}} \sqrt{\iint \kappa(\boldsymbol{\theta})^2 \,\rm d\boldsymbol{\theta}^2},
\end{equation}
where $\sigma_{\epsilon}$ is the mean intrinsic shape noise of the source galaxies and $\rm n_{\rm g}$ the mean density of background sources. Assuming the mass of the cluster follows an NFW distribution \citep[][]{Navarro1996} its density profile is given by%
\begin{equation}
     \rho(r) = \frac{\rho_s}{(r/r_s)\,(1 + r/r_s)^2},
\end{equation}
where $\rho_s$ is the characteristic density and $r_s = R_{200c}/c$ is the scale radius. The surface mass density $\Sigma$, projected along the LOS, can be written as \citep[see][]{Bartelmann2001}
\begin{equation}
    \Sigma(x) = 2 \, r_{\rm s} \, \rho_{\rm s} \, g(x),
\end{equation}
where $x = r/r_{\rm s}$ is a dimensionless radius. The function $g$ is defined with respect to the distance to the halo centre and the concentration, $c$. If $c > 1$, it can be written as \citep{wright00}\begin{equation}
    g(x) = \frac{1}{1+c}\left\{
    \begin{array}{ll}
        - \frac{\sqrt{c^2-x^2}}{(1-x^2)(1+c)} + \frac{1}{(1-x^2)^{3/2}}\rm arccosh \left(\frac{x^2+c}{x(1+c)}\right) & \mbox{(x\,<\,1),}\\
        \frac{\sqrt{c^2+1}}{3}\left(1+\frac{1}{1+c}\right) & \mbox{ (x\,=\,1),} \\
         - \frac{\sqrt{c^2-x^2}} {(1-x^2)(1+c)} - \frac{1}{(x^2-1)^{3/2}}\rm arccos \left( \frac{x^2+c} {x(1+c)} \right) & \mbox{(1<x<c),} \\
        0 & \mbox{(x>c).}
    \end{array}
\right. 
\end{equation}
The concentration parameter $c$ is derived using the semi-analytical model introduced by \cite{Diemer2019}, as implemented in {\tt COLOSSUS} \citep[][]{Diemer2018}.
The S/N of the halo can then be written as \citep{Berge2010}
\begin{equation}
    \nu  =\, <Z>\,2\sqrt{2\,\pi} \,\frac{\sqrt{\rm n_{\rm g}}}{\sigma_{\epsilon}}  \frac{\rm \,r_{\rm s}^2\, \rho_{\rm s}}{D_{\rm d} \Sigma_{\infty}}  \sqrt{\int_0^{\,c} x g(x)^2 dx}. 
\label{Eqn:S/N}
\end{equation}
The weight parameter $Z$ is designed to take into account the impact of the distribution of the sources \citep[e.g.][]{Seitz1997}. the angular-diameter distance to the lens is described by $D_{\rm d}$, and $\Sigma_{\infty}$ is the value of $\Sigma_{\rm crit}$ for a source at a redshift of infinity. 
For each halo in the catalogue, we computed its theoretical S/N ($\nu$) assuming a {\it Euclid}-like redshift source distribution $\rm p_{\rm z_s}$ (derived from the simulation), a galaxy density of $\rm n_{\rm g} = 30 \, gal.{arcmin}^{-2}$ and a shape noise of $\sigma_{\rm \epsilon} = 0.26$.

The left-hand panel of Fig. \ref{FigSNR} shows the resulting ideal weak lensing detection selection function in the mass-redshift plane for the characteristics of the simulation. The figure is obtained by generating a grid in the mass-redshift plane and computing the S/N value for each point of the grid using Eq.~\ref{Eqn:S/N}. The colour map highlights the increase in S/N towards the upper-left corner, corresponding to the detection of massive clusters at low redshift. The solid white lines correspond to S/N contours of level S/N =[0,1,2,3,4,5,6,7,10,12,15,20,25,30,35], which have been interpolated from the $(z, M)$ grid points.

However, the above only yields an average value for the S/N of detection assuming a mean source distribution and an optimal filter. In practice, the filters are never perfectly adapted to the cluster shape as assumed in the theoretical S/N estimation. Furthermore, the detection S/N depends on the redshift distribution of the sources that are behind the cluster. The galaxies behind the cluster being few in number, small variations in their redshift can artificially boost or decrease the weak lensing signal.

\begin{figure*}[!ht]
\centering
\includegraphics[clip, angle=0, keepaspectratio, width=1.\textwidth]{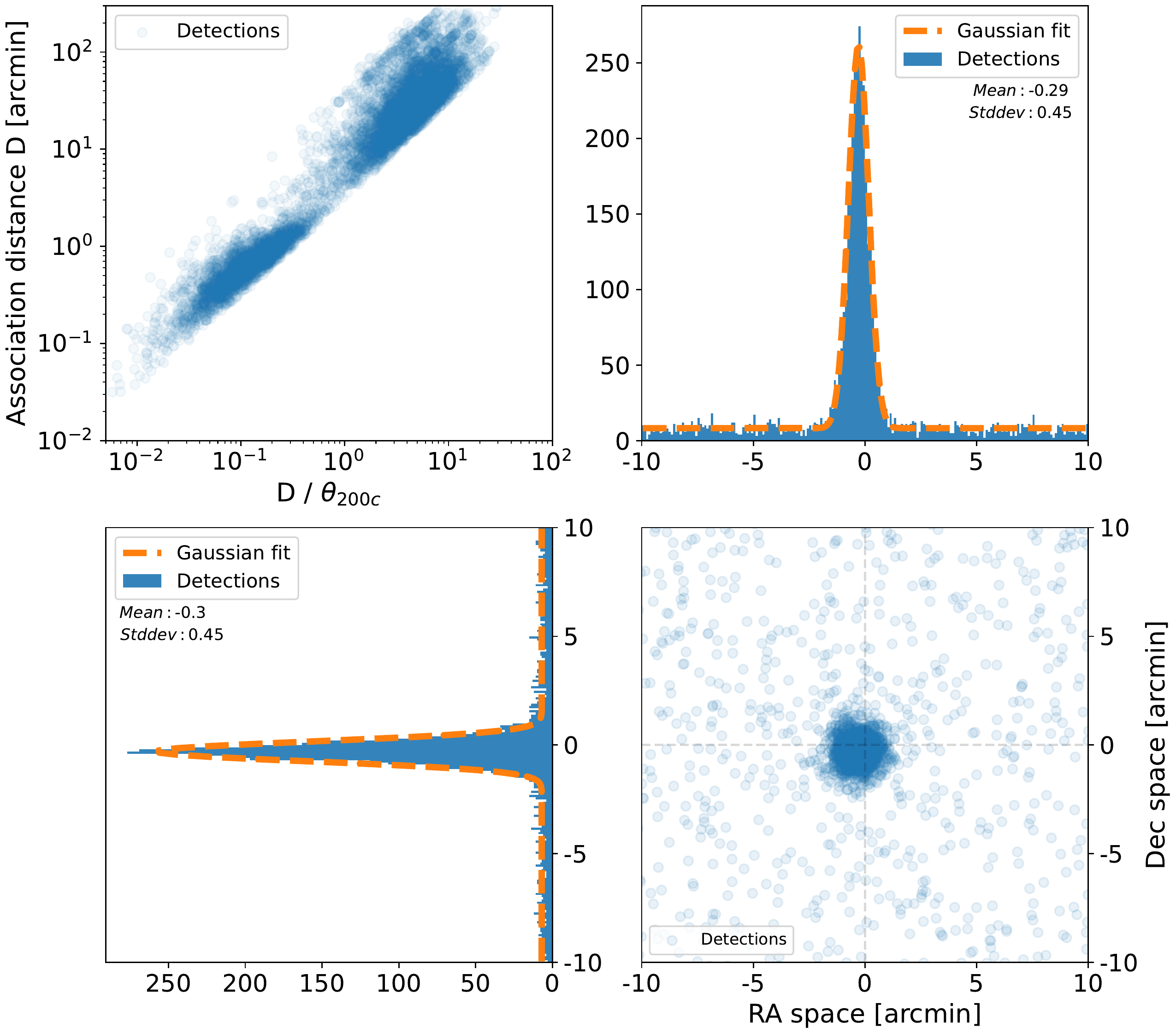} %
\caption{Dispersion of the association distance (without maximum association distance) for the W3 wavelet filter, considering all 256 realisations. In all the panels, the blue dots correspond to an association between a peak detection and its closest halo. {\it Upper left:} Absolute association distance, $D$, as a function of its normalised expression with respect to the characteristic halo radius, $\theta_{200c}$. {\it Lower right:} Zoomed-in view of the central $20\times20$ arcminute distribution of absolute association distances expressed in RA and Dec, to highlight their dispersion. The two remaining panels show the projection of the absolute association distance distribution onto the RA and Dec axes: the projection onto the Dec axis ({\it lower-left panel}) and the projection onto the RA axis ({\it upper-right panel}). The RA and Dec axes are zoomed around the central area. On these two panels, the dashed orange line shows the result of a fit by a Gaussian function whose parameters ($\mu$, $\sigma$) are labelled `Mean' and `Stddev'.} 
\label{Matching radius scale 3}
\end{figure*}

We applied a conservative threshold to the 256 DM halo catalogues by keeping all the haloes with S/N$> 2$. Although a halo with S/N = 2 is unlikely to be detected, we applied this value to take into account the variance in the detection due to the distribution of the sources, and the use of an ideal filter. Below this value we do not expect to detect many clusters. We discuss this choice of pre-selection threshold further in Sect~\ref{subsubsection_unmatched}.
With this pre-selection, in practice, from a typical $5^{\circ} \times 5^{\circ}$ field, only 2-3\% of the $15,000$ haloes in the catalogue are kept. The remaining $300 - 400$ haloes are then the basis for the matching procedure. The right-hand panel of Fig.~\ref{FigSNR} shows the selected haloes in the mass-redshift plane for one $5^{\circ} \times 5^{\circ}$ field.

\begin{figure*}[!ht]
\centering
\includegraphics[ clip, angle=0, keepaspectratio, width=1.\textwidth]{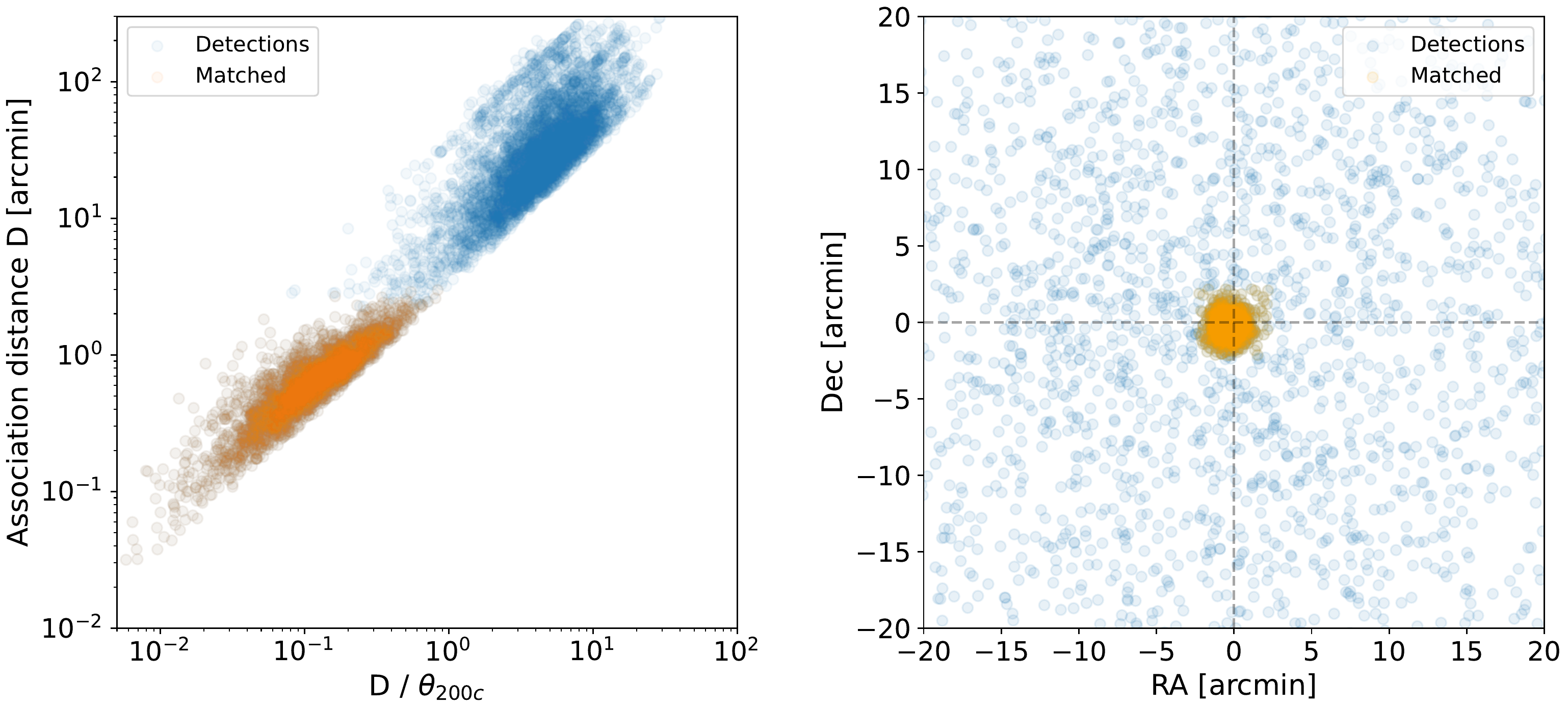} %
\caption{Matched detections for the W3 filter. The two panels are similar to Fig. \ref{Matching radius scale 3}. The blue dots correspond to an association between a peak detection and its closest halo. The orange dots correspond to the detections with an absolute association distance lower than the defined MMD.}
\label{Validation Matching radius scale 3}
\end{figure*}

\subsection{Matching procedure} 
\label{Subsection_Matching_procedure}

The matching procedure between the catalogue haloes and the peak candidates is a complex task. The typical approach consists of evaluating the distance between the candidate peak positions and the closest halo, and to consider these matched if the distance is below a given maximum matching distance (MMD). 

In practice, most matching methods simply draw a circle of a given apparent angular radius around the candidate peak and match the closest halo that falls in the circle \citep[e.g.][]{Hamana2004, Gavazzi2006,Hamana2015, Hamana2020}. However, in some procedures the matching is more refined, drawing a circle of a given comoving distance (in Mpc) around the halo and then matching the detection with highest S/N within this radius \citep[e.g.][]{Miyazaki2017,Oguri2021}.
The latter is more precise, but requires access to the redshift of the haloes in the catalogue.

Our study is further complicated by the characteristics of the different filter functions, and the inherent multi-scale nature of the wavelet approach. We therefore used a different method, drawing on SZ cluster survey matching procedures \citep[see e.g.][]{Planck2014}.
In our study, the matching is undertaken by computing the apparent angular distance $D$ (hereafter the association distance) between each candidate peak and each halo in the DM halo catalogue, after S/N pre-selection. This association distance is then normalised by the apparent halo radius $\theta_{200c}$. This normalised distance $D/\theta_{200c}$, together with the MMD defined below, are our matching criteria. Together, they take into account the filter size and have the effect of favouring the closest and most massive clusters in the matching.

The matching procedure is implemented iteratively. For $N$ detections and $M$ haloes, we obtain an array of $N \times M$ measured distances. We start the iterative process for a given candidate peak by selecting the minimum normalised distance $D/\theta_{200c}$ among all the measured normalised distances. The detection and halo corresponding to this minimum normalised distance are considered as matched and the corresponding row and column are removed from the array. The  procedure is repeated until we reach the defined MMD. This procedure is perfectly suited to cross-match detections with simulated halo catalogues, yielding excellent results compared to typical methods. However, for application to cross-matching with real cluster catalogues (e.g. optical, X-ray, etc.) access to an estimate of the $\rm R_{200c}$ (or $\rm M_{200c}$) of the haloes is required.

\subsection{Maximum matching distance} 
\label{subsection_Matching_radius}

The choice of the MMD is critical, and must take the positional precision of the detection method into account, as larger filter sizes will lead to a loss of precision on the coordinates of the peak centre. At the same time, if the chosen MMD is too large, false associations may be introduced even after pre-selecting the halo catalogue, while associations may be missed if the MMD is too small. Therefore, the MMD must be optimised and adapted to the detection method in question.

For this study, we developed an empirical  method that is able to adapt the MMD to the detection method. Our approach derives the MMD from the uncertainty in the position of the detection introduced by the size of the filter. In practice, this is undertaken by performing an association procedure on all the 256 fields, which corresponds to applying the matching procedure without any limit in the matching distance.  

Figure~\ref{Matching radius scale 3} shows an example of the distribution of the corresponding association distances 
for all the 256 realisations (about $75,000$ haloes after catalogue pre-selection) for the filter W3, for which the filter radius equals $2\farcm$32. 
Each blue dot corresponds to a pairing between a peak detection and its closest halo.
The upper-left panel shows the absolute association distance $D$, expressed in arcminutes, as a function of the association distance $D$ normalised by the angular halo radius $\theta_{200c}$. 
We clearly distinguish two distinct groups of associations. In the upper group, the detections are outside the characteristic halo radius. Therefore, we can consider this group to be false associations. The lower group is composed of associations for which the peak position is within the $\theta_{200c}$ radius of the halo and can thus be considered as true associations.
To isolate the correct associations and define a correct MMD, we use a different  visualisation of the results. In the lower-right panel of Fig. \ref{Matching radius scale 3}, the blue dots represent a unique peak detection-halo pair whose association distance $D$ has been decomposed in terms of RA and Dec (we show only the central $20 \times 20$ arcminutes). The high-density central region corresponds to peak detections that have been correctly matched to an underlying halo. The dispersion in this high-density central region in the RA-Dec plane highlights the uncertainties in the position of the peak detection introduced by the size of the filter.

To highlight the properties of the dispersion, we projected the distribution on both RA and Dec axes, displayed for the W3 filter in the upper-right and lower-left panels of Fig. \ref{Matching radius scale 3}, respectively. The positional uncertainty is well approximated by a Gaussian distribution. 
In practice, we obtain the same parameters if we fit a 2D Gaussian distribution to the RA-Dec plane or a 1D Gaussian distribution separately to the RA and Dec components.
From the fit we extracted the standard deviation $\sigma$, which corresponds to the dispersion of the positional uncertainty of the detection with respect to the true halo position. The Gaussian fit shows a y-offset that can be explained by randomly matched haloes. The Gaussian fit also shows an x-offset corresponding, to the mean of the Gaussian distribution $\mu$, which is due to the pixelisation. From these parameters, we defined the MMD to be $5\sigma\,+\,\lvert \mu \rvert$. We then applied this procedure to all the filters, and the resulting MMDs for each detection method are shown in Table \ref{table:Association distance}. Comparing these values with the values of the filter radii shown in Fig.~\ref{CompareFall}, we see that the MMD is directly proportional to the filter radius, as expected.

\begin{table}
 \caption{\footnotesize Estimated maximum matching distance (MMD) for different filter functions, evaluated using the procedure described in Sect.~\ref{subsection_Matching_radius}.}
 \label{table:Association distance}     
 \centering                          
 \begin{tabular}{c c c c}        
\hline\hline                 
   Filter function & $\sigma$ & $\lvert \mu \rvert$ & MMD [arcmin] \\     
 \hline                        
    S96 & 0.83 & 0.26 & 4.4 \\
    TANH & 0.47 & 0.28 & 2.6 \\
    J04 & 0.87 & 0.31 & 4.7 \\
    M18 & 0.60 & 0.28 & 3.3  \\ 
    W2 & 0.23 & 0.27 & 1.7 \\
    W3 & 0.45 & 0.30 & 2.5 \\
    W4 & 0.89 & 0.29 & 4.7 \\
 \hline
 \end{tabular}
\end{table}
\begin{table*}[!ht]
 \caption{\footnotesize Completeness of the different detection methods for different S/N halo pre-selections, in percent.}   
 \label{Table: Completeness high S/N}      
 \centering                          
 \begin{tabular}{c c c c c }       
\hline\hline                 
   Filter function & Completeness (S/N>2) & Completeness (S/N>5) & Completeness (S/N>7) & Completeness (S/N>10)\\    
 \hline                        
    S96 & $13.3\pm0.3$ & $65.6\pm0.1$ & $89.3\pm0.1$ & $98.1\pm0.1$\\
    TANH & $11.9\pm0.2$ & $62.3\pm0.1$ & $85.3\pm0.1$ & $95.8\pm0.1$\\
    J04 & $12.9\pm 0.2 $ & $59.7\pm0.2$ & $87.7\pm 0.1$ & $97.4\pm0.1$ \\ 
    M18 & $17.1\pm 0.2 $ & $71.5\pm0.1$ & $93.9\pm0.1$ & $98.1\pm0.1$ \\ 
    W4 & $12.7 \pm 0.2$ & $60.0\pm0.2 $ & $89.1\pm 0.1$& $98.2\pm0.1$\\
    W234 & $21.0\pm 0.2 $ & $75.3\pm0.1$ & $95.7\pm0.1$ & $99.2\pm0.1$ \\
 \hline
 \end{tabular}
\tablefoot{The completeness is computed on each field of 5$^{\circ}$ $\times$ 5$^{\circ}$ using different S/N halo pre-selection in the matching procedure, with S/N > 2, 5, 7 and 10. The results give the mean and associated uncertainty in the completeness (in percent), estimated from the 256 realisations.} 
\end{table*}

Once the MMDs are defined, we can apply the matching procedure defined in Sect.~\ref{Subsection_Matching_procedure}.
Figure~\ref{Validation Matching radius scale 3} shows the results of the matching for the W3 filter. 
The blue dots still refer to the associations from all the detections and their closest halo. The orange dots correspond to the detections that have been successfully matched. Considering only the pairs for which the association distance is lower than the MMD allows us to separate the two groups of associations distinctly. This further allows us to build two catalogues: one of matched detections, and another of unmatched detections.

\subsection{Multi-scale matching procedure}
\label{subsubsection_multi-scalematching}

In the multi-scale approach, a same cluster can be detected on several successive scales, complicating the matching analysis. To assess the performance of the multi-scale approach, it is important to recombine the multiple detections and to associate them with one single scale (the finest scale), as explained in Sect.~\ref{subsection_multi-scale_detection_procedure}. 

In practice, the multi-scale matching procedure works as follows. Starting with the finest scale, the matching procedure is performed using the appropriate matching distance defined in Table \ref{table:Association distance}. Associations at that scale are considered to be unique, and these haloes are removed from the catalogue that is used on subsequent scales. This procedure is repeated for each subsequent scale. 

The final result is a catalogue of associated and non-associated detections for each wavelet scale. These are important to measure the individual performance of each wavelet scale, and to compare their overall contribution in terms of detections and associations. The final catalogues of associated and non-associated detections are then obtained by concatenation of the individual single-scale catalogues. 


\section{Results and performance}
\label{section_results}
We now assess and compare the performance of the detection methods. We start by quantifying the completeness and purity of each sample, as commonly done by other studies \citep[e.g.][]{Miyazaki2017,Adam2019,Hamana2020}. 
Then we analyse in more detail our detections by performing an analysis of the distribution of the matched detections in terms of redshift, mass and scale. We then quantify the number of false detections due to the shape noise for each method. Finally, we undertake a characterisation and a classification of the matched and unmatched detections. For the wavelet filters, we undertake the above both on one single scale (W4), and using the full multi-scale approach (W234).

\subsection{Detection method performance}
\label{subsection_Performances}

\subsubsection{Completeness}
\label{subsubsection_Completeness}

To quantify the performance of the different methods and to compare them, we used the {\bf completeness,} C, defined as follows:
\begin{equation}
    \rm C= \frac{Number\,\, of\,\, matched \,\, detections}{Number\,\, of\,\, clusters\,\, in\,\, the\,\, halo\,\, catalogue}.
\end{equation}
We note 
that the completeness depends on the characteristics of the halo catalogue used for matching. Changing the pre-selection S/N in the halo catalogue will change the values of the completeness for each method. However, this does not affect the comparison between the methods. 

To further develop our analysis, we computed the completeness for each detection method using different S/N thresholds in the halo catalogue. For each method, we computed the completeness of the 256 realisations considering only the haloes in the catalogue whose theoretical S/N is above a given value. We repeated this operation for different S/N limits.
The results, averaged over the 256 realisations, are shown in Table \ref{Table: Completeness high S/N}.
We see that considering the haloes with S/N$ > 2$ results in a completeness below 25\% for all the detection methods. As expected, the completeness increases with the halo pre-selection S/N threshold, and almost all of the highest S/N clusters are detected by the different detection methods.
Even if the overall behaviour is comparable, this quantification shows that the detection methods based on the W234 and M18 filters outperform those based on the S96, TANH, and J04 filters in terms of completeness.

\subsubsection{Purity}
\label{subsubsection_purity}

Although the completeness for a given S/N halo selection is a good indicator of the performance of the different detection methods, it does not provide any information on the proportion of unmatched detections. That is why a measure of the purity of the sample is also needed to refine the comparison. The {\bf purity,} P, is defined as%
\begin{equation}
    \rm P = \frac{Number\,\, of\,\, matched \,\,detections}{Total\,\, number\,\, of\,\, detections}.
\end{equation}
This quantity measures the proportion of matched detections in the sample \citep[e.g.][]{Adam2019} and can be used to evaluate the false detection rate.
We computed the purity of the samples obtained from each detection method for a fixed catalogue pre-selection threshold of S/N$>2$. This threshold was chosen so as to not decrease the purity artificially as explained in Sect. \ref{subsection_S/N}. 
The results for the different detection methods, averaged over the 256 realisations, are summarised in Table \ref{table:First Results}. The estimated purity for all detection methods is comparable, at around 85\%. This is somewhat lower than the expected number from the false detections estimated in Sect. \ref{subsection_false_detections}. This led us to perform a case-by-case analysis of the unmatched detections, as  detailed in Sect. \ref{subsubsection_unmatched}. In particular, we observed that for some detections the matching procedure has failed or is too conservative. Although this obviously has an impact on the estimated purity, it does not affect the comparison between the methods because it affects all of them to the same extent.

\subsubsection{Purity versus completeness}

\begin{table*}[!ht]
 \caption{\footnotesize Detection method performance.}
 \label{table:First Results}     
 \centering                          
 \begin{tabular}{c c c c} \toprule  
   Filter function & Number of detections & Purity\, P [\%] &  Completeness C [\%] \\ \midrule   
    S96 & $57.6 \pm 1.0$ & $83.7\pm0.3$& $13.3\pm0.3$\\
    TANH & $50.9 \pm 0.9$ & $84.5\pm0.4$& $11.9\pm0.2$ \\
    J04 & $47.9 \pm 0.5$ & $88.3\pm 0.3$ & $12.9\pm0.2$ \\ 
    M18 & $66.6 \pm 0.7$ & $83.1\pm 0.3$ & $17.1\pm0.2$ \\ 
    W4 & $48.0\pm0.6$ & $88.3\pm0.3$ & $12.7\pm0.2$ \\
    W234 & $83.2 \pm 0.9$ & $82.7\pm 0.3$ & $21.0\pm0.2$\\\bottomrule
\end{tabular}
\tablefoot{ For each detection method, the detections are obtained applying the detection procedure described in Sect. \ref{section_detectionAlgorithms}. 
The purity and the completeness are measured using the matching described in Sect. \ref{section_matching} using a S/N$ > 2$ pre-selection in the halo catalogue.
The results correspond to the mean and its uncertainties (in percent) estimated from the 256 realisations.}
\end{table*}

Comparing the results in Table \ref{table:First Results}, we see that the detection methods based on the S96, TANH, J04, and W4 filters lead to higher purity and lower completeness. Although the completeness is not directly linked to the number of detections, but rather to the number of matched detections, the lower completeness for these filters is consistent with their lower mean number of detections compared to the two other methods, and also with their Fourier space representation, whose band-pass width is narrower. We note that the wavelet filter W4 alone attains similar performance to the single-scale filters S96, TANH and J04. Conversely, the completeness is higher and the purity is lower for the M18 and W234 filters, which use larger band-pass filters. However, the multi-scale approach spans a wider range of possible cluster scales, and therefore allows us to reach a higher completeness. As a consequence, the multi-scale approach results in a higher mean number of detections for the same purity. Compared to the M18 AM filter, for instance, the multi-scale approach yields 25\% more detections at a purity of $\sim85\%$.

Table \ref{table:First Results} also shows that some detection methods are more efficient in terms of completeness, while others are more efficient in terms of purity. Since the purity and completeness cannot be compared separately, a fair comparison between the detection methods requires an analysis of the evolution of the completeness as a function of the purity. We therefore computed this for each detection method using the pre-selected S/N$>2$ catalogue for the matching. In practice, we varied the threshold in the detection procedure described in Sect. \ref{subsection_detectionProcedure_1scale} in steps of $5\%$ and obtained new values of the completeness and purity for each new detection threshold value. 

\begin{figure}[!t]
\centering
\includegraphics[width=\hsize]{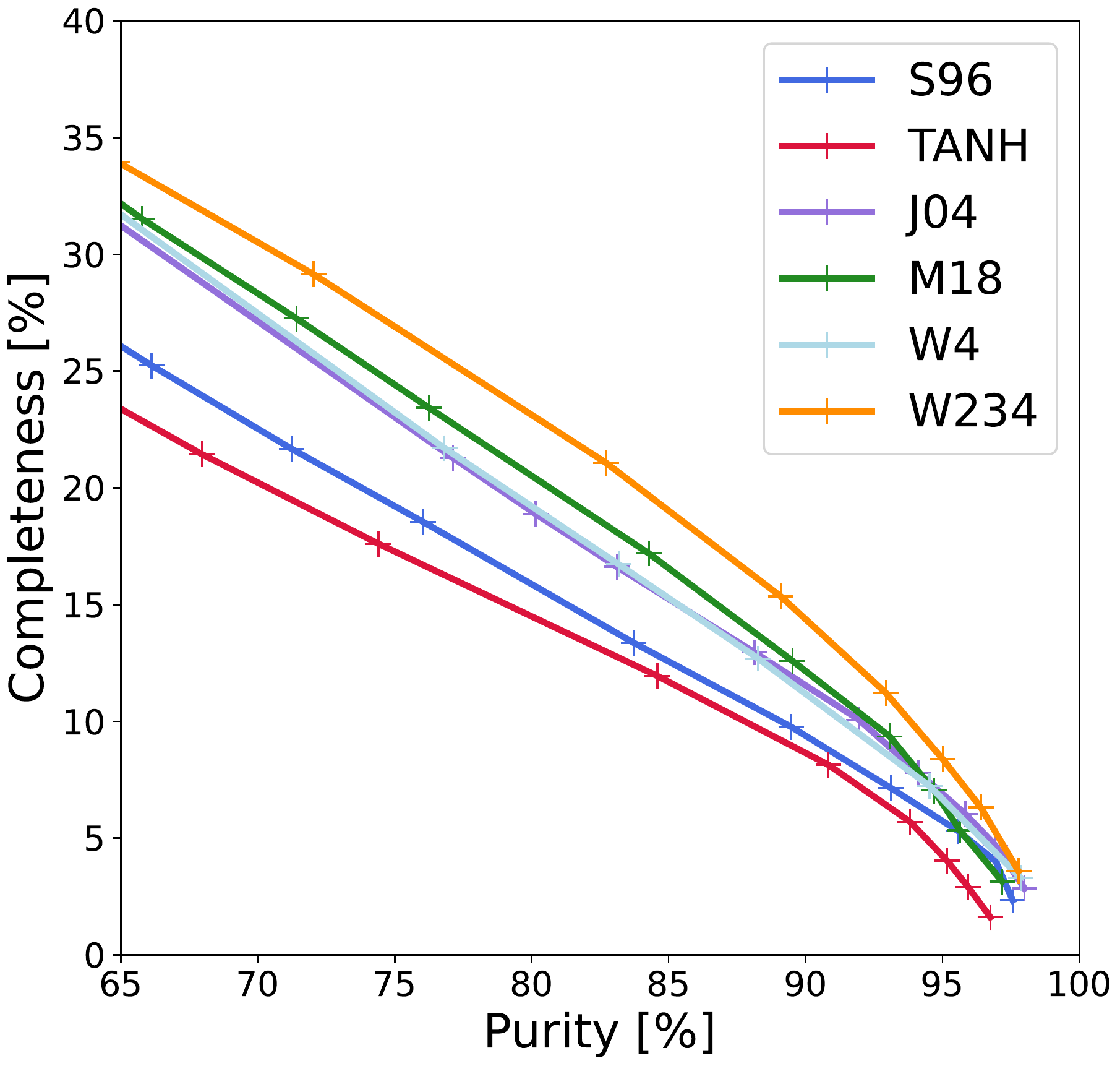}
\caption{Completeness as a function of purity for the detection methods based on the S96 (blue), TANH (red), J04 (purple), and M18 (green) AM filters, the W4 wavelet filter (light blue), and the W234 wavelet filter (orange). The cross markers represent precisely where the methods are evaluated in terms of purity and completeness. The purity and the completeness are computed using a S/N$>2$ halo pre-selection and averaging over the 256 realisations.} 
\label{Fig : Purity vs Completeness}
\end{figure}

The results, averaged over the 256 realisations, are shown in Fig. \ref{Fig : Purity vs Completeness}. We see that the multi-scale approach outperforms the other detection methods at all values of purity. Compared to the detection method based on the TANH filter, there is an increase in the completeness of up to 65\% at low purity. The increase in completeness is less for the detection method based on the M18 filter, but it remains significant in the range of purities studied.
We also note that the performance of the wavelet filter W4 alone and the single-scale filter J04 are comparable.

\subsection{Analysis of the detections}

Here we estimate the number of false detections due to the shape noise, and characterise the matched detections in terms of mass, redshift and scale.

\begin{figure*}[!ht]
\centering
\includegraphics[width=1.\columnwidth]{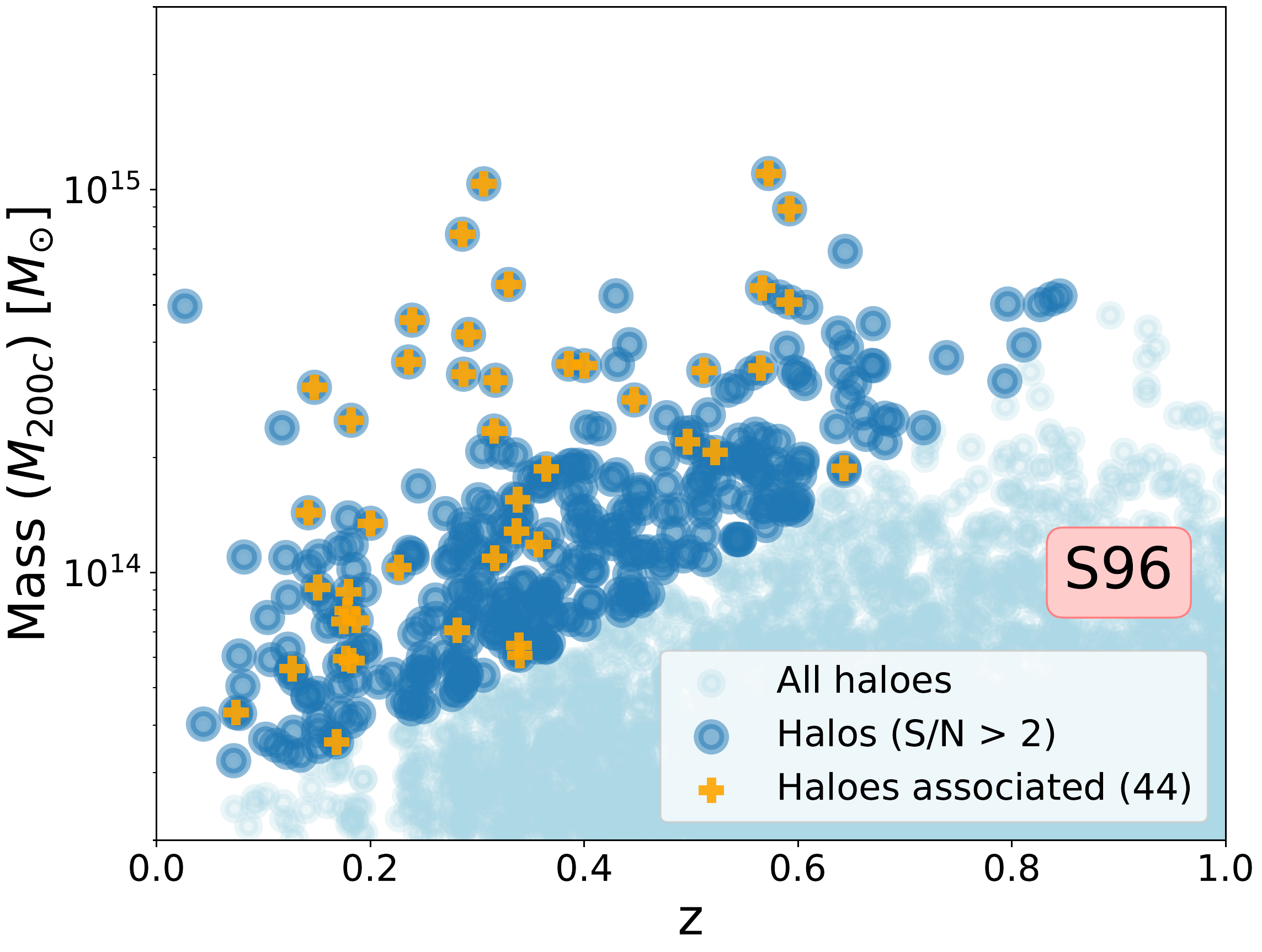}
\hfill
\includegraphics[width=1.\columnwidth]{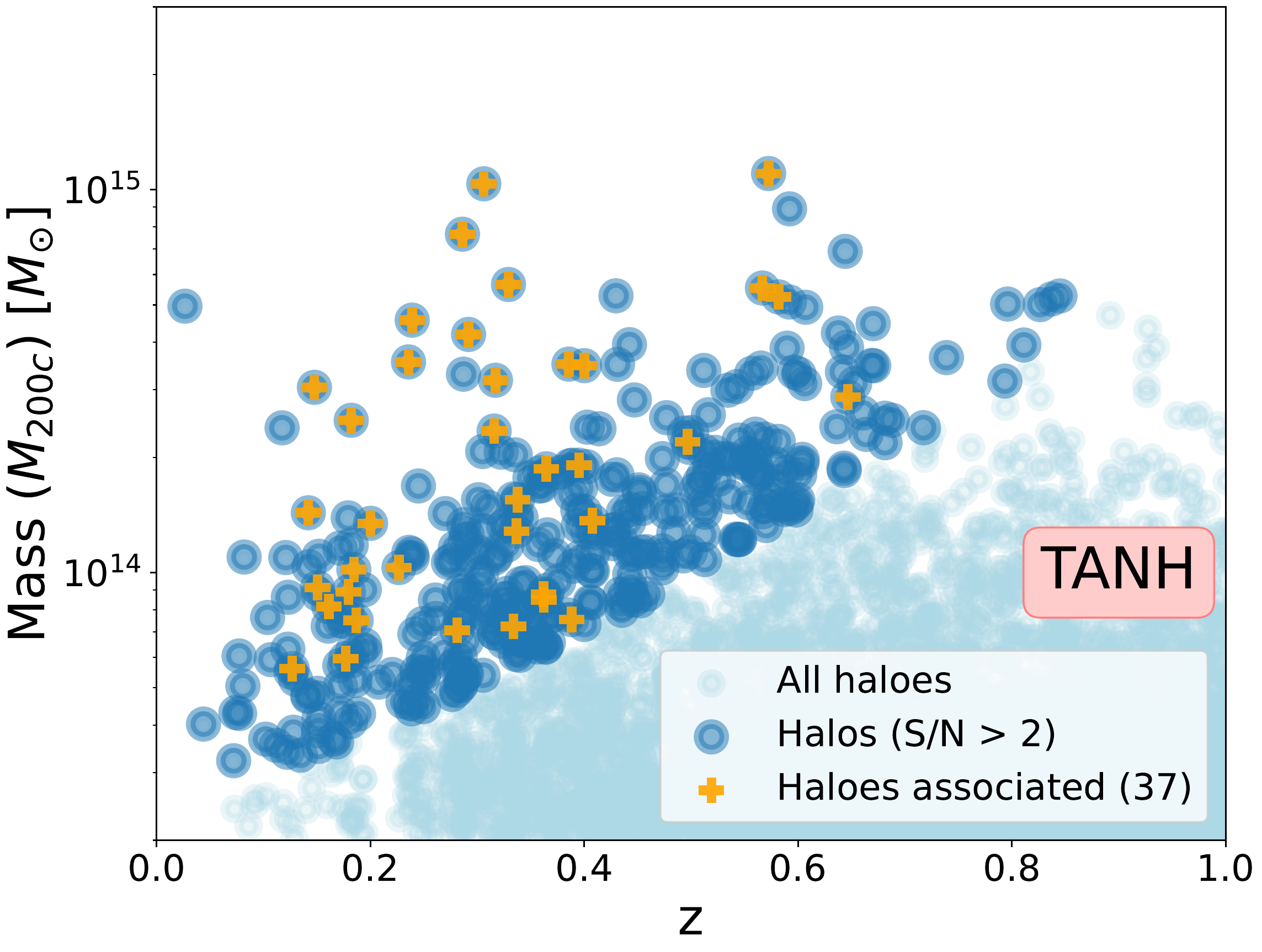}
\vfill
\vfill
\centering
\includegraphics[width=1.\columnwidth]{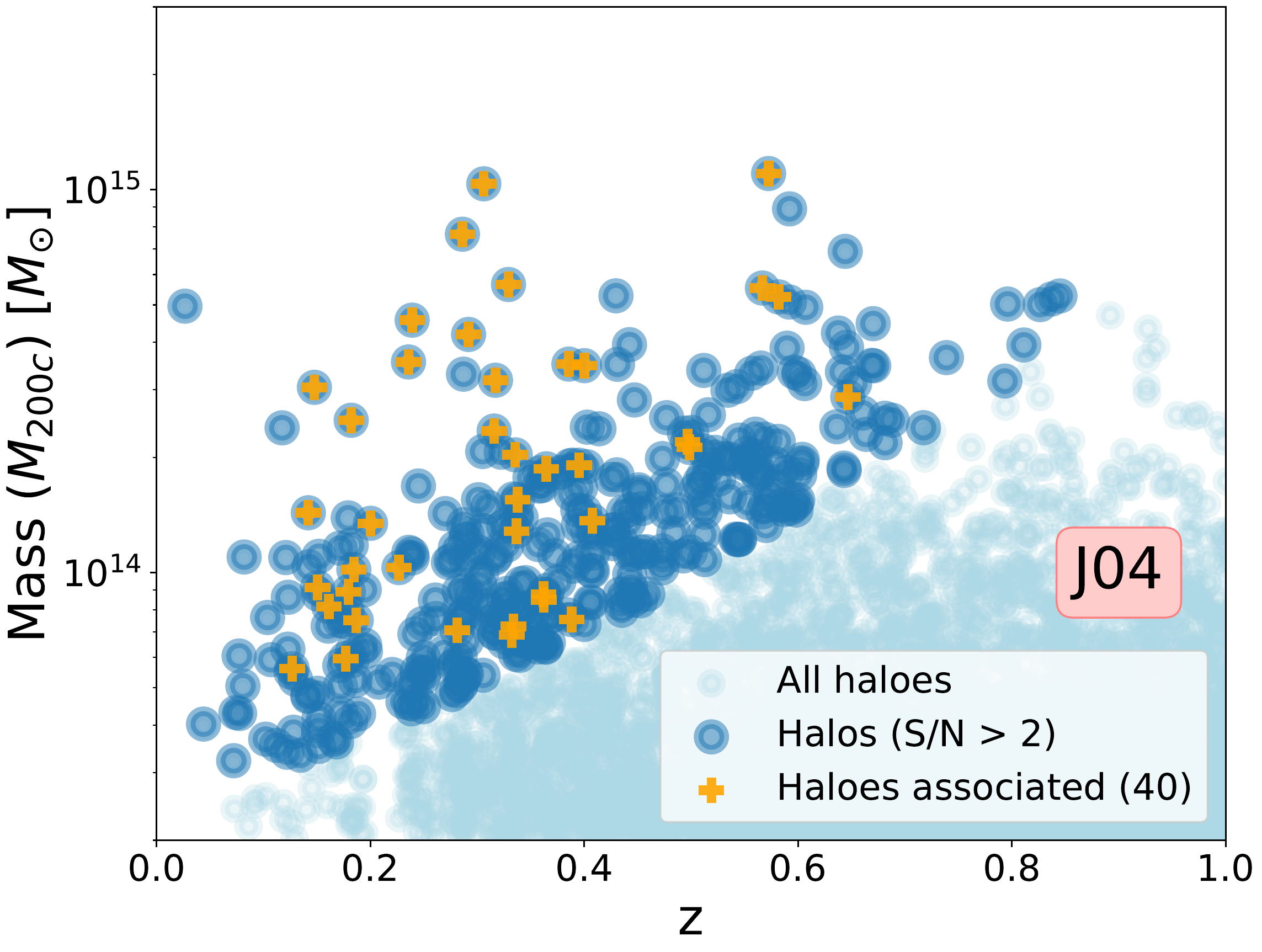}
\hfill
\includegraphics[width=1.\columnwidth]{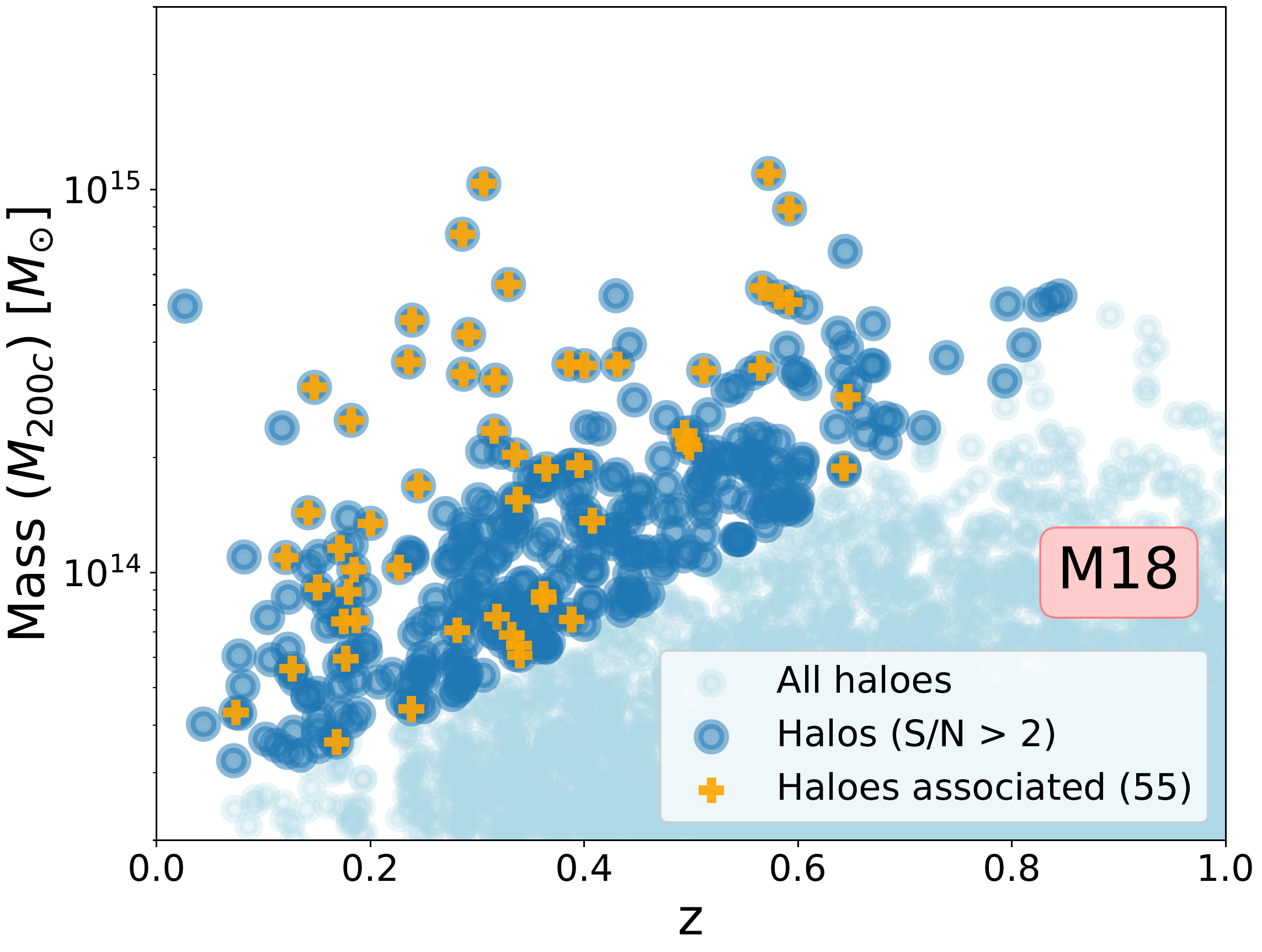}
\vfill
\vfill
\includegraphics[width=1.\columnwidth]{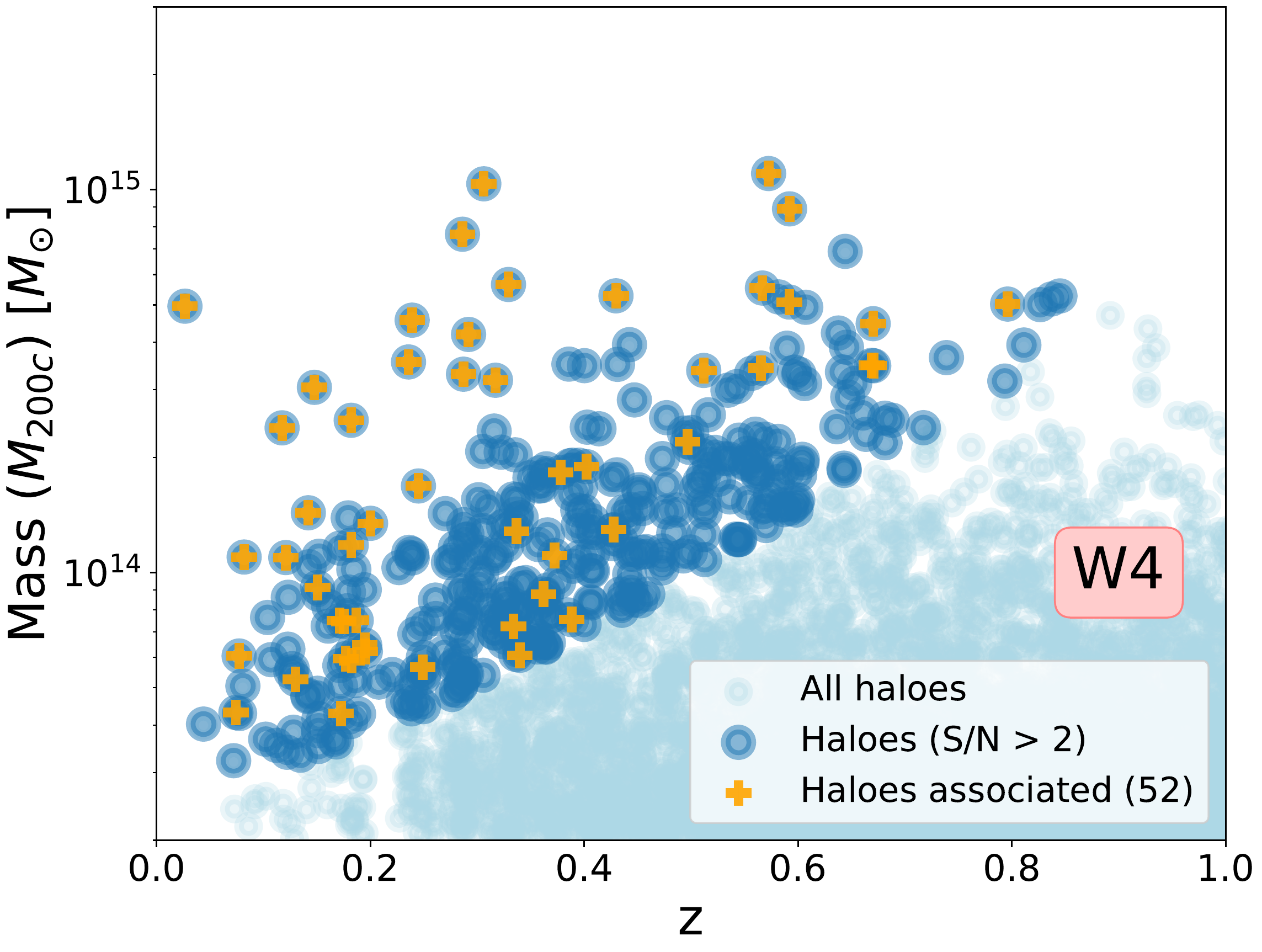}
\hfill
\includegraphics[width=1.\columnwidth]{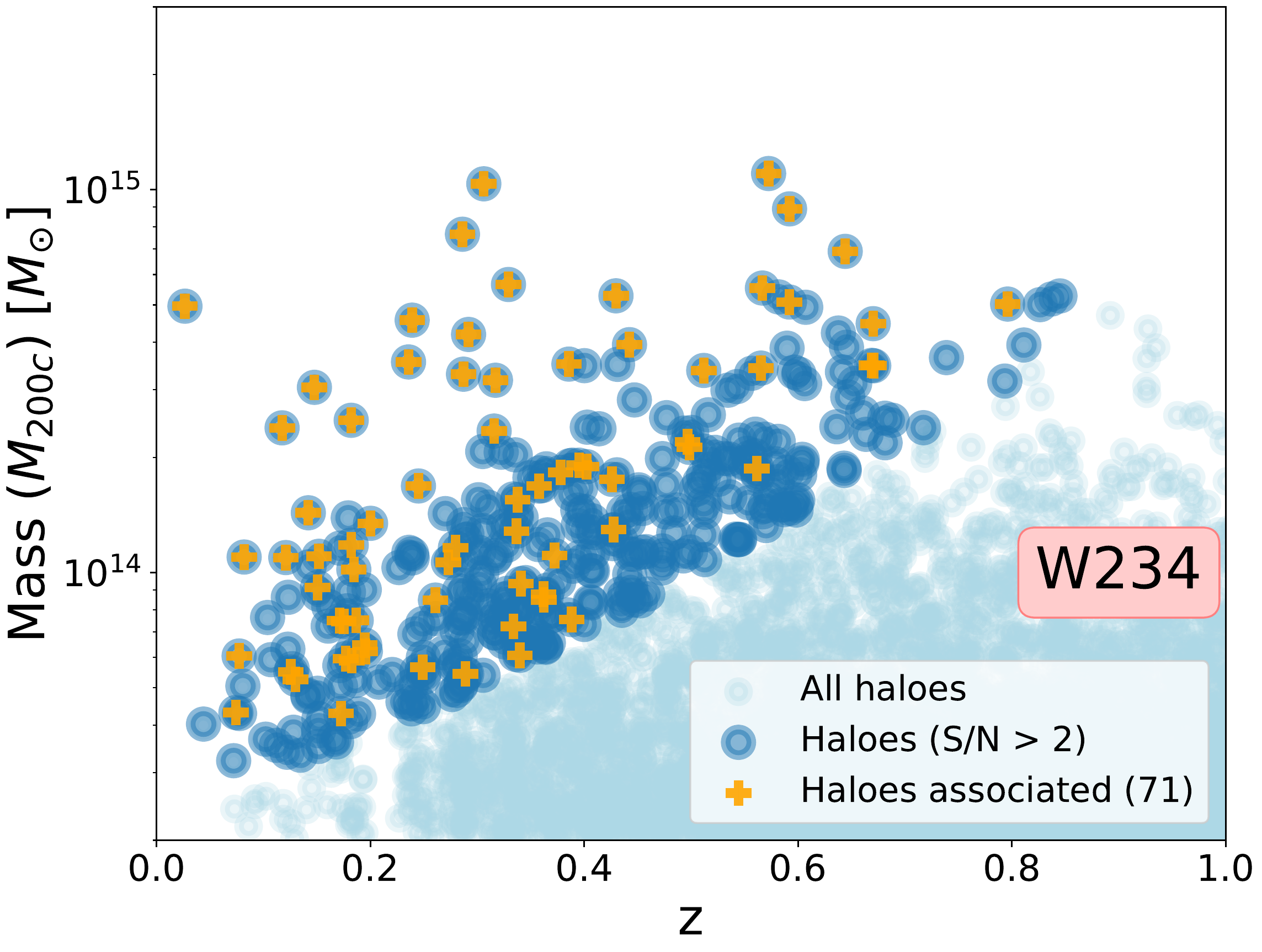}
\caption{Distribution of the matched clusters in the $z-M$ plane for one $5^\circ \times 5^\circ$ patch, using the S96 AM filter in the upper-left panel, the TANH AM filter in the upper-right panel, the J04 AM filter in the middle-left panel, the M18 filter in the middle-right panel, the W4 wavelet filter in the bottom-left panel, and the W234 wavelet filters in the bottom-right panel. The light blue dots correspond to all the haloes within the halo catalogue, and the dark blue dots to the haloes we have pre-selected to have S/N$>2$. The overplotted orange crosses correspond to the haloes that have been matched with a detection, and in parentheses we note the number of matches for each method on one patch. This patch was chosen to be representative of the performance of the different methods in terms of detection number.}
     \label{FIG: z-M all results}
\end{figure*}

\subsubsection{False detections}
\label{subsection_false_detections}
In the following we define false detections as those detections not related to any lensing signal (i.e. overdensity), and therefore due only to the noise. Then, we quantify the number of false detections due to the shape noise for each detection method.
For this purpose, we randomised the orientation of the sources in the noisy galaxy catalogues to suppress the lensing signal and to generate pure noise galaxy catalogues.
We then derived the associated convergence maps as described in Sect.~\ref{section_detectionAlgorithms} and applied the various detection methods.
For each method, we then computed the mean false detection rate averaged over the 256 realisations. This was obtained from the ratio of the number of detections due purely to the shape noise to the number of detections obtained when the signal is not suppressed. The resulting false detection rate due to the shape noise is given, for each detection method, in the second column of Table \ref{table: detections}.

The false detection rate due to shape noise is $1.1\%$ for the J04 detection method, $1.8\%$ for the detection methods based on the S96, TANH and W234 filters, and 3\% for the M18 detection method.
The impact of the shape noise depends on the filter. The noise level is lower at large scales compared to smaller scales, which explains why the J$04$ detection method is less sensitive to the noise compared to the other methods. In contrast, the M$18$ detection method is the most sensitive because it includes small scales. Similarly, the oscillation behaviour of the S96 and TANH filters increases their sensitivity to the noise, by integrating smaller scales. The W$234$ detection method also includes small scales, but the impact of the shape noise is reduced owing to its multi-scale detection procedure.
Despite these differences, the number of false detections due to the shape noise is small, irrespective of detection method.

\subsubsection{Characterisation of the matched detections}

Once the matching is completed, we characterise the matched detections by plotting their distribution in the redshift-mass plane.
Figure \ref{FIG: z-M all results} shows the distribution of the clusters detected by the five different detection methods in the redshift-mass plane.
The figure is obtained from one single realisation, but is representative of the full 256 realisation sample. The matching is performed using only the haloes with S/N$ > 2$.  
 
As expected, all the detection methods are most sensitive to the signal from high mass clusters lying at low redshift. Comparison shows that the multi-scale detection method W234 has more matched detections than the others, in agreement with the completeness results presented in Table \ref{Table: Completeness high S/N}. 
In addition, almost all the clusters detected by the S96, TANH, J04, and M18 detection methods are also detected by the multi-scale detection procedure. Compared to the four other detection methods, the multi-scale approach generally detects more clusters at lower masses, and up to higher redshifts. In Appendix \ref{Appendix B} we further analyse the multi-scale approach by investigating the contribution of each wavelet scale to the total number of matched detections.

A striking aspect of Fig.~\ref{FIG: z-M all results} is that some haloes with a very high predicted S/N are not detected, regardless of the method used. We performed a case-by-case analysis of these haloes and we found several explanations for this. In some cases, the detection is missed because of the shape noise. In other cases, the detection is missed because the lensing signal is decreased by the redshift distribution of the sources behind the cluster (through Eqs.~\ref{Eqn: convergence} and \ref{Sigma_crit}). In some other cases, the cluster is detected but the matching procedure fails because the normalised distance is not always the best criterion.
 
\subsection{Classification of the detections}
\label{subsection_classification}
We now undertake a further analysis of the matched and unmatched detections by classifying them into different categories. In particular, Fig.~\ref{Fig : Pie Chart W234} summarises these categories and the results for the W234 method, after detection and matching of all 256 realisations.

\subsubsection{Classification of the matched detections}
\label{subsubsection_matched}

The matched detections can be classified into two categories.

 \begin{table*}[!h]
  \caption{\footnotesize Classification of the \bf{detections}.}
 \label{table: detections}    
  \centering                          
  \begin{tabular}{ c c c c c c } \toprule     
    Filter function  & Matched [\%] & False detections  [\%] & Boosted objects ($\rm p_z$) [\%] & Blended haloes [\%]  & Others [\%]\\ \midrule               
      S96 & $83.7\pm0.3$ \,(48) & $2.0\pm0.2$\,(2)& $3.5\pm 0.2$ (2)& $5.8\pm0.3$\,(3) & $\sim 5.0 (3)$\\
      TANH &  $84.5\pm0.3$\,(43) & $1.7\pm0.3$\,(1) & $4.3\pm 0.3$\,(2)& $6.2\pm 0.3$\,(3)&$\sim 3.3 (2)$\\
      J04 & $88.3 \pm 0.3$ (42)&$1.1\pm 0.2 $\,(1)& $3.6\pm 0.2$ (2)& $5.9\pm 0.4$ (2)& $\sim1.1$ (1)\\ 
      M18 & $83.1 \pm 0.3$ (56)&$3.0\pm 0.3 $ \,(2)& $4.0\pm 0.1$ (3)&  $5.6 \pm 0.3$ (3)& $\sim4.3$ (3)\\ 
      W234 & $82.7\pm 0.3$ (67)&$1.8\pm 0.3$ \,(1)& $6.4\pm 0.2$ (6)& $5.5\pm 0.2$ (5)& $\sim3.6$ (4) \\ \bottomrule
 \end{tabular} 
 \tablefoot{The detections have been classified in the categories described in Sect.~\ref{subsubsection_matched} and \ref{subsubsection_unmatched}, for each detection method and averaged on each field of size 5$^{\circ}$ $\times$ 5$^{\circ}$. The first column gives the proportion of matched detection using a S/N>2 pre-selected halo catalogue in the matching procedure described in Sect.~\ref{section_matching}. Then, the unmatched detections have been classified in the categories described in Sect. \ref{subsubsection_unmatched}. 
 The false detections due to shape noise are in the second column. The unmatched detections due to haloes whose lensing signal has been boosted are in the third column. The unmatched detections due to blended lensing signal from LOS alignments or the merging of two haloes of S/N>1.5 are in the fourth column. The last column gives the proportion of unmatched detections that remain unclassified. These results correspond to the mean and associated uncertainty of the detections that fall into these categories. The numbers in parentheses correspond to the average number of detections in each category in a 5$^{\circ}$ $\times$ 5$^{\circ}$ realisation (rounded up to an integer number).
}
 \end{table*} 
 
\begin{figure}[!t]
\centering
\includegraphics[width=\columnwidth]{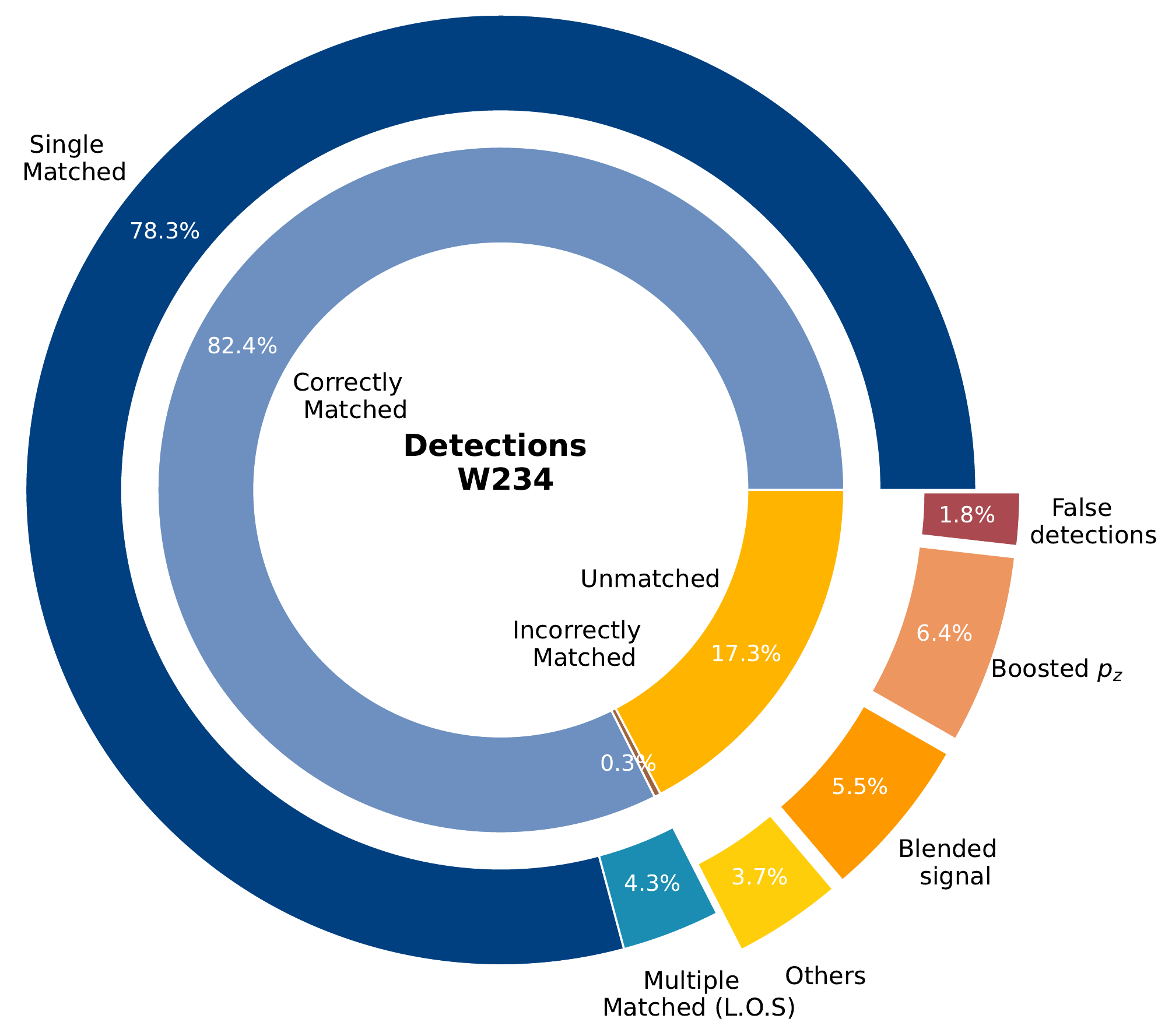}
\caption{Characterisation of the detections from the W234 method. The matched detections (blue segments) are categorised as correctly matched with single or multiple haloes, or incorrectly matched as defined in Sect.~\ref{subsubsection_matched}. The unmatched detections are sub-categorised as blended signal, boosted signal from the source distribution, false detections, or others, as defined in Sect.~\ref{subsubsection_unmatched}. The results are averaged over the 256 realisations.} 
\label{Fig : Pie Chart W234}
\end{figure}

{\it True detections correctly matched}. Almost all the matched detections correspond to correct associations to a single halo of the catalogue (`Single matched' in Fig.~\ref{Fig : Pie Chart W234}). 
However, it is inevitable that some true detections can be correctly matched but should  in fact be associated with two (or possibly more) haloes whose lensing signal is merged.
This may occur when two haloes are aligned in the LOS or when their apparent angular distance is small compared to the size of the filter.
In this case, the matching procedure does not differentiate and the association is undertaken simply with the closest halo. We estimate that about 5\% of the matched detections fall in this category by searching for multiple haloes in the vicinity of the matched detections ('Multiple Matched (LOS)' in Fig.~\ref{Fig : Pie Chart W234}). These  imperfect associations can slightly reduce the estimated completeness, but all the methods are impacted in the same way by this effect.\\

{\it False detections incorrectly matched.}
Among the matched detections, a small fraction are in fact randomly-matched false detections. These are detections due to the shape noise that are matched owing to being close to a halo in the catalogue. Their number 
depends both on the number of false detections (see Sect. \ref{subsection_false_detections}) and on the false association rate (given in Table \ref{table:TFA}). In this study, we intentionally keep this number very small for all the detection methods (less than 0.3 \%). However, one must be careful because both the detection and matching procedures can make this number become large. Thus, it is essential to always compute this number to fully assess the performance of the detection methods.

\subsubsection{Classification of the unmatched detections}
\label{subsubsection_unmatched}

Here we focus on the detections that are not matched with a halo, to fully complete the analysis of the different detection methods.
These unmatched detections can be classified into several categories. 

{\it False detections unmatched correctly. }The most intuitive class for of unmatched detections is that composed of those due to the shape noise (see Sect.~\ref{subsection_false_detections}.) Beyond the small fraction that is incorrectly matched owing to being located close to a halo in the catalogue (see above), almost all of these false detections are not matched. Their number thus corresponds approximately to the false detection rate, and is given in the second column of Table~\ref{table: detections}. This number is very small compared to the estimated purity in Sect. \ref{subsection_Performances}, and other explanations need to be found to explain the remaining $\sim 15\%$ of the unmatched detections.\\

\begin{table}[!t]
\caption{\footnotesize False association rate for different filter functions.}   
\label{table:TFA}     
\centering                          
\begin{tabular}{c c c} \toprule       
    Filter function &  False association rate [\%] \\ \midrule  
     S96 & $12.8\pm0.3$ & \\
     TANH  & $9.0\pm0.3$  &  \\
     J04 & $14.2\pm 0.4$ &\\ 
     M18 & $10.6\pm 0.4$& \\ 
     W234 & $10.8 \pm 0.4$ &\\
     W2 & $2.8 \pm 0.3$ &\\
     W3 & $8.7 \pm 0.4$ &\\
     W4 & $13.6 \pm 0.4$& \\ 
     \bottomrule   
 \end{tabular}
 \tablefoot{The false association rate is computed for each detection method including each individual wavelet filter W2, W3 and W4. The fraction of false associations is estimated by matching random detection positions with the halo catalogues with a S/N$ > 2$ halo pre-selection. The errors are computed using the 256 realisations.}
 \end{table}

{\it Boosted lensing signal unmatched incorrectly.} One possible explanation for the remaining unmatched detections is that some of these detections should be associated with haloes that have been removed from the halo catalogue after the S/N pre-selection. 
This could happen because the S/N estimated for each halo in the catalogue to perform the selection is theoretical. It is computed assuming an average theoretical source density and a theoretical distribution, $\rm p_{\rm z_s}$, for the redshift of the sources. In practice the sources behind a cluster are few in number. As a consequence their number and their redshift distribution can deviate significantly from the theoretical values, and can in some cases boost the lensing signal of a cluster, regardless of its mass and redshift.
At the same time, selecting haloes with a much lower S/N in the halo catalogue serves to increase the false association rate, and so the number of false detections matched incorrectly (see Sect~\ref{subsubsection_matched}). Table \ref{table:TFA} gives the false association rate when using a S/N$ > 2$ pre-selection in the halo catalogue. This remains below 15\% for all detection methods. Thus, the choice of imposing a S/N$ > 2$ pre-selection in the halo catalogue is a trade-off between the false association rate and the missed association rate.

We estimated the fraction of unmatched detections due to a boosted lensing signal by undertaking the matching procedure a second time for all 256 realisations, but with a halo pre-selection of S/N$> 1.5$. The results for each method is given in the third column of Table \ref{table: detections}. \\

{\it Blended lensing signal unmatched incorrectly}. A further explanation for the remaining unmatched detections is the case of two haloes contributing to the same detection: 

\begin{enumerate}
\item Blended LOS: The blended lensing signal can be caused by LOS alignments (e.g. filamentary structures). This means that the detection is coming from two haloes with S/N$ < 2$ that are aligned along the LOS and whose lensing signals cannot be separated without considering the redshift distribution of the background galaxies in the detection step.

\item Blended haloes: The blended lensing signal can also be linked to the size of the filter. Indeed, some unmatched detections can come from two haloes whose apparent distance is small compared to the filter radius. The lensing signal of these two haloes being merged within the AM map, there is a single detection associated with the two haloes, the position of which depends on the mass and redshift of the two haloes. In this case, the detection is unmatched when the theoretical S/N of the two haloes is less than 2, or when the position of the detection is too far away from the centre of the two haloes.
\end{enumerate}

We estimated the proportion of blended lensing signal falling in the two categories by searching for multiple haloes with S/N $> 1.5$ in the vicinity of the unmatched detections. The proportion of unmatched detections due to the blending of the signal coming from two (or possibly more) haloes is given in the fourth column of Table \ref{table: detections}. In this class of unmatched detections, we found that about a third of them are due to intrinsic LOS alignments.

A small fraction of unmatched detections do not fall into the above three categories. The fraction for each method is listed as `others' in the last column of Table~\ref{table: detections}. Most of these remaining detections can be matched with haloes with S/N$<1.5$. But further decreasing the halo pre-selection threshold increases the false association rate, making this classification very uncertain.

\section{Summary and conclusions}

The sensitivity of wide-field optical surveys now allows for the blind detection of galaxy clusters through their weak lensing signal. This situation will vastly improve with the launch of {\it Euclid} and the {\it Roman} Space Telescope. However, the construction of weak-lensing-selected galaxy cluster catalogues requires the improvement or development of new galaxy cluster detection algorithms able to cope with the increasingly large volume of data.

In this paper we have introduced a new, fast, multi-scale approach, based on wavelet filters operating on complementary scales, for detecting galaxy clusters through their weak lensing signal. This new method, W234, was compared to four commonly used approaches (S96, TANH, J04, and M18) based on AM filters. The comparison was performed on the same set of {\it Euclid}-like mocks obtained from the DUSTGRAIN-\textit{pathfinder} simulations, composed of 256 realisations of a field of 5$^\circ$ $\times$ 5$^\circ$, corresponding to a total area of 6\,400 deg$^2$.

We first undertook a thorough examination of the filter characteristics in real and Fourier space. This allowed us to demonstrate the equivalence between some of the commonly used AM filters (S96, TANH, J04, and M18) and wavelet filters at individual scales; moreover, we show that the M18 AM filter is comparable to a combination of two wavelet filters.
We then investigated different options for implementing the detection methods. Applying the detection methods to the 256 realisations, we find: 
\begin{itemize}
    \item Similar results are obtained when applying the detection algorithms to the shear directly at the galaxy positions in real space or to the binned shear in Fourier space. Binning a 5$^\circ$ $\times$ 5$^\circ$ field into $512\times512$ bins, the power spectrum integral of the AM map suffers a maximum 1.5\% loss when passing from real to Fourier space.
    \item The binning scheme that is adopted can have an impact on the detection efficiency in terms of the number and distribution of detections. In particular, binning by averaging the shear in each pixel yields a smaller number of detections than binning by summing the shear, and a significant fraction of these detections ($\sim 28\%$) are different from those detected in the real space approach. In contrast, the summed pixel binning yields very similar results, in terms of both number and distribution, to application to the shear in real space. 
    \item Application of the detection methods to the shear or the convergence yields similar results. In particular, the power spectrum loss on the AM map is negligible when passing from the shear to the convergence, and the number and distribution of detections agree exactly. 
\end{itemize}

To quantitatively compare the results obtained from the different filters, each having different characteristic radii, we developed a new adaptive matching procedure to match the detections to the haloes in the simulations. We find:
\begin{itemize}
    \item The characteristics of the halo catalogue used for the matching can have a dramatic effect on the number of false associations. We thus applied a pre-selection to the halo catalogue based on the theoretical weak lensing S/N expected for an NFW halo, and kept only haloes with S/N$>2$ for the subsequent matching step.
    \item The association distance is a critical parameter when dealing with filters that have different characteristic radii. It must take both the spatial resolution of the filter and the angular size of the haloes into account. We obtained an optimal matching distance for the detections from each filter by considering the absolute association distance, $D$, of all haloes as a function of the association distance normalised by the characteristic radius of each halo, $D/\theta_{200c}$. Then, decomposing the distribution of the absolute association distance in terms of RA and Dec and fitting with a Gaussian, we obtained an optimal MMD for each filter. 
    \item For the multi-scale wavelet detections, a halo might be detected on more than one successive scale. We thus carried out a special matching procedure for the detections from the wavelet filters, building a cumulative catalogue from the finest to the coarsest scales.
\end{itemize}

We then compared the performance of the detection methods, after running them on the 256 simulations and using the matching procedure developed in this paper. We find:
\begin{itemize} 
    \item The completeness is a strong function of the halo catalogue pre-selection threshold. However, irrespective of the exact S/N pre-selection threshold, the S96, TANH, and J04 filters exhibit a lower completeness than the M18 and W234 filters. In contrast, the J04 filter exhibits slightly higher purity at the expense of completeness.
    \item The evolution of the completeness as a function of purity shows that the multi-scale approach outperforms the methods based on AM filters at all values of purity.
    \item The multi-scale method detects practically all the detections from the individual AM approaches, with the addition of supplemental detections, particularly at lower masses and higher redshifts.
\end{itemize}

Finally, we performed an exhaustive analysis of the matched and unmatched detections for each method, using a S/N$>2$ pre-selection on the halo catalogue. We found similar results for all methods. For the W234 method, we find:
\begin{itemize}
    \item More than $82\%$ of the detections are correctly matched, and of these, the overwhelming majority (95\%) are due to a single halo. For our matching procedure, correctly matched detections due to multiple haloes along the LOS account for less than 5\% of the total. Furthermore, because our detection and matching methods aim at high purity, the number of incorrectly matched haloes is negligible ($<1\%$).
    \item About $17\%$ of the detections are unmatched. Only a small fraction of these ($10\%$) are due to false detections. The remainder can be explained mostly by haloes below the S/N threshold whose signal has been boosted because of a fortuitous redshift distribution of background sources ($37\%$), or a blended lensing signal from LOS alignments, or angular separation distances smaller than the filter size ($32\%$).
    The final 20\% of the unmatched detections can likely be associated with haloes with S/N$<1.5$ that do not appear in the catalogue due to the pre-selection.
\end{itemize}

In this work we have shown that a multi-scale approach is faster and gives better results, in terms of purity and completeness, than the currently used AM methods. We have also introduced a new adaptative matching procedure that allows a fair comparison between detection methods that operate on different scales. A significant  advantage of our method is its computational speed. Compared to methods that apply single-scale AM filters to the shear directly at the galaxy positions, our multi-scale method applied to the convergence yields a gain of two to three orders of magnitude in speed. 

Based on these results, we aim to apply our multi-scale approach to current and future large-area survey data, which will allow the construction of large catalogues of clusters selected through their weak lensing signal. This next step also involves new challenges. As detailed in \cite{Hamana2020}, the presence of foreground galaxies that are not part of the cluster can dilute the lensing signal. This dilution effect contaminates the weak lensing dataset and analyses. One way to tackle this issue is to use photometric redshifts to select sources and minimise contamination. By taking the redshift distribution within the dataset into account, the tomographic approach reduces the impact of the dilution effect. This solution has already shown promising results for the detection of galaxy clusters in the HSC survey \citep[e.g.][]{Hamana2020,Oguri2021}. The tomographic approach may also yield coarse information on the redshift of the detected sources. We are confident that a multi-scale approach combined with tomography will improve on the results shown here. Application to upcoming deep, large-area surveys will allow us to develop our method further.


\begin{acknowledgements}
The authors would like to thank Joel Bergé for providing his code to estimate the theoretical S/N of detection of a halo by weak lensing, Jean-Baptiste Melin for his advices and Peter Schneider for his helpful comments on the filters. GL acknowledges funding from CNES, the French space agency. GC acknowledges support from INAF theory Grant 2022: Illuminating Dark Matter using Weak Lensing by Cluster Satellites.

\end{acknowledgements}

\begin{appendix} 

\section{Contribution of each wavelet scale}
\label{Appendix B}

In this section we study the contribution of each wavelet scale to the matched detections made by the multi-scale approach.

\begin{figure*}[!hb]
\centering
\includegraphics[width=1.\columnwidth]{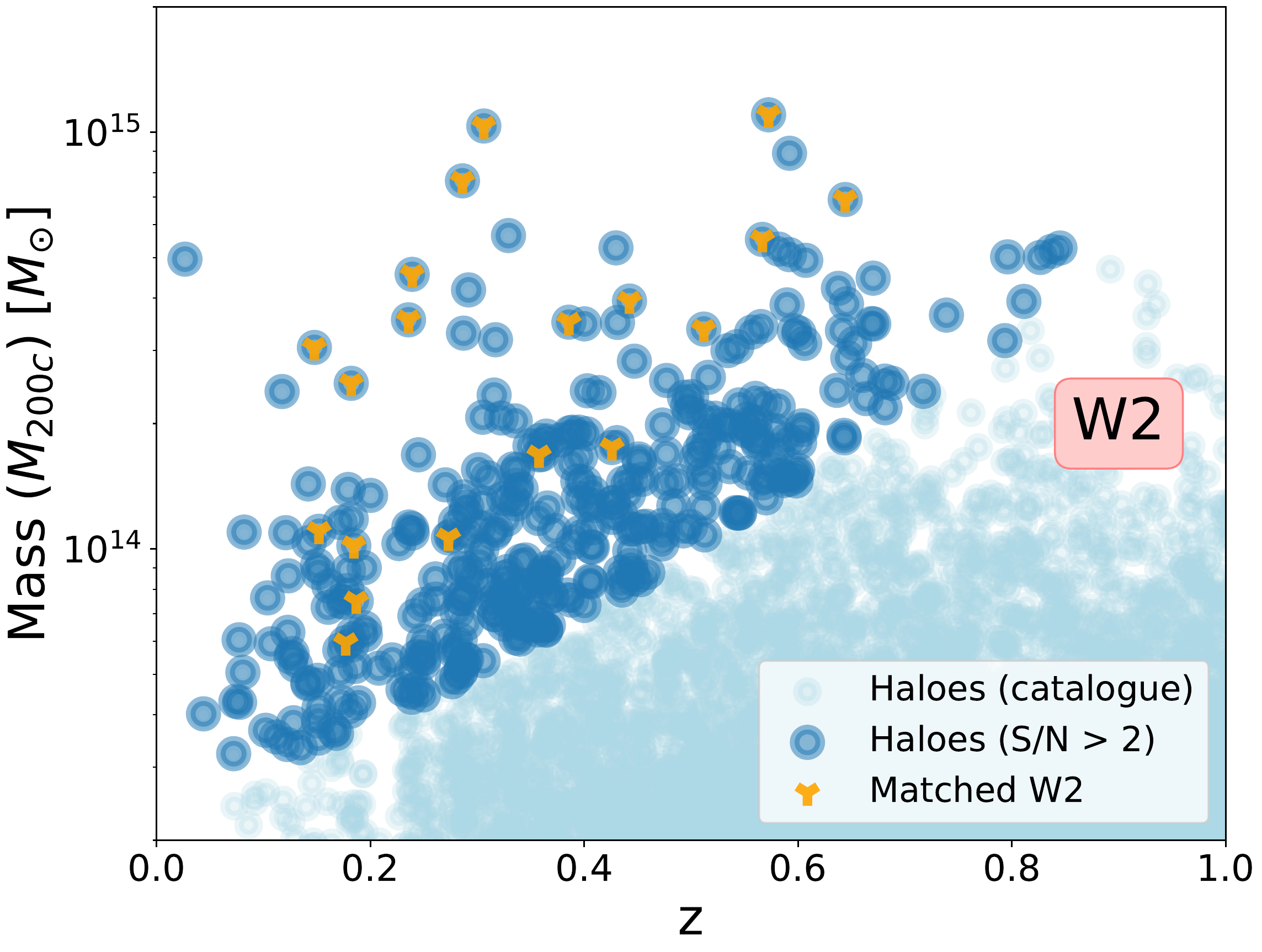}
\hfill
\includegraphics[width=1.\columnwidth]{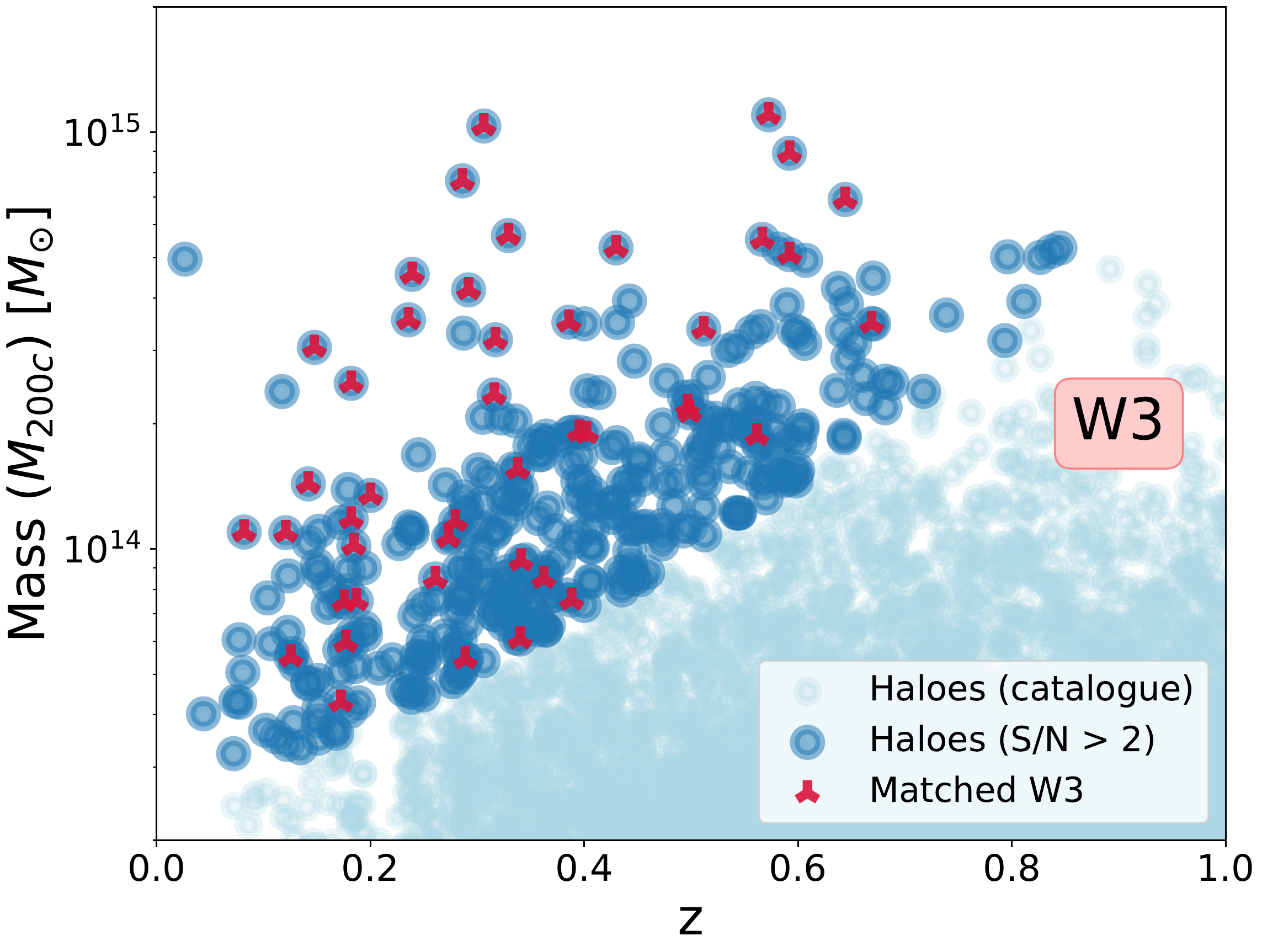}
\vfill
\centering
\includegraphics[width=1.\columnwidth]{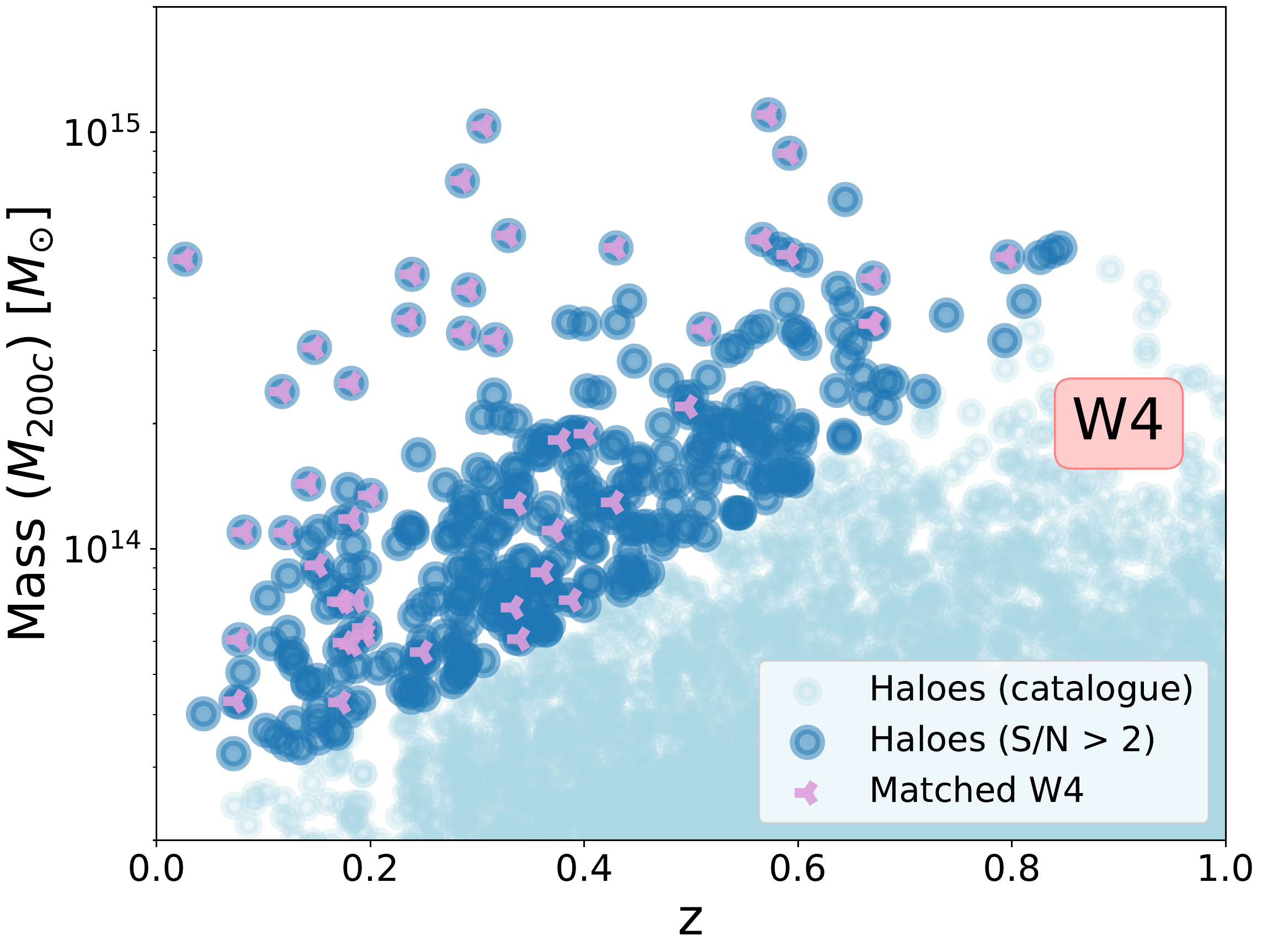}
\hfill
\includegraphics[width=1.\columnwidth]{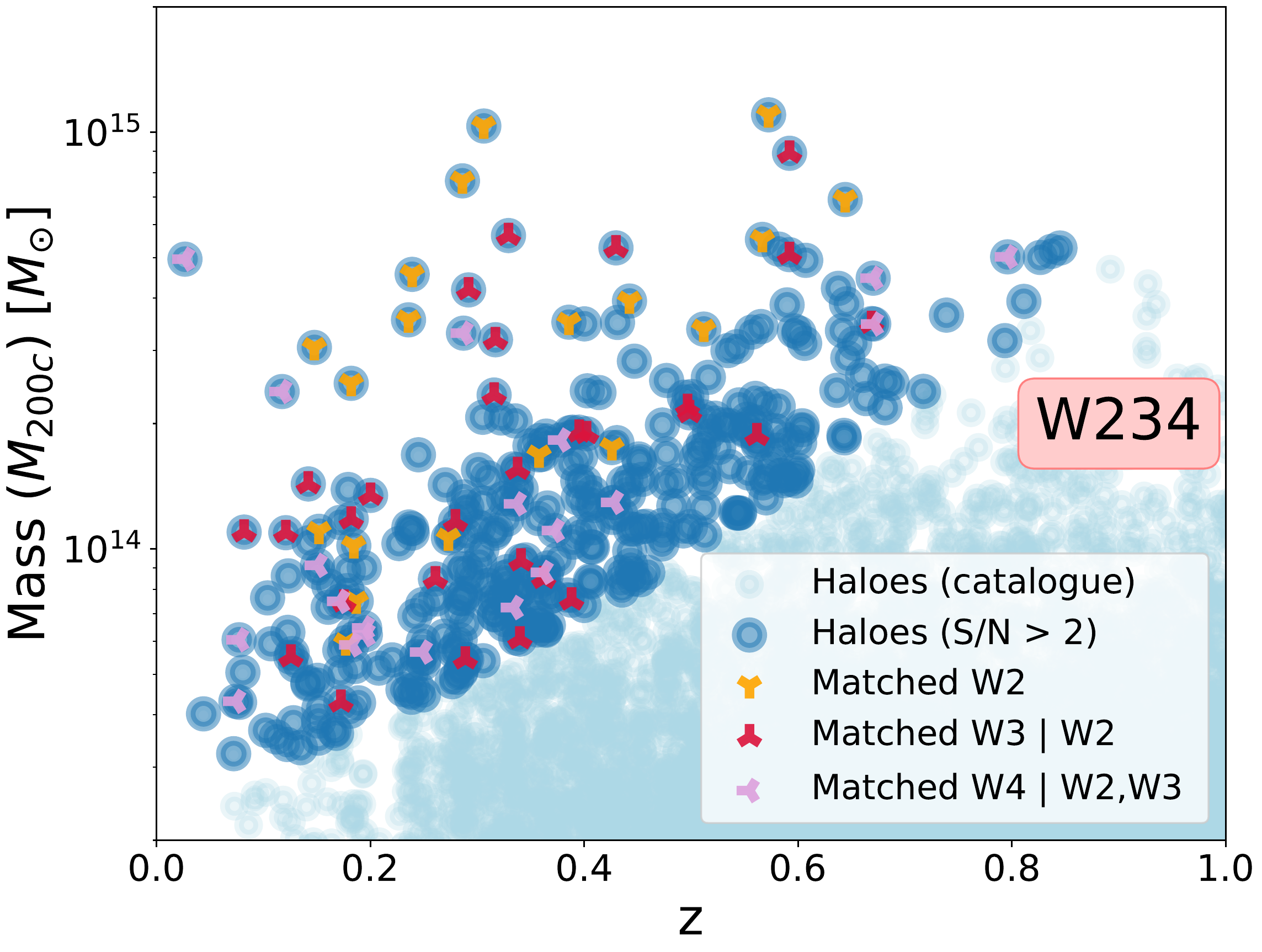}
\caption{Distribution of the matched clusters in the z-M plane for one $5^\circ \times 5^\circ$ patch, obtained using the W2 wavelet filter (upper-left panel), W3 wavelet filter (upper-right panel), W4 wavelet filter (lower-left panel), and the multi-scale approach (lower-right panel).}
\label{FIG: z-M all scales results}
\end{figure*}

Figure~\ref{FIG: z-M all scales results} shows the distribution of the matched clusters in the $z-M$ plane for the W2 wavelet filter (upper-left panel), the W3 wavelet filter (upper-right panel) and the W4 wavelet filter (lower-left panel). The lower-right panel shows the distribution of the matched clusters in the $z-M$ plane for the multi-scale detection method. The colours allow us to see the contribution of each scale after recombination of the multiple detections. The different panels also highlight the differences in terms of targeted clusters by the different wavelet scales. In particular, we see the complementarity of the different wavelet filters.

If we look at the number of matched detections for each wavelet scale after recombination, we see that the proportion of detections made at each scale does not vary much from one realisation to another. On average, the W2 wavelet filter provides up to $29\% \pm 0.39$, W3 around $33\% \pm 0.57,$ and W4 around $38\% \pm 0.42 $ of the matched detections. Hence, the three wavelet scales are useful and contribute similarly to the total number of matched detections. 

\end{appendix}


\begin{thebibliography}{79}
\expandafter\ifx\csname natexlab\endcsname\relax\def\natexlab#1{#1}\fi

\bibitem[{{Aihara} {et~al.}(2018){Aihara}, {Armstrong}, {Bickerton}, {Bosch},
  {Coupon}, {Furusawa}, {Hayashi}, {Ikeda}, {Kamata}, {Karoji}, {Kawanomoto},
  {Koike}, {Komiyama}, {Lang}, {Lupton}, {Mineo}, {Miyatake}, {Miyazaki},
  {Morokuma}, {Obuchi}, {Oishi}, {Okura}, {Price}, {Takata}, {Tanaka},
  {Tanaka}, {Tanaka}, {Uchida}, {Uraguchi}, {Utsumi}, {Wang}, {Yamada},
  {Yamanoi}, {Yasuda}, {Arimoto}, {Chiba}, {Finet}, {Fujimori}, {Fujimoto},
  {Furusawa}, {Goto}, {Goulding}, {Gunn}, {Harikane}, {Hattori}, {Hayashi},
  {He{\l}miniak}, {Higuchi}, {Hikage}, {Ho}, {Hsieh}, {Huang}, {Huang},
  {Imanishi}, {Iwata}, {Jaelani}, {Jian}, {Kashikawa}, {Katayama}, {Kojima},
  {Konno}, {Koshida}, {Kusakabe}, {Leauthaud}, {Lee}, {Lin}, {Lin},
  {Mandelbaum}, {Matsuoka}, {Medezinski}, {Miyama}, {Momose}, {More}, {More},
  {Mukae}, {Murata}, {Murayama}, {Nagao}, {Nakata}, {Niida}, {Niikura},
  {Nishizawa}, {Oguri}, {Okabe}, {Ono}, {Onodera}, {Onoue}, {Ouchi}, {Pyo},
  {Shibuya}, {Shimasaku}, {Simet}, {Speagle}, {Spergel}, {Strauss}, {Sugahara},
  {Sugiyama}, {Suto}, {Suzuki}, {Tait}, {Takada}, {Terai}, {Toba}, {Turner},
  {Uchiyama}, {Umetsu}, {Urata}, {Usuda}, {Yeh}, \& {Yuma}}]{Aihara2018}
{Aihara}, H., {Armstrong}, R., {Bickerton}, S., {et~al.} 2018, \pasj, 70, S8

\bibitem[{Allen {et~al.}(2011)Allen, Evrard, \& Mantz}]{Allen2011}
Allen, S.~W., Evrard, A.~E., \& Mantz, A.~B. 2011, Annual Review of Astronomy
  and Astrophysics, 49, 409–470

\bibitem[{Andreon \& Berg{\'{e} }(2012)}]{Andreon2012}
Andreon, S. \& Berg{\'{e} }, J. 2012, Astronomy {\&} Astrophysics, 547, A117

\bibitem[{Bartelmann \& Schneider(2001)}]{Bartelmann2001}
Bartelmann, M. \& Schneider, P. 2001, Physics Reports, 340, 291–472

\bibitem[{{Berg{\'e}} {et~al.}(2010){Berg{\'e}}, {Amara}, \&
  {R{\'e}fr{\'e}gier}}]{Berge2010}
{Berg{\'e}}, J., {Amara}, A., \& {R{\'e}fr{\'e}gier}, A. 2010, \apj, 712, 992

\bibitem[{{Bocquet} {et~al.}(2019){Bocquet}, {Dietrich}, {Schrabback}, {Bleem},
  {Klein}, {Allen}, {Applegate}, {Ashby}, {Bautz}, {Bayliss}, {Benson},
  {Brodwin}, {Bulbul}, {Canning}, {Capasso}, {Carlstrom}, {Chang}, {Chiu},
  {Cho}, {Clocchiatti}, {Crawford}, {Crites}, {de Haan}, {Desai}, {Dobbs},
  {Foley}, {Forman}, {Garmire}, {George}, {Gladders}, {Gonzalez}, {Grandis},
  {Gupta}, {Halverson}, {Hlavacek-Larrondo}, {Hoekstra}, {Holder}, {Holzapfel},
  {Hou}, {Hrubes}, {Huang}, {Jones}, {Khullar}, {Knox}, {Kraft}, {Lee}, {von
  der Linden}, {Luong-Van}, {Mantz}, {Marrone}, {McDonald}, {McMahon}, {Meyer},
  {Mocanu}, {Mohr}, {Morris}, {Padin}, {Patil}, {Pryke}, {Rapetti},
  {Reichardt}, {Rest}, {Ruhl}, {Saliwanchik}, {Saro}, {Sayre}, {Schaffer},
  {Shirokoff}, {Stalder}, {Stanford}, {Staniszewski}, {Stark}, {Story},
  {Strazzullo}, {Stubbs}, {Vanderlinde}, {Vieira}, {Vikhlinin}, {Williamson},
  \& {Zenteno}}]{Bocquet2019}
{Bocquet}, S., {Dietrich}, J.~P., {Schrabback}, T., {et~al.} 2019, \apj, 878,
  55

\bibitem[{{Closson Ferguson} {et~al.}(2009){Closson Ferguson}, {Armus},
  {Borne}, {Dickinson}, {Gawiser}, {Gilmore}, {Ivezic}, {Margoniner}, {Norman},
  {Obric}, {Obric}, {Rasmussen}, {Roskar}, {Schmidt}, {Seigar}, {Stanford},
  {Strauss}, {Wechsler}, {Newman}, {Tyson}, {Zentner}, \& {LSST Galaxies
  Collaboration}}]{Lsst2009}
{Closson Ferguson}, H., {Armus}, L., {Borne}, K., {et~al.} 2009, in American
  Astronomical Society Meeting Abstracts, Vol. 213, American Astronomical
  Society Meeting Abstracts \#213, 460.07

\bibitem[{{Cropper} {et~al.}(2013){Cropper}, {Hoekstra}, {Kitching}, {Massey},
  {Amiaux}, {Miller}, {Mellier}, {Rhodes}, {Rowe}, {Pires}, {Saxton}, \&
  {Scaramella}}]{Cropper2013}
{Cropper}, M., {Hoekstra}, H., {Kitching}, T., {et~al.} 2013, \mnras, 431, 3103

\bibitem[{{Davis} {et~al.}(1985){Davis}, {Efstathiou}, {Frenk}, \&
  {White}}]{Davis1985}
{Davis}, M., {Efstathiou}, G., {Frenk}, C.~S., \& {White}, S.~D.~M. 1985, \apj,
  292, 371

\bibitem[{{Diemer}(2018)}]{Diemer2018}
{Diemer}, B. 2018, \apjs, 239, 35

\bibitem[{{Diemer} \& {Joyce}(2019)}]{Diemer2019}
{Diemer}, B. \& {Joyce}, M. 2019, \apj, 871, 168

\bibitem[{{Dietrich} \& {Hartlap}(2010)}]{Dietrich2010}
{Dietrich}, J.~P. \& {Hartlap}, J. 2010, \mnras, 402, 1049

\bibitem[{{Euclid Collaboration} {et~al.}(2019){Euclid Collaboration}, {Adam},
  {Vannier}, {Maurogordato}, {Biviano}, {Adami}, {Ascaso}, {Bellagamba},
  {Benoist}, {Cappi}, {D{\'\i}az-S{\'a}nchez}, {Durret}, {Farrens}, {Gonzalez},
  {Iovino}, {Licitra}, {Maturi}, {Mei}, {Merson}, {Munari}, {Pell{\'o}},
  {Ricci}, {Rocci}, {Roncarelli}, {Sarron}, {Amoura}, {Andreon}, {Apostolakos},
  {Arnaud}, {Bardelli}, {Bartlett}, {Baugh}, {Borgani}, {Brodwin}, {Castander},
  {Castignani}, {Cucciati}, {De Lucia}, {Dubath}, {Fosalba}, {Giocoli},
  {Hoekstra}, {Mamon}, {Melin}, {Moscardini}, {Paltani}, {Radovich},
  {Sartoris}, {Schultheis}, {Sereno}, {Weller}, {Burigana}, {Carvalho},
  {Corcione}, {Kurki-Suonio}, {Lilje}, {Sirri}, {Toledo-Moreo}, \&
  {Zamorani}}]{Adam2019}
{Euclid Collaboration}, {Adam}, R., {Vannier}, M., {et~al.} 2019, \aap, 627,
  A23

\bibitem[{{Fan} {et~al.}(2010){Fan}, {Shan}, \& {Liu}}]{Fan2010}
{Fan}, Z., {Shan}, H., \& {Liu}, J. 2010, \apj, 719, 1408

\bibitem[{Gavazzi \& Soucail(2006)}]{Gavazzi2006}
Gavazzi, R. \& Soucail, G. 2006, Astronomy \& Astrophysics, 462, 459–471

\bibitem[{Giocoli {et~al.}(2018)Giocoli, Baldi, \& Moscardini}]{Giocoli2018}
Giocoli, C., Baldi, M., \& Moscardini, L. 2018, Monthly Notices of the Royal
  Astronomical Society, 481, 2813–2828

\bibitem[{{Giocoli} {et~al.}(2017){Giocoli}, {Di Meo}, {Meneghetti}, {Jullo},
  {de la Torre}, {Moscardini}, {Baldi}, {Mazzotta}, \& {Metcalf}}]{Giocoli2017}
{Giocoli}, C., {Di Meo}, S., {Meneghetti}, M., {et~al.} 2017, \mnras, 470, 3574

\bibitem[{{Giocoli} {et~al.}(2015){Giocoli}, {Metcalf}, {Baldi}, {Meneghetti},
  {Moscardini}, \& {Petkova}}]{Giocoli2014}
{Giocoli}, C., {Metcalf}, R.~B., {Baldi}, M., {et~al.} 2015, \mnras, 452, 2757

\bibitem[{{Hamana} {et~al.}(2012){Hamana}, {Oguri}, {Shirasaki}, \&
  {Sato}}]{Hamana2012}
{Hamana}, T., {Oguri}, M., {Shirasaki}, M., \& {Sato}, M. 2012, \mnras, 425,
  2287

\bibitem[{{Hamana} {et~al.}(2015){Hamana}, {Sakurai}, {Koike}, \&
  {Miller}}]{Hamana2015}
{Hamana}, T., {Sakurai}, J., {Koike}, M., \& {Miller}, L. 2015, \pasj, 67, 34

\bibitem[{Hamana {et~al.}(2020)Hamana, Shirasaki, \& Lin}]{Hamana2020}
Hamana, T., Shirasaki, M., \& Lin, Y.-T. 2020, Publications of the Astronomical
  Society of Japan, 72

\bibitem[{{Hamana} {et~al.}(2004){Hamana}, {Takada}, \& {Yoshida}}]{Hamana2004}
{Hamana}, T., {Takada}, M., \& {Yoshida}, N. 2004, \mnras, 350, 893

\bibitem[{{Hasselfield} {et~al.}(2013){Hasselfield}, {Hilton}, {Marriage},
  {Addison}, {Barrientos}, {Battaglia}, {Battistelli}, {Bond}, {Crichton},
  {Das}, {Devlin}, {Dicker}, {Dunkley}, {D{\"u}nner}, {Fowler}, {Gralla},
  {Hajian}, {Halpern}, {Hincks}, {Hlozek}, {Hughes}, {Infante}, {Irwin},
  {Kosowsky}, {Marsden}, {Menanteau}, {Moodley}, {Niemack}, {Nolta}, {Page},
  {Partridge}, {Reese}, {Schmitt}, {Sehgal}, {Sherwin}, {Sievers}, {Sif{\'o}n},
  {Spergel}, {Staggs}, {Swetz}, {Switzer}, {Thornton}, {Trac}, \&
  {Wollack}}]{Hasselfield2013}
{Hasselfield}, M., {Hilton}, M., {Marriage}, T.~A., {et~al.} 2013, \jcap, 2013,
  008

\bibitem[{Hennawi \& Spergel(2005)}]{Hennawi2005}
Hennawi, J. \& Spergel, D. 2005, The Astrophysical Journal, 624

\bibitem[{{Hetterscheidt} {et~al.}(2005){Hetterscheidt}, {Erben}, {Schneider},
  {Maoli}, {van Waerbeke}, \& {Mellier}}]{Hetterscheidt2005}
{Hetterscheidt}, M., {Erben}, T., {Schneider}, P., {et~al.} 2005, \aap, 442, 43

\bibitem[{{Ivezi{\'c}} {et~al.}(2019){Ivezi{\'c}}, {Kahn}, {Tyson}, {Abel},
  {Acosta}, {Allsman}, {Alonso}, {AlSayyad}, {Anderson}, {Andrew}, {Angel},
  {Angeli}, {Ansari}, {Antilogus}, {Araujo}, {Armstrong}, {Arndt}, {Astier},
  {Aubourg}, {Auza}, {Axelrod}, {Bard}, {Barr}, {Barrau}, {Bartlett}, {Bauer},
  {Bauman}, {Baumont}, {Bechtol}, {Bechtol}, {Becker}, {Becla}, {Beldica},
  {Bellavia}, {Bianco}, {Biswas}, {Blanc}, {Blazek}, {Blandford}, {Bloom},
  {Bogart}, {Bond}, {Booth}, {Borgland}, {Borne}, {Bosch}, {Boutigny},
  {Brackett}, {Bradshaw}, {Brandt}, {Brown}, {Bullock}, {Burchat}, {Burke},
  {Cagnoli}, {Calabrese}, {Callahan}, {Callen}, {Carlin}, {Carlson},
  {Chandrasekharan}, {Charles-Emerson}, {Chesley}, {Cheu}, {Chiang}, {Chiang},
  {Chirino}, {Chow}, {Ciardi}, {Claver}, {Cohen-Tanugi}, {Cockrum}, {Coles},
  {Connolly}, {Cook}, {Cooray}, {Covey}, {Cribbs}, {Cui}, {Cutri}, {Daly},
  {Daniel}, {Daruich}, {Daubard}, {Daues}, {Dawson}, {Delgado}, {Dellapenna},
  {de Peyster}, {de Val-Borro}, {Digel}, {Doherty}, {Dubois},
  {Dubois-Felsmann}, {Durech}, {Economou}, {Eifler}, {Eracleous}, {Emmons},
  {Fausti Neto}, {Ferguson}, {Figueroa}, {Fisher-Levine}, {Focke}, {Foss},
  {Frank}, {Freemon}, {Gangler}, {Gawiser}, {Geary}, {Gee}, {Geha}, {Gessner},
  {Gibson}, {Gilmore}, {Glanzman}, {Glick}, {Goldina}, {Goldstein}, {Goodenow},
  {Graham}, {Gressler}, {Gris}, {Guy}, {Guyonnet}, {Haller}, {Harris},
  {Hascall}, {Haupt}, {Hernandez}, {Herrmann}, {Hileman}, {Hoblitt}, {Hodgson},
  {Hogan}, {Howard}, {Huang}, {Huffer}, {Ingraham}, {Innes}, {Jacoby}, {Jain},
  {Jammes}, {Jee}, {Jenness}, {Jernigan}, {Jevremovi{\'c}}, {Johns}, {Johnson},
  {Johnson}, {Jones}, {Juramy-Gilles}, {Juri{\'c}}, {Kalirai}, {Kallivayalil},
  {Kalmbach}, {Kantor}, {Karst}, {Kasliwal}, {Kelly}, {Kessler}, {Kinnison},
  {Kirkby}, {Knox}, {Kotov}, {Krabbendam}, {Krughoff}, {Kub{\'a}nek},
  {Kuczewski}, {Kulkarni}, {Ku}, {Kurita}, {Lage}, {Lambert}, {Lange},
  {Langton}, {Le Guillou}, {Levine}, {Liang}, {Lim}, {Lintott}, {Long},
  {Lopez}, {Lotz}, {Lupton}, {Lust}, {MacArthur}, {Mahabal}, {Mandelbaum},
  {Markiewicz}, {Marsh}, {Marshall}, {Marshall}, {May}, {McKercher}, {McQueen},
  {Meyers}, {Migliore}, {Miller}, {Mills}, {Miraval}, {Moeyens}, {Moolekamp},
  {Monet}, {Moniez}, {Monkewitz}, {Montgomery}, {Morrison}, {Mueller},
  {Muller}, {Mu{\~n}oz Arancibia}, {Neill}, {Newbry}, {Nief}, {Nomerotski},
  {Nordby}, {O'Connor}, {Oliver}, {Olivier}, {Olsen}, {O'Mullane}, {Ortiz},
  {Osier}, {Owen}, {Pain}, {Palecek}, {Parejko}, {Parsons}, {Pease},
  {Peterson}, {Peterson}, {Petravick}, {Libby Petrick}, {Petry},
  {Pierfederici}, {Pietrowicz}, {Pike}, {Pinto}, {Plante}, {Plate}, {Plutchak},
  {Price}, {Prouza}, {Radeka}, {Rajagopal}, {Rasmussen}, {Regnault}, {Reil},
  {Reiss}, {Reuter}, {Ridgway}, {Riot}, {Ritz}, {Robinson}, {Roby}, {Roodman},
  {Rosing}, {Roucelle}, {Rumore}, {Russo}, {Saha}, {Sassolas}, {Schalk},
  {Schellart}, {Schindler}, {Schmidt}, {Schneider}, {Schneider}, {Schoening},
  {Schumacher}, {Schwamb}, {Sebag}, {Selvy}, {Sembroski}, {Seppala}, {Serio},
  {Serrano}, {Shaw}, {Shipsey}, {Sick}, {Silvestri}, {Slater}, {Smith},
  {Smith}, {Sobhani}, {Soldahl}, {Storrie-Lombardi}, {Stover}, {Strauss},
  {Street}, {Stubbs}, {Sullivan}, {Sweeney}, {Swinbank}, {Szalay}, {Takacs},
  {Tether}, {Thaler}, {Thayer}, {Thomas}, {Thornton}, {Thukral}, {Tice},
  {Trilling}, {Turri}, {Van Berg}, {Vanden Berk}, {Vetter}, {Virieux},
  {Vucina}, {Wahl}, {Walkowicz}, {Walsh}, {Walter}, {Wang}, {Wang}, {Warner},
  {Wiecha}, {Willman}, {Winters}, {Wittman}, {Wolff}, {Wood-Vasey}, {Wu},
  {Xin}, {Yoachim}, \& {Zhan}}]{Ivezic2019}
{Ivezi{\'c}}, {\v{Z}}., {Kahn}, S.~M., {Tyson}, J.~A., {et~al.} 2019, \apj,
  873, 111

\bibitem[{Jarvis {et~al.}(2004)Jarvis, Bernstein, \& Jain}]{Jarvis2004}
Jarvis, M., Bernstein, G., \& Jain, B. 2004, Monthly Notices of the Royal
  Astronomical Society, 352, 338–352

\bibitem[{{Kaiser} \& {Squires}(1993)}]{KandS1993}
{Kaiser}, N. \& {Squires}, G. 1993, \apj, 404, 441

\bibitem[{{Lanusse} {et~al.}(2016){Lanusse}, {Starck}, {Leonard}, \&
  {Pires}}]{Lanusse2016}
{Lanusse}, F., {Starck}, J.~L., {Leonard}, A., \& {Pires}, S. 2016, \aap, 591,
  A2

\bibitem[{Laureijs {et~al.}(2011)Laureijs, Amiaux, Arduini, Auguères,
  Brinchmann, Cole, Cropper, Dabin, Duvet, Ealet, Garilli, Gondoin, Guzzo,
  Hoar, Hoekstra, Holmes, Kitching, Maciaszek, Mellier, Pasian, Percival,
  Rhodes, Criado, Sauvage, Scaramella, Valenziano, Warren, Bender, Castander,
  Cimatti, Fèvre, Kurki-Suonio, Levi, Lilje, Meylan, Nichol, Pedersen, Popa,
  Lopez, Rix, Rottgering, Zeilinger, Grupp, Hudelot, Massey, Meneghetti,
  Miller, Paltani, Paulin-Henriksson, Pires, Saxton, Schrabback, Seidel, Walsh,
  Aghanim, Amendola, Bartlett, Baccigalupi, Beaulieu, Benabed, Cuby, Elbaz,
  Fosalba, Gavazzi, Helmi, Hook, Irwin, Kneib, Kunz, Mannucci, Moscardini, Tao,
  Teyssier, Weller, Zamorani, Osorio, Boulade, Foumond, Di~Giorgio, Guttridge,
  James, Kemp, Martignac, Spencer, Walton, Blümchen, Bonoli, Bortoletto,
  Cerna, Corcione, Fabron, Jahnke, Ligori, Madrid, Martin, Morgante, Pamplona,
  Prieto, Riva, Toledo, Trifoglio, Zerbi, Abdalla, Douspis, Grenet, Borgani,
  Bouwens, Courbin, Delouis, Dubath, Fontana, Frailis, Grazian, Koppenhöfer,
  Mansutti, Melchior, Mignoli, Mohr, Neissner, Noddle, Poncet, Scodeggio,
  Serrano, Shane, Starck, Surace, Taylor, Verdoes-Kleijn, Vuerli, Williams,
  Zacchei, Altieri, Sanz, Kohley, Oosterbroek, Astier, Bacon, Bardelli, Baugh,
  Bellagamba, Benoist, Bianchi, Biviano, Branchini, Carbone, Cardone, Clements,
  Colombi, Conselice, Cresci, Deacon, Dunlop, Fedeli, Fontanot, Franzetti,
  Giocoli, Garcia-Bellido, Gow, Heavens, Hewett, Heymans, Holland, Huang,
  Ilbert, Joachimi, Jennins, Kerins, Kiessling, Kirk, Kotak, Krause, Lahav, van
  Leeuwen, Lesgourgues, Lombardi, Magliocchetti, Maguire, Majerotto, Maoli,
  Marulli, Maurogordato, McCracken, McLure, Melchiorri, Merson, Moresco,
  Nonino, Norberg, Peacock, Pello, Penny, Pettorino, Di~Porto, Pozzetti,
  Quercellini, Radovich, Rassat, Roche, Ronayette, Rossetti, Sartoris,
  Schneider, Semboloni, Serjeant, Simpson, Skordis, Smadja, Smartt, Spano,
  Spiro, Sullivan, Tilquin, Trotta, Verde, Wang, Williger, Zhao, Zoubian, \&
  Zucca}]{Laureijs2011}
Laureijs, R., Amiaux, J., Arduini, S., {et~al.} 2011, Euclid Definition Study
  Report

\bibitem[{{Leauthaud} {et~al.}(2007){Leauthaud}, {Massey}, {Kneib}, {Rhodes},
  {Johnston}, {Capak}, {Heymans}, {Ellis}, {Koekemoer}, {Le F{\`e}vre},
  {Mellier}, {R{\'e}fr{\'e}gier}, {Robin}, {Scoville}, {Tasca}, {Taylor}, \&
  {Van Waerbeke}}]{Leauthaud2007}
{Leauthaud}, A., {Massey}, R., {Kneib}, J.-P., {et~al.} 2007, \apjs, 172, 219

\bibitem[{{Leonard} {et~al.}(2015){Leonard}, {Lanusse}, \&
  {Starck}}]{Leonard2015}
{Leonard}, A., {Lanusse}, F., \& {Starck}, J.-L. 2015, \mnras, 449, 1146

\bibitem[{Leonard {et~al.}(2012)Leonard, Pires, \& Starck}]{Leonard2012}
Leonard, A., Pires, S., \& Starck, J.-L. 2012, Monthly Notices of the Royal
  Astronomical Society, 423, 3405–3412

\bibitem[{{Lesci} {et~al.}(2022){Lesci}, {Marulli}, {Moscardini}, {Sereno},
  {Veropalumbo}, {Maturi}, {Giocoli}, {Radovich}, {Bellagamba}, {Roncarelli},
  {Bardelli}, {Contarini}, {Covone}, {Ingoglia}, {Nanni}, \&
  {Puddu}}]{Lesci2022}
{Lesci}, G.~F., {Marulli}, F., {Moscardini}, L., {et~al.} 2022, \aap, 659, A88

\bibitem[{{Lin} {et~al.}(2016){Lin}, {Kilbinger}, \& {Pires}}]{Lin2016}
{Lin}, C.-A., {Kilbinger}, M., \& {Pires}, S. 2016, \aap, 593, A88

\bibitem[{Maturi {et~al.}(2005)Maturi, Meneghetti, Bartelmann, Dolag, \&
  Moscardini}]{Maturi2005}
Maturi, M., Meneghetti, M., Bartelmann, M., Dolag, K., \& Moscardini, L. 2005,
  Astronomy \& Astrophysics, 442, 851–860

\bibitem[{{Maturi} {et~al.}(2007){Maturi}, {Schirmer}, {Meneghetti},
  {Bartelmann}, \& {Moscardini}}]{Maturi2007}
{Maturi}, M., {Schirmer}, M., {Meneghetti}, M., {Bartelmann}, M., \&
  {Moscardini}, L. 2007, \aap, 462, 473

\bibitem[{{Miyazaki} {et~al.}(2007){Miyazaki}, {Hamana}, {Ellis}, {Kashikawa},
  {Massey}, {Taylor}, \& {Refregier}}]{Miyazaki2007}
{Miyazaki}, S., {Hamana}, T., {Ellis}, R.~S., {et~al.} 2007, \apj, 669, 714

\bibitem[{{Miyazaki} {et~al.}(2002){Miyazaki}, {Hamana}, {Shimasaku},
  {Furusawa}, {Doi}, {Hamabe}, {Imi}, {Kimura}, {Komiyama}, {Nakata}, {Okada},
  {Okamura}, {Ouchi}, {Sekiguchi}, {Yagi}, \& {Yasuda}}]{Miyazaki2002}
{Miyazaki}, S., {Hamana}, T., {Shimasaku}, K., {et~al.} 2002, \apjl, 580, L97

\bibitem[{Miyazaki {et~al.}(2018)Miyazaki, Oguri, Hamana, Shirasaki, Koike,
  Komiyama, Umetsu, Utsumi, Okabe, More, Medezinski, Lin, Miyatake, Murayama,
  Ota, \& Mitsuishi}]{Miyazaki2017}
Miyazaki, S., Oguri, M., Hamana, T., {et~al.} 2018, Publications of the
  Astronomical Society of Japan, 70

\bibitem[{{Navarro} {et~al.}(1996){Navarro}, {Frenk}, \& {White}}]{Navarro1996}
{Navarro}, J.~F., {Frenk}, C.~S., \& {White}, S. D.~M. 1996, \apj, 462, 563

\bibitem[{Oguri {et~al.}(2021)Oguri, Miyazaki, Li, Luo, Mitsuishi, Miyatake,
  More, Nishizawa, Okabe, Ota, Plazas~Malagón, \& Utsumi}]{Oguri2021}
Oguri, M., Miyazaki, S., Li, X., {et~al.} 2021, Publications of the
  Astronomical Society of Japan, 73, 817–829

\bibitem[{{Pacaud} {et~al.}(2006){Pacaud}, {Pierre}, {Refregier}, {Gueguen},
  {Starck}, {Valtchanov}, {Read}, {Altieri}, {Chiappetti}, {Gandhi}, {Garcet},
  {Gosset}, {Ponman}, \& {Surdej}}]{Pacaud2006}
{Pacaud}, F., {Pierre}, M., {Refregier}, A., {et~al.} 2006, \mnras, 372, 578

\bibitem[{{Pace} {et~al.}(2007){Pace}, {Maturi}, {Meneghetti}, {Bartelmann},
  {Moscardini}, \& {Dolag}}]{Pace2007}
{Pace}, F., {Maturi}, M., {Meneghetti}, M., {et~al.} 2007, \aap, 471, 731

\bibitem[{{Perrenod}(1980)}]{Perrenod1980}
{Perrenod}, S.~C. 1980, \apj, 236, 373

\bibitem[{{Pires} {et~al.}(2012){Pires}, {Leonard}, \& {Starck}}]{Pires2012}
{Pires}, S., {Leonard}, A., \& {Starck}, J.-L. 2012, \mnras, 423, 983

\bibitem[{Pires {et~al.}(2020)Pires, Vandenbussche, Kansal, Bender, Blot,
  Bonino, Boucaud, Brinchmann, Capobianco, Carretero, Castellano, Cavuoti,
  Cl{\'{e} }dassou, Congedo, Conversi, Corcione, Dubath, Fosalba, Frailis,
  Franceschi, Fumana, Grupp, Hormuth, Kermiche, Knabenhans, Kohley, Kubik,
  Kunz, Ligori, Lilje, Lloro, Maiorano, Marggraf, Massey, Meylan, Padilla,
  Paltani, Pasian, Poncet, Potter, Raison, Rhodes, Roncarelli, Saglia,
  Schneider, Secroun, Serrano, Stadel, Cresp{\'{\i}}, Tereno, Toledo-Moreo, \&
  Wang}]{Pires2020}
Pires, S., Vandenbussche, V., Kansal, V., {et~al.} 2020, Astronomy {\&}
  Astrophysics, 638, A141

\bibitem[{{Planck Collaboration} {et~al.}(2014){Planck Collaboration}, {Ade},
  {Aghanim}, {Armitage-Caplan}, {Arnaud}, {Ashdown}, {Atrio-Barandela},
  {Aumont}, {Aussel}, {Baccigalupi}, {Banday}, {Barreiro}, {Barrena},
  {Bartelmann}, {Bartlett}, {Battaner}, {Benabed}, {Beno{\^\i}t},
  {Benoit-L{\'e}vy}, {Bernard}, {Bersanelli}, {Bielewicz}, {Bikmaev}, {Bobin},
  {Bock}, {B{\"o}hringer}, {Bonaldi}, {Bond}, {Borrill}, {Bouchet}, {Bridges},
  {Bucher}, {Burenin}, {Burigana}, {Butler}, {Cardoso}, {Carvalho}, {Catalano},
  {Challinor}, {Chamballu}, {Chary}, {Chen}, {Chiang}, {Chiang}, {Chon},
  {Christensen}, {Churazov}, {Church}, {Clements}, {Colombi}, {Colombo},
  {Comis}, {Couchot}, {Coulais}, {Crill}, {Curto}, {Cuttaia}, {Da Silva},
  {Dahle}, {Danese}, {Davies}, {Davis}, {de Bernardis}, {de Rosa}, {de Zotti},
  {Delabrouille}, {Delouis}, {D{\'e}mocl{\`e}s}, {D{\'e}sert}, {Dickinson},
  {Diego}, {Dolag}, {Dole}, {Donzelli}, {Dor{\'e}}, {Douspis}, {Dupac},
  {Efstathiou}, {Eisenhardt}, {En{\ss}lin}, {Eriksen}, {Feroz}, {Finelli},
  {Flores-Cacho}, {Forni}, {Frailis}, {Franceschi}, {Fromenteau}, {Galeotta},
  {Ganga}, {G{\'e}nova-Santos}, {Giard}, {Giardino}, {Gilfanov},
  {Giraud-H{\'e}raud}, {Gonz{\'a}lez-Nuevo}, {G{\'o}rski}, {Grainge},
  {Gratton}, {Gregorio}, {Groeneboom}, {Gruppuso}, {Hansen}, {Hanson},
  {Harrison}, {Hempel}, {Henrot-Versill{\'e}}, {Hern{\'a}ndez-Monteagudo},
  {Herranz}, {Hildebrandt}, {Hivon}, {Hobson}, {Holmes}, {Hornstrup}, {Hovest},
  {Huffenberger}, {Hurier}, {Hurley-Walker}, {Jaffe}, {Jaffe}, {Jones},
  {Juvela}, {Keih{\"a}nen}, {Keskitalo}, {Khamitov}, {Kisner}, {Kneissl},
  {Knoche}, {Knox}, {Kunz}, {Kurki-Suonio}, {Lagache}, {L{\"a}hteenm{\"a}ki},
  {Lamarre}, {Lasenby}, {Laureijs}, {Lawrence}, {Leahy}, {Leonardi},
  {Le{\'o}n-Tavares}, {Lesgourgues}, {Li}, {Liddle}, {Liguori}, {Lilje},
  {Linden-V{\o}rnle}, {L{\'o}pez-Caniego}, {Lubin}, {Mac{\'\i}as-P{\'e}rez},
  {MacTavish}, {Maffei}, {Maino}, {Mandolesi}, {Maris}, {Marshall}, {Martin},
  {Mart{\'\i}nez-Gonz{\'a}lez}, {Masi}, {Massardi}, {Matarrese}, {Matthai},
  {Mazzotta}, {Mei}, {Meinhold}, {Melchiorri}, {Melin}, {Mendes}, {Mennella},
  {Migliaccio}, {Mikkelsen}, {Mitra}, {Miville-Desch{\^e}nes}, {Moneti},
  {Montier}, {Morgante}, {Mortlock}, {Munshi}, {Murphy}, {Naselsky}, {Nati},
  {Natoli}, {Nesvadba}, {Netterfield}, {N{\o}rgaard-Nielsen}, {Noviello},
  {Novikov}, {Novikov}, {O'Dwyer}, {Olamaie}, {Osborne}, {Oxborrow}, {Paci},
  {Pagano}, {Pajot}, {Paoletti}, {Pasian}, {Patanchon}, {Pearson}, {Perdereau},
  {Perotto}, {Perrott}, {Perrotta}, {Piacentini}, {Piat}, {Pierpaoli},
  {Pietrobon}, {Plaszczynski}, {Pointecouteau}, {Polenta}, {Ponthieu}, {Popa},
  {Poutanen}, {Pratt}, {Pr{\'e}zeau}, {Prunet}, {Puget}, {Rachen}, {Reach},
  {Rebolo}, {Reinecke}, {Remazeilles}, {Renault}, {Ricciardi}, {Riller},
  {Ristorcelli}, {Rocha}, {Rosset}, {Roudier}, {Rowan-Robinson},
  {Rubi{\~n}o-Mart{\'\i}n}, {Rumsey}, {Rusholme}, {Sandri}, {Santos},
  {Saunders}, {Savini}, {Schammel}, {Scott}, {Seiffert}, {Shellard},
  {Shimwell}, {Spencer}, {Stanford}, {Starck}, {Stolyarov}, {Stompor},
  {Sudiwala}, {Sunyaev}, {Sureau}, {Sutton}, {Suur-Uski}, {Sygnet}, {Tauber},
  {Tavagnacco}, {Terenzi}, {Toffolatti}, {Tomasi}, {Tristram}, {Tucci},
  {Tuovinen}, {T{\"u}rler}, {Umana}, {Valenziano}, {Valiviita}, {Van Tent},
  {Vibert}, {Vielva}, {Villa}, {Vittorio}, {Wade}, {Wandelt}, {White}, {White},
  {Yvon}, {Zacchei}, \& {Zonca}}]{Planck2014}
{Planck Collaboration}, {Ade}, P.~A.~R., {Aghanim}, N., {et~al.} 2014, \aap,
  571, A29

\bibitem[{{Planck Collaboration XIII}(2016)}]{Planck2016}
{Planck Collaboration XIII}. 2016, \aap, 594, A13

\bibitem[{Pratt {et~al.}(2019)Pratt, Arnaud, Biviano, Eckert, Ettori, Nagai,
  Okabe, \& Reiprich}]{Pratt2019}
Pratt, G.~W., Arnaud, M., Biviano, A., {et~al.} 2019, Space Science Reviews,
  215

\bibitem[{{Pyne} \& {Birkinshaw}(1993)}]{Pyne1993}
{Pyne}, T. \& {Birkinshaw}, M. 1993, \apj, 415, 459

\bibitem[{{Rozo} {et~al.}(2010){Rozo}, {Wechsler}, {Rykoff}, {Annis}, {Becker},
  {Evrard}, {Frieman}, {Hansen}, {Hao}, {Johnston}, {Koester}, {McKay},
  {Sheldon}, \& {Weinberg}}]{Rozo2010}
{Rozo}, E., {Wechsler}, R.~H., {Rykoff}, E.~S., {et~al.} 2010, \apj, 708, 645

\bibitem[{{Salvati} {et~al.}(2021){Salvati}, {Saro}, {Bocquet}, {Costanzi},
  {Ansarinejad}, {Benson}, {Bleem}, {Calzadilla}, {Carlstrom}, {Chang},
  {Chown}, {Crites}, {deHaan}, {Dobbs}, {Everett}, {Floyd}, {Grandis},
  {George}, {Halverson}, {Holder}, {Holzapfel}, {Hrubes}, {Lee}, {Luong-Van},
  {McDonald}, {McMahon}, {Meyer}, {Millea}, {Mocanu}, {Mohr}, {Natoli},
  {Omori}, {Padin}, {Pryke}, {Reichardt}, {Ruhl}, {Ruppin}, {Schaffer},
  {Schrabback}, {Shirokoff}, {Staniszewski}, {Stark}, {Vieira}, \&
  {Williamson}}]{Salvati2021}
{Salvati}, L., {Saro}, A., {Bocquet}, S., {et~al.} 2021, arXiv e-prints,
  arXiv:2112.03606

\bibitem[{{Schirmer}(2004)}]{Schirmer2004Thesis}
{Schirmer}. 2004, PhD thesis, Rheinische Friedrich-Wilhelms-Universität Bonn

\bibitem[{{Schirmer} {et~al.}(2007){Schirmer}, {Erben}, {Hetterscheidt}, \&
  {Schneider}}]{Schirmer2007}
{Schirmer}, M., {Erben}, T., {Hetterscheidt}, M., \& {Schneider}, P. 2007,
  \aap, 462, 875

\bibitem[{{Schirmer} {et~al.}(2004){Schirmer}, {Erben}, {Schneider}, {Wolf}, \&
  {Meisenheimer}}]{Schirmer2004}
{Schirmer}, M., {Erben}, T., {Schneider}, P., {Wolf}, C., \& {Meisenheimer}, K.
  2004, \aap, 420, 75

\bibitem[{Schneider(1996)}]{Schneider1996}
Schneider, P. 1996, Monthly Notices of the Royal Astronomical Society, 283,
  837–853

\bibitem[{{Schneider} \& {Bartelmann}(1997)}]{Schneider1997}
{Schneider}, P. \& {Bartelmann}, M. 1997, \mnras, 286, 696

\bibitem[{{Schneider} {et~al.}(1992){Schneider}, {Ehlers}, \&
  {Falco}}]{Schneider1992}
{Schneider}, P., {Ehlers}, J., \& {Falco}, E.~E. 1992, {Gravitational Lenses}

\bibitem[{{Schneider} {et~al.}(1998){Schneider}, {van Waerbeke}, {Jain}, \&
  {Kruse}}]{Schneider1998}
{Schneider}, P., {van Waerbeke}, L., {Jain}, B., \& {Kruse}, G. 1998, \mnras,
  296, 873

\bibitem[{{Schneider} {et~al.}(2002){Schneider}, {van Waerbeke}, {Kilbinger},
  \& {Mellier}}]{Schneider2002}
{Schneider}, P., {van Waerbeke}, L., {Kilbinger}, M., \& {Mellier}, Y. 2002,
  \aap, 396, 1

\bibitem[{{Schrabback} {et~al.}(2015){Schrabback}, {Hilbert}, {Hoekstra},
  {Simon}, {van Uitert}, {Erben}, {Heymans}, {Hildebrandt}, {Kitching},
  {Mellier}, {Miller}, {Van Waerbeke}, {Bett}, {Coupon}, {Fu}, {Hudson},
  {Joachimi}, {Kilbinger}, \& {Kuijken}}]{Schrabback2015}
{Schrabback}, T., {Hilbert}, S., {Hoekstra}, H., {et~al.} 2015, \mnras, 454,
  1432

\bibitem[{{Schrabback} {et~al.}(2018){Schrabback}, {Schirmer}, {van der Burg},
  {Hoekstra}, {Buddendiek}, {Applegate}, {Brada{\v{c}}}, {Eifler}, {Erben},
  {Gladders}, {Hern{\'a}ndez-Mart{\'\i}n}, {Hildebrandt}, {Hoag}, {Klaes}, {von
  der Linden}, {Marchesini}, {Muzzin}, {Sharon}, \&
  {Stefanon}}]{Schrabback2018}
{Schrabback}, T., {Schirmer}, M., {van der Burg}, R. F.~J., {et~al.} 2018,
  \aap, 610, A85

\bibitem[{{Seitz} \& {Schneider}(1997)}]{Seitz1997}
{Seitz}, C. \& {Schneider}, P. 1997, \aap, 318, 687

\bibitem[{{Shan} {et~al.}(2012){Shan}, {Kneib}, {Tao}, {Fan}, {Jauzac},
  {Limousin}, {Massey}, {Rhodes}, {Thanjavur}, \& {McCracken}}]{Shan2012}
{Shan}, H., {Kneib}, J.-P., {Tao}, C., {et~al.} 2012, \apj, 748, 56

\bibitem[{{Shan} {et~al.}(2018){Shan}, {Liu}, {Hildebrandt}, {Pan}, {Martinet},
  {Fan}, {Schneider}, {Asgari}, {Harnois-D{\'e}raps}, {Hoekstra}, {Wright},
  {Dietrich}, {Erben}, {Getman}, {Grado}, {Heymans}, {Klaes}, {Kuijken},
  {Merten}, {Puddu}, {Radovich}, \& {Wang}}]{Shan2018}
{Shan}, H., {Liu}, X., {Hildebrandt}, H., {et~al.} 2018, \mnras, 474, 1116

\bibitem[{Spergel {et~al.}(2015)Spergel, Gehrels, Baltay, Bennett,
  Breckinridge, Donahue, Dressler, Gaudi, Greene, Guyon, Hirata, Kalirai,
  Kasdin, Macintosh, Moos, Perlmutter, Postman, Rauscher, Rhodes, Wang,
  Weinberg, Benford, Hudson, Jeong, Mellier, Traub, Yamada, Capak, Colbert,
  Masters, Penny, Savransky, Stern, Zimmerman, Barry, Bartusek, Carpenter,
  Cheng, Content, Dekens, Demers, Grady, Jackson, Kuan, Kruk, Melton, Nemati,
  Parvin, Poberezhskiy, Peddie, Ruffa, Wallace, Whipple, Wollack, \&
  Zhao}]{Spergel2015}
Spergel, D., Gehrels, N., Baltay, C., {et~al.} 2015, Wide-Field InfrarRed
  Survey Telescope-Astrophysics Focused Telescope Assets WFIRST-AFTA 2015
  Report

\bibitem[{{Springel} {et~al.}(2001){Springel}, {White}, {Tormen}, \&
  {Kauffmann}}]{Springel2001}
{Springel}, V., {White}, S. D.~M., {Tormen}, G., \& {Kauffmann}, G. 2001,
  \mnras, 328, 726

\bibitem[{Starck {et~al.}(1998)Starck, Murtagh, \& Bijaoui}]{starck1998image}
Starck, J., Murtagh, F., \& Bijaoui, A. 1998, Image Processing and Data
  Analysis: The Multiscale Approach (Cambridge University Press)

\bibitem[{{Starck} {et~al.}(2006){Starck}, {Pires}, \&
  {R{\'e}fr{\'e}gier}}]{Starck2006}
{Starck}, J.~L., {Pires}, S., \& {R{\'e}fr{\'e}gier}, A. 2006, \aap, 451, 1139

\bibitem[{{Tang} \& {Fan}(2005)}]{Tang2005}
{Tang}, J.~Y. \& {Fan}, Z.~H. 2005, \apj, 635, 60

\bibitem[{{Van Waerbeke}(1998)}]{Waerbeke1998}
{Van Waerbeke}, L. 1998, \aap, 334, 1

\bibitem[{{van Waerbeke}(2000)}]{VanWaerbeke2000}
{van Waerbeke}, L. 2000, \mnras, 313, 524

\bibitem[{{Vikhlinin} {et~al.}(2009){Vikhlinin}, {Kravtsov}, {Burenin},
  {Ebeling}, {Forman}, {Hornstrup}, {Jones}, {Murray}, {Nagai}, {Quintana}, \&
  {Voevodkin}}]{Vikhlinin2009}
{Vikhlinin}, A., {Kravtsov}, A.~V., {Burenin}, R.~A., {et~al.} 2009, \apj, 692,
  1060

\bibitem[{Voit(2005)}]{Voit2005}
Voit, G.~M. 2005, Reviews of Modern Physics, 77, 207–258

\bibitem[{{White} {et~al.}(2002){White}, {van Waerbeke}, \&
  {Mackey}}]{White2002}
{White}, M., {van Waerbeke}, L., \& {Mackey}, J. 2002, \apj, 575, 640

\bibitem[{{White} \& {Rees}(1978)}]{White1978}
{White}, S.~D.~M. \& {Rees}, M.~J. 1978, \mnras, 183, 341

\bibitem[{Wittman {et~al.}(2006)Wittman, Dell’Antonio, Hughes, Margoniner,
  Tyson, Cohen, \& Norman}]{Wittman2006}
Wittman, D., Dell’Antonio, I.~P., Hughes, J.~P., {et~al.} 2006, The
  Astrophysical Journal, 643, 128–143

\bibitem[{{Wright} \& {Brainerd}(2000)}]{wright00}
{Wright}, C.~O. \& {Brainerd}, T.~G. 2000, \apj, 534, 34

\end{thebibliography}
\end{document}